\documentclass[aps,pra,reprint,amsmath,amssymb,groupedaddress,showpacs]{revtex4-1}
\usepackage{graphicx}%Include figure files
\usepackage{float}
\usepackage{dcolumn} %Align table columns on decimal point
\usepackage{bm}      %bold math
\usepackage[bookmarks=false]{hyperref} %Hyperlinks
\usepackage[font={small}]{caption}     %FIXES FIGURE CAPTION PROBLEM
\usepackage{enumitem}%For setting up list indents
\usepackage{textcomp}%for good checks over rho in reduced rho
\hypersetup{
    unicode=false,          % non-Latin characters in Acrobat's bookmarks
    pdftoolbar=true,        % show Acrobat's toolbar?
    pdfmenubar=true,        % show Acrobat's menu?
    pdffitwindow=true,      % window fit to page when opened
    pdfstartview={},        % fits the width of the page to the window FitH
    pdftitle={Ent: A Multipartite Entanglement Measure, and Parameterization of Entangled States},                  % title
    pdfauthor={Samuel R. Hedemann},    % author
    pdfsubject={Quantum Entanglement}, % subject of the document
    pdfcreator={},                     % creator of the document
    pdfproducer={},                    % producer of the document
    pdfkeywords={Ent,} {TGX States,} {Theta States,} {13-Step Algorithm,} {Multipartite Entanglement,} {Quantum Entanglement,} {Entanglement Measure,} {Schmidt Decomposition},    % list of keywords
    pdfnewwindow=true,      % links in new window
    colorlinks=true,        % false: boxed links; true: colored links
    linkcolor=black,        % color of internal links (change box color with linkbordercolor)
    citecolor=black,        % color of links to bibliography
    filecolor=black,        % color of file links
    urlcolor=black          % color of external links
}%All links are clickable black text->least obtrusive, easy to find with mouse
\newcommand{\Sec}[1]{\hyperref[sec:#1]{Sec.{\kern 2pt}\ref*{sec:#1}}}
\newcommand{\Section}[1]{\hyperref[sec:#1]{Section~\ref*{sec:#1}}}
\newcommand{\Fig}[2][]{\hyperref[fig:#2]{Fig.{\kern 2pt}\ref*{fig:#2}#1}}%#1=a,b,etc. is the optional argument, in [] at call. Ex: \Fig[c]{2} = Fig. 2c
\newcommand{\Figure}[2][]{\hyperref[fig:#2]{Figure~\ref*{fig:#2}#1}}%#1=a,b,etc. is the optional argument, in [] at call. Ex: \Fig[c]{2} = Fig. 2c
\newcommand{\App}[1]{\hyperref[sec:#1]{App.{\kern 2pt}\ref*{sec:#1}}}
\newcommand{\Appendix}[1]{\hyperref[sec:#1]{Appendix~\ref*{sec:#1}}}
\newcommand{\Eq}[1]{\hyperref[eq:#1]{(\ref*{eq:#1})}}
\newcommand{\Eqs}[2]{\hyperref[eq:#1]{(\ref*{eq:#1}--\ref*{eq:#2})}}%Ex: \Eqs{1}{4}=(1--4)
\newcommand{\Table}[2][]{\hyperref[tab:#2]{Table~\ref*{tab:#2}#1}}
\newcommand{\mbar}[0]{\mathop {m}\limits^{{\kern -3.5pt}~_{\overline{{\kern 7pt}}}}}
\newcommand{\mbarsub}[0]{{\kern 0.0pt}\mathop {m}\limits^{{\kern -0.3pt}{\overline{{\kern 5.5pt}}}}{\kern 0.0pt}}
\newcommand{\redrho}[1]{{\rho}{\kern -5.3pt}{{\textnormal{\raisebox{-1.2pt}{\scalebox{1.2}{\textasciicaron}}}}}{\kern -4.2pt}{~}^{(#1)}}
\newcommand{\redrhotiny}[1]{{\rho}{\kern -4.5pt}{{\textnormal{\raisebox{-1.0pt}{\scalebox{1.2}{\textasciicaron}}}}}{\kern -4.2pt}{~}^{(#1)}}
\newcommand{\redP}{{P}{\kern -6.2pt}{{\textnormal{\raisebox{1.4pt}{\scalebox{1.2}{\textasciicaron}}}}}}
\newcommand{\vertsp}[1]{#1_{~_{~_{~_{~}}}}^{~^{~^{~^{~}}}}\!\!\!\!\!\!\!\!\!}
\newcommand{\topsp}[1]{#1^{~^{~^{~^{~}}}}\!\!\!\!\!\!\!\!\!}
\newcommand{\squareZero}{\framebox[12pt][c]{\scalebox{0.6}{\textcolor[rgb]{1,1,1}{$\blacksquare$}}}{\kern -13pt}\textcolor[rgb]{1,1,1}{\rule[6pt]{14pt}{2pt}}}
\newcommand{\squareOne}{\framebox[12pt][c]{\scalebox{0.6}{$\blacksquare$}}{\kern -13pt}\textcolor[rgb]{1,1,1}{\rule[6pt]{14pt}{2pt}}}
\newcommand{\squareTwo}{\framebox[12pt][c]{\scalebox{0.6}{$\blacksquare${\kern -7.7pt}\raisebox{12pt}{$\blacksquare$}}}{\kern -13pt}\textcolor[rgb]{1,1,1}{\rule[13.5pt]{14pt}{2pt}}}
\newcommand{\nmaxnot}{n_{\,\overline{{\kern -1.8pt}\max^{~^{~^{~}}}\!\!\!\!\!\!\!\!\!\!}}\,}
\newcommand{\oneoversqrttwoinline}[0]{\scalebox{0.8}{$\frac{1}{\rule{0pt}{7.5pt}\sqrt{{\kern 0.5pt}\rule{0pt}{5pt}2}}$}}
\newcommand{\LowerExponentOrSubScript}[2]{\raisebox{#1}{${\scriptstyle #2}$}}
\makeatletter
\newcommand{\mydotfill}{\leavevmode \cleaders \hb@xt@ .65em{\hss .\hss }\hfill \kern \z@}%.44em is standard for \dotfill
\makeatother
\begin{document}
%*******************************************************************************
%                                   TITLE
\title{Ent:~A~Multipartite~Entanglement~Measure,~and~Parameterization~of~Entangled~States}
%*******************************************************************************
%*******************************************************************************
%                                  BYLINES
\author{Samuel R. Hedemann}
\affiliation{P.O.{\kern 2.5pt}Box 72, Freeland, MD 21053, USA}
\date{\today}
%*******************************************************************************
%*******************************************************************************
\begin{abstract}%                 0. ABSTRACT
A multipartite entanglement measure called the ent is presented and shown to be an entanglement monotone, with the special property of automatic normalization.  Necessary and sufficient conditions are developed for constructing maximally entangled states in every multipartite system such that they are true-generalized X states (TGX) states, a generalization of the Bell states, and are extended to general nonTGX states as well.  These results are then used to prove the existence of maximally entangled basis (MEB) sets in all systems. A parameterization of general pure states of all ent values is given, and proposed as a multipartite Schmidt decomposition.  Finally, we develop an ent vector and ent array to handle more general definitions of multipartite entanglement, and the ent is extended to general mixed states, providing a general multipartite entanglement measure. 
\end{abstract}
%*******************************************************************************
%*******************************************************************************
%                             PACS & TITLE COMMAND
 %Entanglement,Nonlocality, EPR Paradox, Bell's Inequalities
\maketitle
%*******************************************************************************
%*******************************************************************************
%                              I. INTRODUCTION
\section{\label{sec:I}Introduction}
%_______________________________________________________________________________
\begin{figure}[H]%Not a figure; Puts hypertarget at top of column to fix problem
\centering
\vspace{-12pt}
\setlength{\unitlength}{0.01\linewidth}
\begin{picture}(100,0)
\put(1,25){\hypertarget{Sec:I}{}}
\end{picture}
\end{figure}
\vspace{-39pt}
%_______________________________________________________________________________
Coined by Schr{\"o}dinger \cite{Sch1}, \textit{entanglement} in pure quantum states is when two or more particles have multiple coincidence outcomes \textit{in superposition} that do not factor, so no particle has a definite pure state of its own, all exhibiting strong correlations with each other over time and space.  For example, for two qubits, each in a generic basis $\{|1\rangle,|2\rangle\}$ (our convention in this paper), \smash{$|\Phi^{+}\rangle\!=\!\oneoversqrttwoinline(|1\rangle\otimes|1\rangle\!+\!|2\rangle\otimes|2\rangle)$} is a maximally entangled state, whereas $\rule{0pt}{10pt}|\psi\rangle\!=\!ac|1\rangle\otimes|1\rangle\!+\!ad|1\rangle\otimes|2\rangle\!+\!bc|2\rangle\otimes|1\rangle\!+\!bd|2\rangle\otimes|2\rangle$ is separable because it can be factored as $|\psi\rangle\!=\!(a|1\rangle\!+\!{\kern 1pt}b|2\rangle)\otimes(c|1\rangle\!+\!{\kern 1pt}d|2\rangle)$, so each qubit has its own pure state, $|\psi^{\LowerExponentOrSubScript{-1.5pt}{(1)}}\rangle\!=\!a|1\rangle\!+\!{\kern 1pt}b|2\rangle$ and $|\psi^{\LowerExponentOrSubScript{-1.5pt}{(2)}}\rangle\!=\!c|1\rangle\!+\!{\kern 1pt}d|2\rangle$.

Einstein, Podolsky, and Rosen \cite{EPR1} acknowledged the strangeness of correlations so strong that individual identities depend on the joint state, asking if that meant that quantum mechanics itself were incomplete, suggesting the need for \textit{hidden variables}. Later, Bell \cite{Bell} showed that even if hidden variables were used, they would need to be \textit{nonlocal} in general, and thus apparently the nonlocality of entanglement is indeed a part of our reality.

As the quest for quantum computation \cite{Feyn,DiVi} has intensified, entanglement has been identified as an important resource for many tasks where a quantum system could outperform its classical counterpart. Furthermore, entanglement has novel applications like quantum teleportation \cite{BBCJ,BPM1,BPM2}, and increasing evidence shows that entanglement can play a significant role in many biological \cite{ScSw,GuCa} and chemical processes \cite{DTBB}.  Therefore, there is a practical need to quantify \textit{how much} entanglement a system has, so that we can both understand it and determine which states have the most of this resource.  This is the purpose of \textit{entanglement measures} \cite{PlVi}.

The more general problem of quantifying entanglement of \textit{mixed states} $\rho\equiv\sum\nolimits_{j}p_{j}|\psi_j\rangle\langle\psi_j|;\;p_j\in[0,1];\;\sum\nolimits_{j}p_{j}=1$ is much harder, generally requiring nonlinear optimization.  In fact, the only computable measures of mixed-state entanglement, such as \cite{HiWo,Woot} and \cite{Pere,ViWe}, are for $2\times 2$ and $2\times 3$ (where $n_1 \times\cdots\times n_N$ means an $N$-partite system where subsystem (mode) $m$ has $n_m\equiv\text{dim}(\mathcal{H}^{(m)})$ levels where the Hilbert space of $\rho$ is $\mathcal{H}=\mathcal{H}^{(1)}\otimes\cdots\otimes\mathcal{H}^{(N)}$).  Thus, an entanglement measure that can cope with mixed states \textit{and} multipartite states is needed.

As generalized by several authors based on Werner's definition \cite{Wern,ZHSL,PlVi}, mixed states expressible as
%===============================================================================
\begin{equation}%                  Equation 1
\rho  = \sum\nolimits_j {p_j \rho _j^{(1)}  \otimes  \cdots  \otimes \rho _j^{(N)} } ,
\label{eq:1}
\end{equation}
%===============================================================================
are \textit{separable} (``classically correlated''), and any state that does not admit such a decomposition is \textit{entangled}, where{\kern 1.5pt} \smash{$\rho _{\LowerExponentOrSubScript{-2pt}{j}}\rule{0pt}{8pt}^{{\kern -3.5pt}\LowerExponentOrSubScript{-2pt}{(m)}}\!\equiv|\psi^{(m)}\rangle\langle \psi^{(m)}| $} is a pure state of mode $m$.

In{\kern -1.5pt} this paper, we focus on the Werner-inspired definition, calling it $N$-partite separability and using $N$-partite entanglement as our primary definition of \textit{multipartite entanglement}.  Furthermore, we focus on \textit{pure}-state entanglement, though we give examples of how to approximately handle mixed states as well.  We then treat more general views of multipartite entanglement and compare our measure to existing measures.  The main sections are
%_______________________________________________________________________________
\begin{table}[H]
\begin{tabular}[b]{l p{0.83\linewidth}@{}p{0.05\linewidth}}%0.84
\hyperlink{Sec:I}{\textbf{I.}}&\hyperlink{Sec:I}{\textbf{Introduction}}\mydotfill\!\! &\hspace{\stretch{1}}\hyperlink{Sec:I}{\textbf{\pageref*{sec:I}}}\\
\hyperlink{Sec:II}{\textbf{II.}}&\hyperlink{Sec:II}{\textbf{The Ent for Pure States}}\mydotfill\!\! &\hspace{\stretch{1}}\hyperlink{Sec:II}{\textbf{\pageref*{sec:II}}}\\
\hyperlink{Sec:III}{\textbf{III.}}&\hyperlink{Sec:III}{\textbf{Construction of Maximally Entangled States in All Systems}}\mydotfill\!\! &\hspace{\stretch{1}}\raisebox{-10.8pt}{\hyperlink{Sec:III}{\textbf{\pageref*{sec:III}}}}\\
\hyperlink{Sec:IV}{\textbf{IV.}}&\hyperlink{Sec:IV}{\textbf{Maximally Entangled Basis Theorem}}\mydotfill\!\! &\hspace{\stretch{1}}\hyperlink{Sec:IV}{\textbf{\pageref*{sec:IV}}}\\
\hyperlink{Sec:V}{\textbf{V.}}&\hyperlink{Sec:V}{\textbf{$\theta$ States: Pure States of Any Entanglement}}\mydotfill\!\! &\hspace{\stretch{1}}\hyperlink{Sec:V}{\textbf{\pageref*{sec:V}}}\\
\hyperlink{Sec:VI}{\textbf{VI.}}&\hyperlink{Sec:VI}{\textbf{Ent{\kern -1pt} Vector,{\kern -1pt} Ent{\kern -1pt} Array,{\kern -1pt} and{\kern -1pt} Mixed-State{\kern -1pt} Ent}}\mydotfill\!\! &\hspace{\stretch{1}}{\hyperlink{Sec:VI}{\textbf{\pageref*{sec:VI}}}}\\
\hyperlink{Sec:VII}{\textbf{VII.}}&\hyperlink{Sec:VII}{\textbf{Conclusions}}\mydotfill\!\! &\hspace{\stretch{1}}\hyperlink{Sec:VII}{\textbf{\pageref*{sec:VII}}}\\
\hyperlink{Sec:VIII}{\textbf{App.}}&\hyperlink{Sec:VIII}{\textbf{Appendices}}\mydotfill\!\! &\hspace{\stretch{1}}\hyperlink{Sec:VIII}{\textbf{\pageref*{sec:App.A}}}\\
\hyperlink{Sec:App.A}{{\kern 7pt}\textbf{A.}}&\hyperlink{Sec:App.A}{\textbf{Identifying the Relevant Reductions}}\mydotfill\!\! &\hspace{\stretch{1}}{\hyperlink{Sec:App.A}{\textbf{\pageref*{sec:App.A}}}}\\
\hyperlink{Sec:App.B}{{\kern 7pt}\textbf{B.}}&\hyperlink{Sec:App.B}{\textbf{Review of Reduced States}}\mydotfill\!\! &\hspace{\stretch{1}}{\hyperlink{Sec:App.B}{\textbf{\pageref*{sec:App.B}}}}\\
\hyperlink{Sec:App.C}{{\kern 7pt}\textbf{C.}}&\hyperlink{Sec:App.C}{\textbf{Proof: Ent is an Entanglement Monotone}}\mydotfill\!\! &\hspace{\stretch{1}}{\hyperlink{Sec:App.C}{\textbf{\pageref*{sec:App.C}}}}\\
\hyperlink{Sec:App.D}{{\kern 7pt}\textbf{D.}}&\hyperlink{Sec:App.D}{\textbf{Derivation of Conditions for Maximal Entanglement and Normalization of the Ent}}\mydotfill\!\! &\hspace{\stretch{1}}\raisebox{-10.8pt}{\hyperlink{Sec:App.D}{\textbf{\pageref*{sec:App.D}}}}\\
\hyperlink{Sec:App.E}{{\kern 7pt}\textbf{E.}}&\hyperlink{Sec:App.E}{\textbf{Ent for Two-Mode Squeezed States}}\mydotfill\!\! &\hspace{\stretch{1}}{\hyperlink{Sec:App.E}{\textbf{\pageref*{sec:App.E}}}}\\
\hyperlink{Sec:App.F}{{\kern 7pt}\textbf{F.}}&\hyperlink{Sec:App.F}{\textbf{Application: Ent Provides a Gauge for Logarithmic Negativity}}\mydotfill\!\! &\hspace{\stretch{1}}\raisebox{-10.8pt}{\hyperlink{Sec:App.F}{\textbf{\pageref*{sec:App.F}}}}\\
\hyperlink{Sec:App.G}{{\kern 7pt}\textbf{G.}}&\hyperlink{Sec:App.G}{\textbf{The 13-Step Algorithm}}\mydotfill\!\! &\hspace{\stretch{1}}{\hyperlink{Sec:App.G}{\textbf{\pageref*{sec:App.G}}}}\\
\hyperlink{Sec:App.H}{{\kern 7pt}\textbf{H.}}&\hyperlink{Sec:App.H}{\textbf{Maximally{\kern -1pt} Entangled{\kern -1pt} TGX{\kern -1pt} State{\kern -1pt} Examples}}\mydotfill\!\! &\hspace{\stretch{1}}{\hyperlink{Sec:App.H}{\textbf{\pageref*{sec:App.H}}}}\\
\hyperlink{Sec:App.I}{{\kern 7pt}\textbf{I.}}&\hyperlink{Sec:App.I}{\textbf{Schmidt Decomposition and Reversal}}\mydotfill\!\! &\hspace{\stretch{1}}{\hyperlink{Sec:App.I}{\textbf{\pageref*{sec:App.I}}}}\\
\hyperlink{Sec:App.J}{{\kern 7pt}\textbf{J.}}&\hyperlink{Sec:App.J}{\textbf{Decomposition Freedom of {\normalsize $\rho$}}}\mydotfill\!\! &\hspace{\stretch{1}}{\hyperlink{Sec:App.J}{\textbf{\pageref*{sec:App.J}}}}\\
\end{tabular}{\kern -10pt}
\end{table}{\kern -10pt}
%_______________________________________________________________________________
{\noindent}In general, all derivations and details are in the \hyperlink{Sec:VIII}{Appendices} to keep the presentation of main results compact.
%                                  END of I
%*******************************************************************************
%*******************************************************************************
%                        II. The Ent for Pure States
\section{\label{sec:II}The Ent for Pure States}
%_______________________________________________________________________________
\begin{figure}[H]%Not a figure; Puts hypertarget at top of column to fix problem
\centering
\vspace{-12pt}
\setlength{\unitlength}{0.01\linewidth}
\begin{picture}(100,0)
\put(1,25){\hypertarget{Sec:II}{}}
\end{picture}
\end{figure}
\vspace{-39pt}
%_______________________________________________________________________________
%-------------------------------------------------------------------------------
%   II.A. Pure-State Ent for Discrete, Finite-Dimensional $N$-Body Systems
\subsection{\label{sec:II.A}Pure-State Ent for Discrete, Finite-Dimensional $N$-Body Systems}
For an $N$-partite system, meaning a system of $N$ subsystems (modes), each with no internal coincidences (see \App{App.A}), and each of possibly different size, where mode $m$ has $n_m$ levels and the total system has $n = n_1 \cdots n_N$ levels, a measure for $N$-partite entanglement of pure-state input $\rho$ is the \textit{ent} \cite{HedD}, defined as
%===============================================================================
\begin{equation}%                  Equation 2
\Upsilon (\rho ) \equiv \frac{1}{M}\left( {1 - \frac{1}{N}\sum\limits_{m = 1}^N {\frac{{n_m P(\redrho{m}) - 1}}{{n_m  - 1}}} } \right)\!,
\label{eq:2}
\end{equation}
%===============================================================================
where $P(\sigma ) \equiv \text{tr}(\sigma ^2 )$ is the purity of state $\sigma$, $\redrho{m}$ is the $n_m$-level single-mode reduction of $\rho$ for mode $m$ (see \App{App.B}), and the proper normalization factor is
%===============================================================================
\begin{equation}%                  Equation 3
M\equiv M(L_{*}) \equiv 1 - \frac{1}{N}\sum\limits_{m = 1}^N {\frac{{n_m P_{\text{MP}}^{(m)}(L_{*}) - 1}}{{n_m  - 1}}},
\label{eq:3}
\end{equation}
%===============================================================================
where, given a pure parent state $\rho$, $P_{\text{MP}}^{(m)}  \equiv P_{\text{MP}}^{(m)} (L_* ) \equiv \min(P (\redrho{m}))$ is the minimum physical purity of $\redrho{m}$,
%===============================================================================
\begin{equation}%                  Equation 4
\begin{array}{*{20}l}
   {P_{\text{MP}}^{(m)} (L_* ) = } &\!\! {\bmod (L_* ,n_m )\left( {\frac{{1 + \text{floor}(L_* /n_m )}}{{L_* }}} \right)^2 }  \\
   {} &\!\! { + (n_m  - \bmod (L_* ,n_m ))\left( {\frac{{\text{floor}(L_* /n_m )}}{{L_* }}} \right)^2 ,}  \\
\end{array}
\label{eq:4}
\end{equation}
%===============================================================================
where $\bmod (a,b) \equiv a - \text{floor}(a/b)b$, and $L_*$ is any number of levels of $\rho$ with equal nonzero probabilities that can support maximal entanglement, and by convention we use the smallest of these in \Eq{4}, as $L_*  = \min \{ \mathbf{L}_* \}$, where $\mathbf{L}_*  \equiv \{ L_* \}$ is the list of values of $L$ that satisfy
%===============================================================================
\begin{equation}%                  Equation 5
\mathop {\min }\limits_{L \in 2, \ldots, \nmaxnot } (1 - M(L)),
\label{eq:5}
\end{equation}
%===============================================================================
where \smash{$\nmaxnot \equiv \frac{n}{{n_{\max } }}$} is the product of all $n_m$ \textit{except} $n_{\max }$, where \smash{$n_{\max } \equiv \max (\mathbf{n})$} and \smash{$\mathbf{n} \equiv (n_1 , \ldots ,n_N )$}.  See \Eq{G.2} and \Eq{H.3} for examples of how to compute $L_*$.

Note that \Eq{2} is \textit{automatically normalized}; $M$ does not require a maximally entangled state.  Ironically, this led to the discovery of a method for constructing maximally entangled states in all multipartite systems, one of the major results of this paper, presented in \Sec{III}.

\begin{figure}[H]%         FIGURE 1 (made with entPaperFig1.m)
\centering
\includegraphics[width=1.00\linewidth]{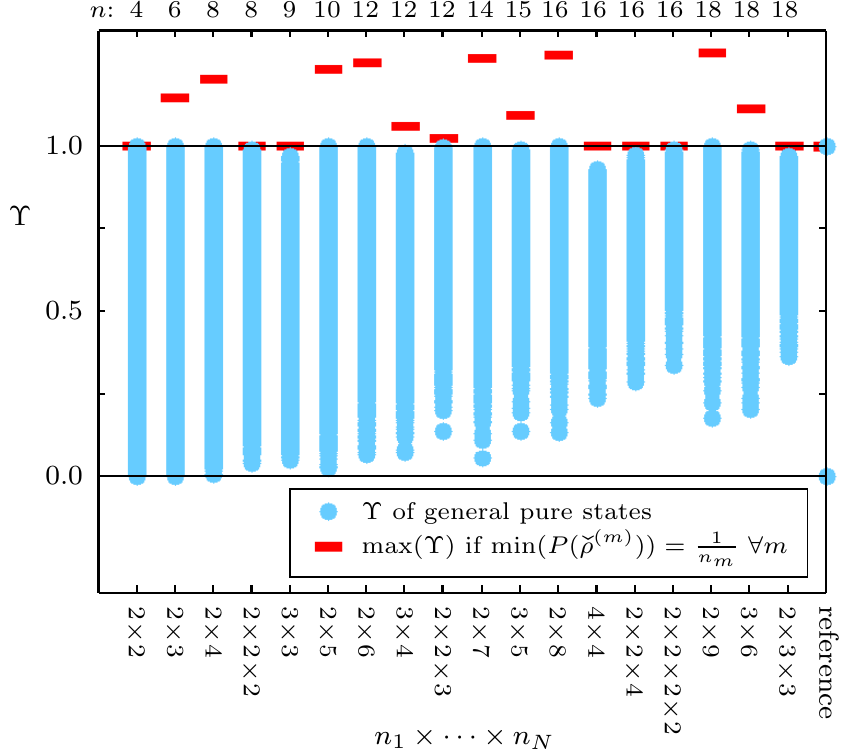}%
\vspace{-6pt}
\caption[]{(color online) Normalization test of \Eq{2}, the ent $\Upsilon$ of arbitrary general quantum pure states $\rho$ for all discrete multipartite systems up to $n=18$ levels, $30,000$ states for each system. This shows a total of $540,000$ \textit{consecutive} examples that do not produce states with higher ent than $1$, providing strong evidence that $\Upsilon$ is properly normalized. (For \textit{proof} that the ent itself is necessary and sufficient to measure entanglement, see \App{App.C} and \App{App.D}.) Red bars show the maximum ent if{\kern 1pt} \smash{$\min(P(\redrho{m})){\kern 1pt}={\kern 1pt}\frac{1}{n_{m}}\;\forall m$} were physical instead of \smash{$P_{\text{MP}}^{(m)} (L_* )$} from{\kern -1.5pt} \Eq{4}{\kern -1.5pt} (see{\kern -1.5pt} \App{App.D}). If $\Upsilon$ were improperly normalized, blue dots would be able to reach red bars above $1$. The lack of points near the separable bottom as $n$ grows is due to the naturally higher density of entangled states in larger systems.  See \Fig{5} for examples of entanglement in these regions.}
\label{fig:1}
\end{figure}

While \App{App.C} and \App{App.D} prove \Eqs{2}{5} explicitly, \Fig{1} gives a numerical demonstration to show that \Eq{2} is correctly normalized by showing $540,000$ \textit{consecutive} examples (no trials discarded) where no ent values exceed $1$.

\textit{The physics of the ent} is that maximal $N$-partite entanglement is when all $N$ reduced states are the most mixed they can be given a pure parent state.

However, there is more to multipartite entanglement than just $N$-partite correlations.  To see all possible entanglement resources of a state, we need an \textit{entanglement vector} listing all multi-body entanglement.  The ent is a simple tool to achieve this, as we will see in \Sec{VI}.
%_______________________________________________________________________________
%_______________________________________________________________________________
\vspace{-15pt}
%                                 End of II.A
%-------------------------------------------------------------------------------
%-------------------------------------------------------------------------------
%   II.B. Pure-State Ent for Discrete, Infinite-Dimensional $N$-Body Systems 
\subsection{\label{sec:II.B}Pure-State Ent for Discrete, Infinite-Dimensional $N$-Body Systems}
For $N$-partite systems where $n_m =\infty\;\forall m$, the ent is
%===============================================================================
\begin{equation}%                  Equation 6
\Upsilon (\rho ) = 1 - \frac{1}{N}\sum\limits_{m = 1}^N {P(\redrho{m})} ,
\label{eq:6}
\end{equation}
%===============================================================================
where, since each $\redrho{m}$ has infinite levels, $P(\redrho{m})\!\in\![0,1]$.\vspace{-7pt}
%...............................................................................
%              II.B.1 Example: Ent for Two-Mode Squeezed States
\subsubsection*{\label{sec:II.B.1}Example: Ent for Two-Mode Squeezed States}
From \cite{Cave,Milb,BaKn,CaSc,ScCa,GeKn}, the two-mode squeezed vacuum state is
%===============================================================================
\begin{equation}%                  Equation 7
|\xi \rangle _2  \equiv S_2 (\xi )|0,0\rangle  = {\textstyle{1 \over {\cosh (r)}}}\sum\limits_{n = 0}^\infty  {( - 1)^n e^{in\theta } \tanh ^n (r)|n,n\rangle },
\label{eq:7}
\end{equation}
%===============================================================================
where $|n,n\rangle  \equiv |n\rangle  \otimes |n\rangle$ and $|n\rangle$ are Fock states \cite{Dira}, and \smash{$S_2 (\xi )\! \equiv\!{\kern 0.5pt} e^{\xi ^* a_1 a_2  - \xi a_1 ^{\dag}  a_2 ^{\dag}  }$} is the unitary two-mode squeezing operator where \smash{$\xi  \equiv re^{i\theta }$}, with \smash{$r \in [0,\infty )$}, \smash{$\theta \in [0,2\pi )$}, and $a_m$ is the mode-$m$ annihilation operator.

\begin{figure}[H]%         FIGURE 2 (made with entPaperFig2.m)
\centering
\includegraphics[width=1.00\linewidth]{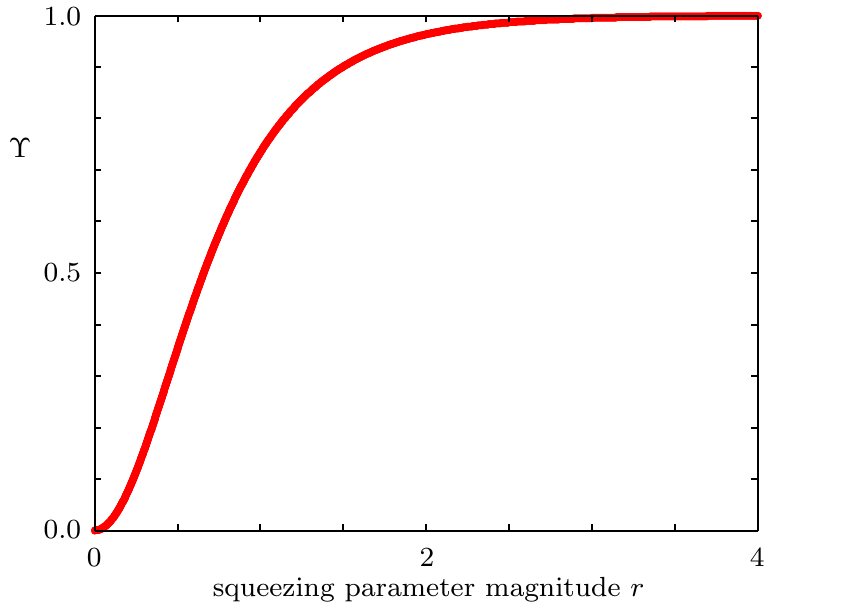}%
\vspace{-3pt}
\caption[]{(color online) Plot of \Eq{8}, the ent $\Upsilon\equiv\Upsilon (\rho _{|\xi \rangle _2 } )$ for a two-mode squeezed vacuum state $|\xi \rangle _2$. When squeezing parameter $r=0$, $|\xi \rangle _2$ is separable and $\Upsilon=0$.  When $r=\infty$, $|\xi \rangle _2$ is maximally entangled and $\Upsilon=1$. The plot shows that \textit{near}-maximal entanglement happens starting around $r\approx 3$.}
\label{fig:2}
\end{figure}

As shown in \App{App.E}, the ent of $|\xi \rangle _2$ is
%===============================================================================
\begin{equation}%                  Equation 8
\Upsilon (\rho _{|\xi \rangle _2 } ) = 1 - {\textstyle{1 \over {\rule{0pt}{7pt}2\cosh ^2 (r) - 1}}},
\label{eq:8}
\end{equation}
%===============================================================================
where $\rho _{|\xi \rangle _2 }  \equiv |\xi \rangle _2 {}_2\langle \xi |$, and which is plotted in \Fig{2}.
%_______________________________________________________________________________
%_______________________________________________________________________________

See \App{App.F} for an application of the ent as a gauge for infinite-range measures such as logarithmic negativity.
%                                End of II.B.1
%...............................................................................
%                                 End of II.B
%-------------------------------------------------------------------------------
%                                  END of II
%*******************************************************************************
%*******************************************************************************
%       III. CONSTRUCTION OF MAXIMALLY ENTANGLED STATES IN ALL SYSTEMS
\section{\label{sec:III}CONSTRUCTION OF MAXIMALLY ENTANGLED STATES IN ALL SYSTEMS}
%_______________________________________________________________________________
\begin{figure}[H]%Not a figure; Puts hypertarget at top of column to fix problem
\centering
\vspace{-12pt}
\setlength{\unitlength}{0.01\linewidth}
\begin{picture}(100,0)
\put(1,30){\hypertarget{Sec:III}{}}
\end{picture}
\end{figure}
\vspace{-39pt}
%_______________________________________________________________________________
Here, we merely \textit{summarize} multipartite maximally entangled state construction; see \App{App.D} for derivations.
%-------------------------------------------------------------------------------
%  III.A. Construction of Maixmally Entanlged TGX States: The 13-Step Algorithm
\subsection{\label{sec:III.A}Construction of Maximally Entangled TGX States: The 13-Step Algorithm}
As defined in \cite{HedX} and \App{App.D.1}, a TGX state is a state for which all parent-state matrix elements appearing in the off-diagonals of all the reductions are identically zero. For example, Bell states are TGX states.  TGX states never have $n$-level superposition, but nonTGX states can.

The 13-step algorithm \smash{$\mathcal{A}_{13}$} can be represented as
%===============================================================================
\begin{equation}%                  Equation 9
\mathcal{A}_{13} (\mathbf{n},S_L ) = \{ |\Phi _j \rangle \},
\label{eq:9}
\end{equation}
%===============================================================================
where \smash{$S_L \in 1,\ldots,n$} is a \textit{starting level}, meaning any level to definitely be in each output state, and \smash{$\{ |\Phi _j \rangle \}$} are the maximally entangled TGX states generated by \smash{$\mathcal{A}_{13}$}. The \textit{explicit steps} of \smash{$\mathcal{A}_{13}$} are listed in \App{App.G}.

\begin{figure}[H]%         FIGURE 3 (made with entPaperFig3.m)
\centering
\includegraphics[width=1.00\linewidth]{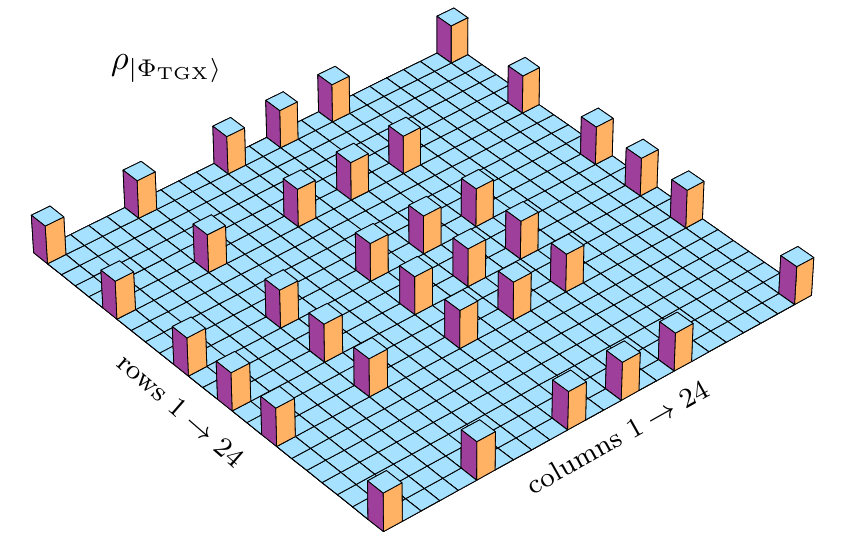}%
\vspace{-3pt}
\caption[]{(color online) Plot of the density matrix of one maximally entangled TGX state $\rho_{|\Phi_{\text{TGX}}\rangle}$ for $2\times 3\times 4$ as produced by \smash{$\mathcal{A}_{13}$}, with starting level $S_{L}=1$. This state has reduction purities \smash{$P(\redrho{1})=P_{\text{MP}}^{(1)}=\frac{1}{2}$}, \smash{$P(\redrho{2})=P_{\text{MP}}^{(2)}=\frac{1}{3}$}, and \smash{$P(\redrho{3})=P_{\text{MP}}^{(3)}=\frac{5}{18}$}, which is the lowest \textit{combination} of reduction purities physically possible for pure states in $2\times 3\times 4$.}
\label{fig:3}
\end{figure}

To illustrate \smash{$\mathcal{A}_{13}$}, \Fig{3} plots one example state, and \Table{1} gives all possible output sets for a few systems, where, for example, in $2\times 2\times 2$, the first four sets are GHZ states \cite{GHZ,GHSZ,Merm} such as $\{1,8\}$ which represents $|\Phi ^ +  \rangle \! =$ \smash{$ \!{\textstyle{1 \over {\sqrt 2 }}}(|1\rangle \! +\! |8\rangle ) \!=\! {\textstyle{1 \over {\sqrt 2 }}}(|1,1,1\rangle \! +\! |2,2,2\rangle )$}, where ket labels start on $1$ and are not Fock states, $|a,b\rangle\equiv|a\rangle |b\rangle \equiv|a\rangle \otimes |b\rangle $, and \Eq{H.1} converts to the coincidence basis.
%_______________________________________________________________________________
%_______________________________________________________________________________
%_______________________________________________________________________________
%                                    TABLE 1
\begin{table}[H]
\caption{\label{tab:1}Sample of results from the 13-step algorithm showing the unique verified maximally entangled TGX sets \smash{$\{L_{\text{ME}}\}_{u}$} generated from all starting levels $S_L$.  For example, the first set $\{1,4\}$ is to be read as \smash{$|\Phi _1 \rangle  = {\textstyle{1 \over {\sqrt 2 }}}(|1\rangle  + |4\rangle )$} as in \Eq{G.21}. Continued in \Table{4} in \App{App.H.1}, with other tables.}
\begin{ruledtabular}
\begin{tabular}{|c|c|c|}
%Column Titles:
$n$ & $n_{1}\times\cdots\times n_{N}$& $\{L_{\text{ME}}\}_{u}$\\[0.5mm]
\hline 
%Table Elements:
$4_{~_{~_{~_{~}}}}^{~^{~^{~^{~}}}}\!\!\!\!\!\!\!\!\!$ & $2\times 2$ & $\{1,4\}, \{2,3\}\,\,$ \\
\hline 
$6_{~_{~_{~_{~}}}}^{~^{~^{~^{~}}}}\!\!\!\!\!\!\!\!\!$ & $2\times 3$ & $\begin{array}{*{20}l}
   {\{1,5\}, \{1,6\}, \{2,4\}, \{2,6\}, \{3,4\}, \{3,5\}\,\,}  \\
\end{array}$ \\
\hline
$\begin{array}{*{20}c}
   {\topsp{8}}  \\
   {}  \\
\end{array}$ & $\begin{array}{*{20}c}
   {\topsp{2\times 4}}  \\
   {}  \\
\end{array}$ & $\begin{array}{*{20}l}
   {\{1,6\}, \{1,7\}, \{1,8\}, \{2,5\}, \{2,7\}, \{2,8\},}  \\
   {\{3,5\}, \{3,6\}, \{3,8\}, \{4,5\}, \{4,6\}, \{4,7\} }  \\
\end{array}$ \\
\hline
$\begin{array}{*{20}c}
   {\topsp{8}}  \\
   {}  \\
\end{array}$ & $\begin{array}{*{20}c}
   {\topsp{2\times 2\times 2}}  \\
   {}  \\
\end{array}$ & $\begin{array}{*{20}l}
   {\{1,8\}, \{2,7\}, \{3,6\}, \{4,5\}, \{1,4,6,7\},}  \\
   {\{2,3,5,8\} }  \\
\end{array}$\,\,\,\,\,\, \\
\end{tabular}
\end{ruledtabular}
\end{table}
%_______________________________________________________________________________
%                                 End of III.A
%-------------------------------------------------------------------------------
%-------------------------------------------------------------------------------
% III.B. Special Case: Construction of Maximally Entangled States in Systems of Equal Mode Sizes
\subsection{\label{sec:III.B}Special Case: Construction of Maximally Entangled States in Systems of Equal Mode Sizes}
For multiqu$d$it systems, where all $N$ modes have $d$ levels so that $n_1  =  \cdots  = n_N  = d$ and $n = n_1  \cdots n_N  = d^N$, the symmetry permits a simple closed form for maximally entangled TGX states with the \textit{fewest levels} as
%===============================================================================
\begin{equation}%                  Equation 10
\begin{array}{*{20}l}
   {|\Phi \rangle} &\!\!\! {\equiv\! \frac{1}{{\sqrt d }}\sum\limits_{k = 1}^d {|1+(k - 1){\textstyle{{n - 1} \over {d - 1}}}\rangle }}  \\
   {} &\!\!\! {=\!{\textstyle{1 \over {\sqrt d }}}(|1^{(1)} \rangle \! \otimes  \cdots  \otimes \! |1^{(N)} \rangle  \!+{\kern -2.5pt}\cdots{\kern -2.5pt}+\! |d^{(1)} \rangle \! \otimes  \cdots  \otimes \! |d^{(N)} \rangle \!{\kern 1pt}).}  \\
\end{array}\!\!
\label{eq:10}
\end{equation}
%===============================================================================

As seen in \Table{2} in \App{App.H.1}, multiqudit systems in TGX states can have any integer multiples of $d$ nonzero levels, from $d$ to $d^{N-1}$, so \Eq{10} is just the $d$-level case, and it is generally not the only possible $d$-level maximally entangled TGX state in a given system.

Thus, \Eq{10} is a simple way to generate a single example for $N$-qudit systems without the 13-step algorithm.
%                                End of III.B
%-------------------------------------------------------------------------------
%-------------------------------------------------------------------------------
%     III.C. Making Maximally Entangled nonTGX States with Multipartite Reverse-Schmidt Decomposition
\subsection{\label{sec:III.C}Making Maximally Entangled nonTGX States with Multipartite Reverse-Schmidt Decomposition}
Here, we apply entanglement-preserving unitary (EPU) operators to maximally entangled TGX states $|\Phi _{\text{TGX}} \rangle$ to reach all general nonTGX maximally entangled states $|\Phi _G \rangle$, because all pure states are unitarily equivalent, so entanglement-specific pure states are as well.  Thus, starting with some phaseless $|\Phi _{\text{TGX}} \rangle$ as the \textit{core} state, such as the output of \smash{$\mathcal{A}_{13}$} from \Eq{9}, we obtain
%===============================================================================
\begin{equation}%                  Equation 11
|\Phi _G \rangle  \equiv U_{\text{EPU}}|\Phi _{\text{TGX}} \rangle,
\label{eq:11}
\end{equation}
%===============================================================================
where $U_{\text{EPU}}$ is an EPU operator of hypothetical form
%===============================================================================
\begin{equation}%                  Equation 12
U_{\text{EPU}}\equiv (U^{(1)}  \otimes  \cdots  \otimes U^{(N)} )D,
\label{eq:12}
\end{equation}
%===============================================================================
where $D \equiv \text{diag}\{ e^{i\eta _1 } ,e^{i\eta _2 } , \ldots ,e^{i\eta _n }\}$ with independent real phase angles $\eta_k$, and $U^{(m)}$ are independent $n_m$-level unitaries.  We include $D$ in \Eq{12} because maximally entangled TGX states can have arbitrary relative phases. However, $D$ is generally \textit{not} a tensor product, so it is \textit{nonlocal}, so $U_{\text{EPU}}$ is also (see App.{\kern 2.5pt}A.1 in \cite{HedX} for a proof).

\begin{figure}[H]%         FIGURE 4 (made with entPaperFig4.m)
\centering
\includegraphics[width=1.00\linewidth]{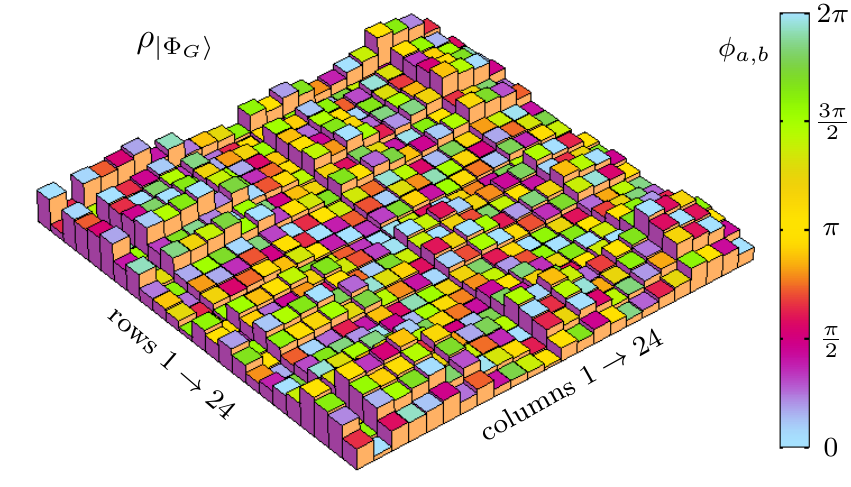}%
\vspace{-4pt}
\caption[]{(color online) Plot of a maximally entangled \textit{nonTGX} density matrix in $2\times 3\times 4$ using the multipartite reverse-Schmidt decomposition of \Eq{11}. Bar heights indicate matrix-element magnitudes $|\rho_{a,b}|$ and the color on each bar top shows phase angle $\phi_{a,b}\equiv\text{arg}(\rho_{a,b})$. Phase angles of upper-triangular matrix elements $\rho_{a,b>a}$ are the negative of the phase angle of their corresponding lower-triangular elements $\rho_{b,a}$ but are shown with the same color to make the pattern easier to see. As in \Fig{3}, reduction purities are their lowest simultaneous values of \smash{$P(\redrho{1})\!=\!\frac{1}{2}$}, \smash{$P(\redrho{2})\!=\!\frac{1}{3}$}, and \smash{$P(\redrho{3})\!=\!\frac{5}{18}$}.}
\label{fig:4}
\end{figure}

\Figure{4} shows one example of \Eq{11}, which we call the \textit{multipartite reverse-Schmidt decomposition} (MRSD), discussed more generally in \Sec{V.B}.
%_______________________________________________________________________________
%_______________________________________________________________________________
The form of $U_{\text{EPU}}$ in \Eq{12} only gives a \textit{hypothetical} multipartite Schmidt decomposition in \Eq{11} because a more general $U_{\text{EPU}}$ may exist.  However, in $2\!\times\! 2$, we now show that using \Eq{12} in \Eq{11} \textit{is} the Schmidt decomposition \cite{Schm}.

To compare methods, the usual Schmidt decomposition in reverse (see \App{App.I}) gives
%===============================================================================
\begin{equation}%                  Equation 13
|\psi \rangle  = {\textstyle{1 \over {\sqrt 2 }}}\!\left(\! {\begin{array}{*{20}r}
   {a_1 a_2  - b_1 b_2 ^* }  \\
   {a_1 b_2  + b_1 a_2 ^* }  \\
   { - b_1 ^* a_2  - a_1 ^* b_2 ^* }  \\
   { - b_1 ^* b_2  + a_1 ^* a_2 ^* }  \\
\end{array}} \right)\!,
\label{eq:13}
\end{equation}
%===============================================================================
which is maximally entangled for all parameter values where \smash{$|a_1 |^2  + |b_1 |^2  = 1$}, \smash{$|a_2 |^2  + |b_2 |^2  = 1$}.  Yet, applying \Eq{12} to Bell state \smash{$|\Phi _{\text{TGX}} \rangle  \equiv {\textstyle{1 \over {\sqrt 2 }}}(|1,1\rangle  + |2,2\rangle )$} yields
%===============================================================================
\begin{equation}%                  Equation 14
|\Phi _G \rangle\!  =\! {\textstyle{e^{i\eta_{1}} \over {\sqrt 2 }}}\!{\kern -0.5pt}\left(\!\! {\begin{array}{*{20}r}
   {c_1 c_2  - d_1 d_2 ^* e^{i\eta } }  \\
   {c_1 d_2  + d_1 c_2 ^* e^{i\eta } }  \\
   { - d_1 ^* c_2  - c_1 ^* d_2 ^* e^{i\eta } }  \\
   { - d_1 ^* d_2  + c_1 ^* c_2 ^* e^{i\eta } }  \\
\end{array}}\! \right)\!=\!{\textstyle{e^{i\phi} \over {\sqrt 2 }}}\!{\kern -0.5pt}\left(\!\! {\begin{array}{*{20}r}
   {a_1 a_2  - b_1 b_2 ^* }  \\
   {a_1 b_2  + b_1 a_2 ^* }  \\
   { - b_1 ^* a_2  - a_1 ^* b_2 ^* }  \\
   { - b_1 ^* b_2  + a_1 ^* a_2 ^* }  \\
\end{array}}\! \right)\!,
\label{eq:14}
\end{equation}
%===============================================================================
agreeing with \Eq{13} exactly, up to a global phase involving \smash{$\phi\equiv\eta_{1}+\frac{\eta}{2}$} where $\eta\equiv\eta_4 -\eta_1$ and \smash{$|c_1 |^2 \! +\! |d_1 |^2 \! =\! 1$}, \smash{$|c_2 |^2 \! +\! |d_2 |^2  \! =\! 1$}. Crucially, $\{a_{1},b_{1},a_{2},b_{2}\}$ are independent of $\{\eta_{k}\}$ (see \App{App.I}). Thus, \Eq{11} and \Eq{12} are \textit{equivalent} to the Schmidt decomposition up to global phase.

It may seem like we can just absorb $D$ into the local-unitary $U^{(1)}  \otimes  \cdots  \otimes U^{(N)} $ in \Eq{12}, but since $D$ is generally nonlocal, that is not possible. However, the zeros in $|\Phi _{\text{TGX}} \rangle$ \textit{do} make it possible in the context of \Eq{11}, but we keep $D$ in \Eq{12} because $U_{\text{EPU}}$ is \textit{not} generally local by itself.  Also, even though $\eta$ becomes a global phase in \Eq{11}, its role in the TGX core is one of \textit{relative phase}.  Alternatively, we could think of $D|\Phi _{\text{TGX}} \rangle$ as a ``fully-phased'' TGX core, however, $U_{\text{EPU}}$ alone could \textit{still} have the form of \Eq{12}, so that is why we keep $D$ separate.

Thus, we have shown that \Eq{11} and \Eq{12} are equivalent to the Schmidt decomposition for parameterizing all $2\times 2$ maximally entangled pure states.
%                                 End of III.C
%-------------------------------------------------------------------------------
%                                 END of III
%*******************************************************************************
%*******************************************************************************
%                  IV. MAXIMALLY ENTANGLED BASIS THEOREM
\section{\label{sec:IV}MAXIMALLY ENTANGLED BASIS THEOREM}
%_______________________________________________________________________________
\begin{figure}[H]%Not a figure; Puts hypertarget at top of column to fix problem
\centering
\vspace{-12pt}
\setlength{\unitlength}{0.01\linewidth}
\begin{picture}(100,0)
\put(1,30){\hypertarget{Sec:IV}{}}
\end{picture}
\end{figure}
\vspace{-39pt}
%_______________________________________________________________________________
First presented in \cite{HedX}, the maximally entangled basis (MEB) conjecture is that \textit{there are always enough maximally entangled (ME) states to form a complete or overcomplete basis in any multipartite system}.

The 13-step algorithm \smash{$\mathcal{A}_{13}$} of \Eq{9} and \App{App.G} \textit{proves} that MEBs exist because \smash{$\mathcal{A}_{13}(\mathbf{n},S_{L})$} makes \textit{at least one} ME TGX state for every starting-level value $S_L  \in 1, \ldots ,n$.  Thus, concatenating all such output sets as
%===============================================================================
\begin{equation}%                  Equation 15
\{ |\Phi _j \rangle \}  \equiv \{ \mathcal{A}_{13} (\mathbf{n},s)\} |_{s = 1}^n  = \{ \{ |\Phi _j \rangle _{(1)} \} , \ldots ,\{ |\Phi _j \rangle _{(n)} \} \} ,
\label{eq:15}
\end{equation}
%===============================================================================
gives a set of at least $n$ ME TGX states, each definitely involving a \textit{different} level as one of the computational basis states $\{ |k\rangle \}  \equiv \{ |1\rangle , \ldots ,|n\rangle \} $, where \smash{$\{ |\Phi _j \rangle _{(s)} \}$} is the set of ME TGX states involving level $s$, and we generally omit the $s$ subscript.  Thus there are always enough ME TGX states to form a complete or overcomplete basis.

For \textit{complete} MEBs, simply \textit{form sets of ME states that contain all of the $n$ computational basis states exactly once}.  Generally, several such sets exist for each system.

For example, in $3\times 3$, there are \textit{two} such sets, found by first identifying the two \textit{generating sets}, which are, in the level-notation of \Table{4} in \App{App.H.1},
%===============================================================================
\begin{equation}%                  Equation 16
\begin{array}{*{20}l}
   {\{ |\Phi _j^1 \rangle \} } &\!\! { \equiv \{ \{ 1,5,9\} ,\{ 2,6,7\} ,\{ 3,4,8\} \}, }  \\
   {\{ |\Phi _j^2 \rangle \} } &\!\! { \equiv \{ \{ 1,6,8\} ,\{ 2,4,9\} ,\{ 3,5,7\} \},\rule{0pt}{10pt} }  \\
\end{array}
\label{eq:16}
\end{equation}
%===============================================================================
where the \smash{$\{ L_{\text{ME}} \} _{|\Phi _j^x \rangle }$} of \Table{4} are the levels of $|\Phi _j^x \rangle $ where $x$ labels the sets.  For instance, set 1 expands as
%===============================================================================
\begin{equation}%                  Equation 17
\{ |\Phi _j^1 \rangle \}  \equiv \left\{\! \begin{array}{l}
 |\Phi _1^1 \rangle , \\ 
 |\Phi _2^1 \rangle ,\rule{0pt}{10pt} \\ 
 |\Phi _3^1 \rangle  \rule{0pt}{10pt}\\ 
 \end{array}\!\!\! \right\} \equiv \left\{\! \begin{array}{l}
 {\textstyle{1 \over {\sqrt 3 }}}(|1\rangle  + |5\rangle  + |9\rangle ), \\ 
 {\textstyle{1 \over {\sqrt 3 }}}(|2\rangle  + |6\rangle  + |7\rangle ), \\ 
 {\textstyle{1 \over {\sqrt 3 }}}(|3\rangle  + |4\rangle  + |8\rangle ) \\ 
 \end{array}\! \right\}\!,
\label{eq:17}
\end{equation}
%===============================================================================
where the ordered basis corresponds to the coincidence basis by \Eq{H.1} as $\{ |1\rangle , \ldots ,|9\rangle \}  = \{ |1,1\rangle ,|1,2\rangle ,|1,3\rangle ,|2,1\rangle ,$ $|2,2\rangle ,|2,3\rangle ,|3,1\rangle ,|3,2\rangle ,|3,3\rangle \} $.

A set \smash{$\{ |\Phi _j^x \rangle \}$} is \textit{not} an MEB by itself, but rather it is a \textit{generating set} for an MEB.  A complete MEB is formed by converting each state of a generating set into a subset of orthonormal states with the same levels, yielding a total of $n$ orthonormal ME TGX states.  If each generating state \smash{$|\Phi _j^x \rangle$} has $L_{*}$ levels, the number of orthonormal states with those same levels is $L_{*}$ so the number of generating states in \smash{$\{ |\Phi _j^x \rangle \}$} is
%===============================================================================
\begin{equation}%                  Equation 18
G\equiv {\textstyle{n \over {L_{*}}}}.
\label{eq:18}
\end{equation}
%===============================================================================
The orthonormal subset of states for each \smash{$|\Phi _j^x \rangle$} is then
%===============================================================================
\begin{equation}%                  Equation 19
|\Phi _{j|l}^x \rangle  \equiv\!\! \sum\limits_{k = \{ L_{\text{ME}} \} _{|\Phi _j^x \rangle } }\!\!\!\!\!\! {F_{l,k}^{[L_{*}]} |k\rangle } ,
\label{eq:19}
\end{equation}
%===============================================================================
for{\kern -1.5pt} $l \!\in\! 1, \ldots ,{\kern -1pt}L_{*}$,{\kern -2pt} where{\kern -2pt} \smash{$\{ L_{\text{ME}} \} _{|\Phi _j^x \rangle }$}{\kern -1pt} are{\kern -1pt} the{\kern -1pt} levels{\kern -1pt} corresponding{\kern -1pt} to{\kern -1pt} \smash{$|\Phi _j^x \rangle$}{\kern -1pt} as{\kern -1pt} in{\kern -1pt} \Table{4}{\kern -1pt},{\kern -1.5pt} $\{|k\rangle\}\rule{0pt}{9.5pt}${\kern -1pt} is{\kern -1pt} the{\kern -1pt} computational{\kern -1pt} basis,{\kern -1pt} and{\kern -1pt} $F${\kern -1pt} is{\kern -1pt} the{\kern -1pt} unitary{\kern -1pt} Fourier{\kern -1pt} matrix{\kern -1pt} with{\kern -1pt} elements
%===============================================================================
\begin{equation}%                  Equation 20
F_{j,k}^{[L_{*}]}  \equiv {\textstyle{1 \over {\sqrt {L_{*}} }}}e^{ - i(j - 1)(k - 1){\textstyle{{2\pi } \over {L_{*}}}}} ,
\label{eq:20}
\end{equation}
%===============================================================================
with $L_{*}$ levels. Thus, in a given generating set $x$, for each of the $G$ generating states we get $L_{*}$ orthonormal states that are automatically orthonormal to the other generating states and their generated subsets, yielding $n$ orthonormal ME TGX states, forming a complete MEB.

In the above example, applying \Eq{19} to set 1 yields
%===============================================================================
\begin{equation}%                  Equation 21
\begin{array}{*{20}r}
   {|\Phi _{1|1}^1 \rangle } &\!\! { =\! {\textstyle{1 \over {\sqrt 3 }}}(}\!\! &\!\! {|1\rangle  + } &\!\!\! {|5\rangle  + } &\!\!\! {|9\rangle } &\!\!\! ){\kern 2.5pt}  \\
   {|\Phi _{1|2}^1 \rangle } &\!\! { =\! {\textstyle{1 \over {\sqrt 3 }}}(}\!\! &\!\! {|1\rangle  + } &\!\!\! {\omega |5\rangle  + } &\!\!\! {\omega ^2 |9\rangle } &\!\!\! ){\kern 2.5pt}  \\
   {|\Phi _{1|3}^1 \rangle } &\!\! { =\! {\textstyle{1 \over {\sqrt 3 }}}(}\!\! &\!\! {|1\rangle  + } &\!\!\! {\omega ^2 |5\rangle  + } &\!\!\! {\omega ^4 |9\rangle } &\!\!\! ){\kern 2.5pt}  \\
   {|\Phi _{2|1}^1 \rangle } &\!\! { =\! {\textstyle{1 \over {\sqrt 3 }}}(}\!\! &\!\! {|2\rangle  + } &\!\!\! {|6\rangle  + } &\!\!\! {|7\rangle } &\!\!\! ){\kern 2.5pt}  \\
   {|\Phi _{2|2}^1 \rangle } &\!\! { =\! {\textstyle{1 \over {\sqrt 3 }}}(}\!\! &\!\! {|2\rangle  + } &\!\!\! {\omega |6\rangle  + } &\!\!\! {\omega ^2 |7\rangle } &\!\!\! ){\kern 2.5pt}  \\
   {|\Phi _{2|3}^1 \rangle } &\!\! { =\! {\textstyle{1 \over {\sqrt 3 }}}(}\!\! &\!\! {|2\rangle  + } &\!\!\! {\omega ^2 |6\rangle  + } &\!\!\! {\omega ^4 |7\rangle } &\!\!\! ){\kern 2.5pt}  \\
   {|\Phi _{3|1}^1 \rangle } &\!\! { =\! {\textstyle{1 \over {\sqrt 3 }}}(}\!\! &\!\! {|3\rangle  + } &\!\!\! {|4\rangle  + } &\!\!\! {|8\rangle } &\!\!\! ){\kern 2.5pt}  \\
   {|\Phi _{3|2}^1 \rangle } &\!\! { =\! {\textstyle{1 \over {\sqrt 3 }}}(}\!\! &\!\! {|3\rangle  + } &\!\!\! {\omega |4\rangle  + } &\!\!\! {\omega ^2 |8\rangle } &\!\!\! ){\kern 2.5pt}  \\
   {|\Phi _{3|3}^1 \rangle } &\!\! { =\! {\textstyle{1 \over {\sqrt 3 }}}(}\!\! &\!\! {|3\rangle  + } &\!\!\! {\omega ^2 |4\rangle  + } &\!\!\! {\omega ^4 |8\rangle } &\!\!\! ),  \\
\end{array}
\label{eq:21}
\end{equation}
%===============================================================================
where \smash{$\omega\!  \equiv\! e^{ - i{2\pi/3}}\!\!$}, and which are orthonormal, forming an{\kern -1pt} MEB{\kern -1pt} since{\kern -1pt} \smash{$\sum\nolimits_{j = 1}^G \!{\sum\nolimits_{l = 1}^{L_{*}  } \!{|\Phi _{j|l}^1 \rangle \!\langle \Phi _{j|l}^1 |} } \! =\! I^{[n]} $}{\kern -1pt} and{\kern -1pt} \smash{$\Upsilon(\rho_{|\Phi _{j|l}^1 \rangle})\!=\!1$}.

Thus,\rule{0pt}{13pt} the completeness of the Bell states is also found in all larger systems for maximally entangled TGX states, another reason to consider them the true generalization of the Bell states with respect to entanglement.
%                                 END of IV
%*******************************************************************************
%*******************************************************************************
%            V. $\theta$ STATES: PURE STATES OF ANY ENTANGLEMENT
\section{\label{sec:V}$\theta$ STATES: PURE STATES OF ANY ENTANGLEMENT}
%_______________________________________________________________________________
\begin{figure}[H]%Not a figure; Puts hypertarget at top of column to fix problem
\centering
\vspace{-12pt}
\setlength{\unitlength}{0.01\linewidth}
\begin{picture}(100,0)
\put(1,30){\hypertarget{Sec:V}{}}
\end{picture}
\end{figure}
\vspace{-39pt}
%_______________________________________________________________________________
It is often useful to have a family of pure states that can be continuously varied from separable to maximally entangled.  Here we give general constructions of these state families called \textit{$\theta$ states} in all discrete systems.
%-------------------------------------------------------------------------------
%                   V.A. Single-Parameter $\theta$ States
\subsection{\label{sec:V.A}Single-Parameter $\theta$ States}
Let a canonical family of single-parameter $\theta$ states be
%===============================================================================
\begin{equation}%                  Equation 22
|\Phi _j (\theta )\rangle  \equiv c_\theta  |(L_{\text{ME}} )_{j,1} \rangle  + s_\theta  {\textstyle{1 \over {\sqrt {L_*  - 1} }}}\sum\limits_{k = 2}^{L_* } {|(L_{\text{ME}} )_{j,k} \rangle } ,
\label{eq:22}
\end{equation}
%===============================================================================
where \smash{$\theta  \in [0,\text{acos}({\textstyle{1 \over {\sqrt {L_* } }}}){\kern 1pt}]$}, $c_{\theta}\!\equiv\!\cos(\theta)$, $s_{\theta}\!\equiv\!\sin(\theta)$, \smash{$L_{\text{ME}}$} is the{\kern -1pt} matrix{\kern -1pt} of{\kern -1pt} \Eq{G.19},{\kern -1pt} and \smash{$L_* =\text{dim}((L_{\text{ME}})_{j,\cdots})$}.

At{\kern -1.5pt} \smash{$\theta\!=\!\theta _{\max } \! \equiv\! \text{acos} (1{\kern -1pt}/\!\sqrt{L_{*}\!})$},{\kern -2pt} then{\kern -2pt} \smash{$c_{\theta _{\max } } \!\! =\! 1{\kern -1pt}/\!\sqrt{L_{*}\!}\,$},{\kern -2pt} and{\kern -2pt} since \rule{0pt}{9.7pt}$\sin (\text{acos}(x))\! =$ $\! \sqrt {1\! -\! x^2 }$,{\kern -1pt}\rule{0pt}{9.5pt} then $s_{\theta _{\max } } \! =\!\sqrt {L_* \! -\! 1} /\!\sqrt {L_{*}\!}\,$, yielding \smash{$|\Phi _j (\theta _{\max } )\rangle \! =\!{\textstyle{1 \over {\sqrt {L_* } }}}\!\sum\nolimits_{k = 1}^{L_* }\! {|(L_{\text{ME}} )_{j,k} \rangle }\!  =\! |\Phi _j \rangle$}, which is maximally{\kern -1pt} entangled.\rule{0pt}{10pt}  When $\theta=0$, \smash{$|\Phi _j (0)\rangle  \equiv |(L_{\text{ME}} )_{j,1} \rangle$} which is one of the \textit{separable} computational basis states. Since $|\Phi _j (\theta )\rangle$ represents a continuum of pure states between $|\Phi _j (0)\rangle$ and $|\Phi _j (\theta_{\max} )\rangle$, and since $|\Phi _j (0)\rangle$ has no entanglement and $|\Phi _j (\theta_{\max} )\rangle$ has maximal entanglement, then the family $|\Phi _j (\theta )\rangle$ contains states of \textit{all} entanglement values, yielding the desired parameterized family.

\begin{figure}[H]%         FIGURE 5 (made with entPaperFig5.m)
\centering
\includegraphics[width=1.00\linewidth]{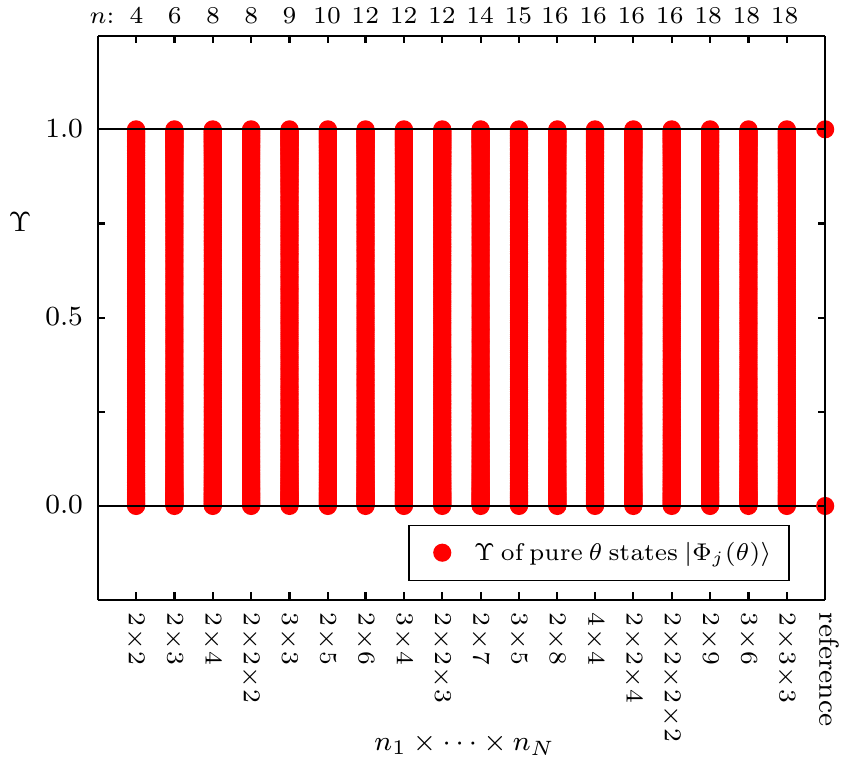}%
\vspace{-3pt}
\caption[]{(color online) Plot of the ent $\Upsilon$ from \Eq{2} of $\theta$ states from \Eq{22} for all discrete multipartite systems up to $n=18$ levels, for $300$ divisions of $\theta$ each.  This shows that $\theta$ states span $\Upsilon\in[0,1]$ for \smash{$\theta  \in [0,\text{acos}({\textstyle{1 \over {\sqrt {L_* } }}})]$}, making them useful for studying pure states of any entanglement, regardless of how difficult it would be to make them randomly (as in \Fig{1}). Identical results were also found up to $n=28$ (not shown).}
\label{fig:5}
\end{figure}

As \Fig{5} shows, $\theta$ states do indeed cover all ent values.  However, they do \textit{not} parameterize all entangled pure states; therefore we show how to do that next.
%_______________________________________________________________________________
%_______________________________________________________________________________
%                                 End of V.A
%-------------------------------------------------------------------------------
%-------------------------------------------------------------------------------
%    V.B. Multi-Parameter $\bm{\theta}$ States and Parameterization of All Entangled Pure States
\subsection{\label{sec:V.B}Multi-Parameter $\bm{\theta}$ States and Parameterization of All Entangled Pure States}
While \Eq{22} allows \textit{one} computational basis state to vary from being in balanced superposition with the other $L_{*}-1$ levels of the TGX set to being the only one left, in general we could arrange more complicated scenarios where multiple levels are remaining but separable.

Thus, we define multi-parameter TGX $\bm{\theta}$ states as
%===============================================================================
\begin{equation}%                  Equation 23
|\Phi _j (\bm{\theta} )\rangle  \equiv \sum\limits_{k = 1}^{L_* } {x_k^{[L_* ]} (\bm{\theta} )|(L_{\text{ME}} )_{j,k} \rangle },
\label{eq:23}
\end{equation}
%===============================================================================
where \smash{$x_k^{[L_* ]} (\bm{\theta})$} are{\kern -1.5pt} Schl{\"a}fli's hyperspherical coordinates{\kern -1.5pt} \cite{Schl,HedU} for $L_*$ dimensions, and $\bm{\theta}\! \equiv\! (\theta _1 , \ldots ,\theta _{L_*  - 1} )\!\in\! [0,\!{\textstyle{\pi  \over 2}}]$. 

The \smash{$|\Phi _j (\bm{\theta} )\rangle$} are maximally entangled for any combination of $\{\theta_u\}$ such that $x_1 =\cdots=x_{L_{*}}$ (balanced superposition).  They \textit{are} separable when only one \smash{$x_k^{[L_* ]} (\bm{\theta})$} is 1, but they are also separable when any superpositions of multiple levels factor as a tensor product, so \smash{$|\Phi _j (\bm{\theta} )\rangle$} describes a wider family of TGX states than \smash{$|\Phi _j (\theta )\rangle$}.

We can hypothesize that the most general way to parameterize pure states of any entanglement is
%===============================================================================
\begin{equation}%                  Equation 24
|\Phi_{G}\rangle\equiv|\Phi_{G} (\bm{\theta} )\rangle  \equiv U_{\text{EPU}} |\Phi _j (\bm{\theta})\rangle ,
\label{eq:24}
\end{equation}
%===============================================================================
where $U_{\text{EPU}}$ is entanglement-preserving unitary (EPU), so the ent of $|\Phi_{G} \rangle$ is the ent of $|\Phi _j (\bm{\theta})\rangle$ as in \Eq{11}. Again we can hypothesize that the most general form of this EPU is $U_{\text{EPU}}  \equiv (U^{(1)}  \otimes  \cdots  \otimes U^{(N)} )D$ as in \Eq{12}.

Applying the usual Schmidt method in reverse, analogously to \Eq{I.4}, but with $\Sigma\equiv\text{diag}\{c_{\theta},s_{\theta}\}$, gives
%===============================================================================
\begin{equation}%                  Equation 25
|\psi (\theta )\rangle  =\! \left(\! {\begin{array}{*{20}r}
   {c_\theta  a_1 a_2  - s_\theta  b_1 b_2 ^* }  \\
   {c_\theta  a_1 b_2  + s_\theta  b_1 a_2 ^* }  \\
   { - c_\theta  b_1 ^* a_2  - s_\theta  a_1 ^* b_2 ^* }  \\
   { - c_\theta  b_1 ^* b_2  + s_\theta  a_1 ^* a_2 ^* }  \\
\end{array}} \right)\!.
\label{eq:25}
\end{equation}
%===============================================================================
To compare \Eq{24} to \Eq{25}, using $|\Phi _j (\bm{\theta})\rangle \equiv|\Phi _j (\theta)\rangle = c_\theta  |1,1\rangle  + s_\theta  |2,2\rangle$ and the $U_{\text{EPU}}$ of \Eq{I.6} with the same substitutions and the conversion of \Eqs{I.8}{I.15}, we get
%===============================================================================
\begin{equation}%                  Equation 26
|\Phi _G (\theta )\rangle  =\!e^{i(\eta_{1}+\frac{\eta}{2})}\! \left(\! {\begin{array}{*{20}r}
   {c_\theta  a_1 a_2  - s_\theta  b_1 b_2 ^* }  \\
   {c_\theta  a_1 b_2  + s_\theta  b_1 a_2 ^* }  \\
   { - c_\theta  b_1 ^* a_2  - s_\theta  a_1 ^* b_2 ^* }  \\
   { - c_\theta  b_1 ^* b_2  + s_\theta  a_1 ^* a_2 ^* }  \\
\end{array}} \right)\!,
\label{eq:26}
\end{equation}
%===============================================================================
showing that \Eq{24} agrees with the usual Schmidt method, up to global phase.  In fact, \Eq{26} can parameterize \textit{any} pure state with \textit{any} entanglement, as (dropping $e^{i(\eta_{1}+\frac{\eta}{2})}$)
%===============================================================================
\begin{equation}%                  Equation 27
|\Phi _G (\Upsilon )\rangle \! =\!{\textstyle{1 \over {\sqrt 2 }}}{\kern -1pt}\! \left(\!\! {\begin{array}{*{20}r}
   {\sqrt{\!1\!+\!\sqrt{{\kern -1pt}1\!-\!\rule{0pt}{8.27pt}\Upsilon\,}}  a_1 a_2  -\! \sqrt{\!1\!-\!\sqrt{{\kern -1pt}1\!-\!\rule{0pt}{8.27pt}\Upsilon\,}}  b_1 b_2 ^* }  \\
   {\sqrt{\!1\!+\!\sqrt{{\kern -1pt}1\!-\!\rule{0pt}{8.27pt}\Upsilon\,}}  a_1 b_2  +\! \sqrt{\!1\!-\!\sqrt{{\kern -1pt}1\!-\!\rule{0pt}{8.27pt}\Upsilon\,}}  b_1 a_2 ^*  }  \\
   { - \sqrt{\!1\!+\!\sqrt{{\kern -1pt}1\!-\!\rule{0pt}{8.27pt}\Upsilon\,}}  b_1 ^* a_2  -\! \sqrt{\!1\!-\!\sqrt{{\kern -1pt}1\!-\!\rule{0pt}{8.27pt}\Upsilon\,}}  a_1 ^* b_2 ^*  }  \\
   { - \sqrt{\!1\!+\!\sqrt{{\kern -1pt}1\!-\!\rule{0pt}{8.27pt}\Upsilon\,}}  b_1 ^* b_2  +\! {\kern -1pt}\sqrt{\!1\!-\!\sqrt{{\kern -1pt}1\!-\!\rule{0pt}{8.27pt}\Upsilon\,}}  a_1 ^* a_2 ^*  }  \\
\end{array}}\! \right){\kern -1pt}\!,
\label{eq:27}
\end{equation}
%===============================================================================
\textit{regardless} of the values of $a_1$, $b_1$, $a_2$, $b_2$, as long as $|a_{k}|^2 + |b_{k}|^2 =1$ for $k=1,2$, and $\Upsilon\in[0,1]$.

This expression of any pure $2\times 2$ state in terms of its entanglement is possible because the ent of \smash{$|\Phi _G (\theta )\rangle$} and its TGX core are equal since \smash{$U_{\text{EPU}}$} preserves entanglement. Thus, \smash{$\Upsilon(\rho_{|\Phi _G (\theta)\rangle})=\Upsilon(\rho _{|\Phi _j (\theta )\rangle })=s_{2\theta }^2$}, yielding \smash{$\theta  \equiv {\textstyle{1 \over 2}}\text{asin} (\sqrt{\Upsilon^{{\kern -14pt}~^{~^{~^{~}}}}}  )$} to put into \Eq{26} to get \Eq{27}, where \smash{$\rho_{|A\rangle} \!\equiv\!|A\rangle\langle A|$}.  In terms of concurrence $C$ \cite{Woot}, set \smash{$\Upsilon\! =\!C^2$} in the right side of \Eq{27} to get \smash{$|\Phi _G (C)\rangle$}.  Note that \Eq{27} has ``Schmidt-diagonal form'' with matched indices as
%===============================================================================
\begin{equation}%                  Equation 28
\begin{array}{*{20}l}
   {|\Phi _G (\theta )\rangle } &\!\! {=\!\sqrt {{\kern -1.3pt}{\textstyle{{1 + {{\kern -14.5pt}~^{~^{~^{~}}}}\sqrt {1 - \Upsilon{{\kern -14pt}~^{~^{~^{~}}}} } } \over 2}}} \binom{\,\,\,\,a_{1}}{-b_{1}^{*}} \!\otimes\! \binom{a_{2}}{b_{2}} \!+\! {\kern -1pt}\sqrt {{\kern -1.3pt}{\textstyle{{1 - {{\kern -14.5pt}~^{~^{~^{~}}}}\sqrt {1 - \Upsilon{{\kern -14pt}~^{~^{~^{~}}}} } } \over 2}}} \binom{\,b_{1}}{a_{1}^{*}} \!\otimes\! \binom{-b_{2}^{*}}{\,\,\,a_{2}^{*}}}  \\
   {} &\!\! {=\rule{0pt}{12pt}c_\theta  |U_{:,1} ^{(1)} \rangle  \otimes |U_{:,1} ^{(2)} \rangle \! +\! s_\theta  |U_{:,2} ^{(1)} \rangle  \otimes |U_{:,2} ^{(2)} \rangle, }  \\
\end{array}\!\!
\label{eq:28}
\end{equation}
%===============================================================================
where $|U_{:,k}^{(m)}\rangle$ is column $k$ of some $n_m$-level unitary $U^{(m)}$, and TGX coefficients $c_\theta$ and $s_\theta$ are the \textit{Schmidt numbers}.

As another example, in $2\!\times\! 2\!\times\! 3$, one maximally entangled TGX state is, from \Eq{H.23},{\kern -1pt} \smash{$|\Phi _1 \rangle  {\kern -1pt}={\kern -1pt} {\textstyle{1 \over {\sqrt 4 }}}{\kern 1.5pt}({\kern 1.0pt}|1,1,1\rangle {\kern 2pt} +$} \smash{$ |1,2,2\rangle \! +\! |2,1,2\rangle \! +\! |2,2,3\rangle )$} where $(L_{\text{ME}})_{1,\cdots}\!=\!({\kern -0.5pt}1,5,8,12)$, so \Eq{23} and \Eq{24} give the entangled \textit{nonTGX} state,
%===============================================================================
\begin{equation}%                  Equation 29
\begin{array}{*{20}r}
   {|\Phi _G (\bm{\theta})\rangle  = } &\!\!\! {c_{\theta _1 } |U_{:,1} ^{(1)} \rangle  \otimes |U_{:,1} ^{(2)} \rangle  \otimes |U_{:,1} ^{(3)} \rangle\,\, }  \\
   {} &\!\!\! { + s_{\theta _1 } c_{\theta _2 } |U_{:,1} ^{(1)} \rangle  \otimes |U_{:,2} ^{(2)} \rangle  \otimes |U_{:,2} ^{(3)} \rangle\,\, }  \\
   {} &\!\!\! { + s_{\theta _1 } s_{\theta _2 } c_{\theta _3 } |U_{:,2} ^{(1)} \rangle  \otimes |U_{:,1} ^{(2)} \rangle  \otimes |U_{:,2} ^{(3)} \rangle\,\, }  \\
   {} &\!\!\! { + s_{\theta _1 } s_{\theta _2 } s_{\theta _3 } |U_{:,2} ^{(1)} \rangle  \otimes |U_{:,2} ^{(2)} \rangle  \otimes |U_{:,3} ^{(3)} \rangle,{\kern -1.2pt} }  \\
\end{array}
\label{eq:29}
\end{equation}
%===============================================================================
which shows that instead of ``Schmidt-diagonal,'' we should say \textit{TGX-indexed}, since the coincidence indices of the TGX core yield the indexing in \Eq{29}, rather than matching indices (though that is sufficient for $N$ qudits), and the coefficients \smash{$x_k^{[L_* ]} (\bm{\theta})$} are the Schmidt numbers.

We can \textit{always} absorb phases of $D$ from $U_{\text{EPU}}$ into the $U^{(m)}$ in \Eq{11} and \Eq{24} because there is always a bipartition of size $n_{\max}\times\nmaxnot$, letting the $e^{i\eta_k}$ factor to the $\nmaxnot$ side of the SVD as in \App{App.I}, since $|\Phi_{\text{TGX}}(\theta) \rangle$ picks out only $L_*$ phase factors from $D$, and $\max(L_*)=\nmaxnot$.

Fortunately, we need only single-parameter $\theta$ states to parameterize all entangled pure states. For example, in $2\times2\times 3$, applying $U_{\text{EPU}}$ to $|\Phi _1 (\theta )\rangle$ yields
%===============================================================================
\begin{equation}%                  Equation 30
\begin{array}{*{20}l}
   {|\Phi _G (\theta )\rangle  = } &\!\! {c_\theta  |U_{:,1} ^{(1)} \rangle  \otimes |U_{:,1} ^{(2)} \rangle  \otimes |U_{:,1} ^{(3)} \rangle }  \\
   {} &\!\! { + s_\theta  {\textstyle{1 \over {\sqrt {3} }}}\!\left(\! \begin{array}{l}
 |U_{:,1} ^{(1)} \rangle  \otimes |U_{:,2} ^{(2)} \rangle  \otimes |U_{:,2} ^{(3)} \rangle  \\ 
  + |U_{:,2} ^{(1)} \rangle  \otimes |U_{:,1} ^{(2)} \rangle  \otimes |U_{:,2} ^{(3)} \rangle  \\ 
  + |U_{:,2} ^{(1)} \rangle  \otimes |U_{:,2} ^{(2)} \rangle  \otimes |U_{:,3} ^{(3)} \rangle  \\ 
 \end{array}\!\! \right)\!\!,}  \\
\end{array}
\label{eq:30}
\end{equation}
%===============================================================================
so this \textit{non}TGX state has $\Upsilon  = {\textstyle{{16} \over {423}}}(75s_\theta  ^2  - 53s_\theta  ^4 )$, where from \Eq{22},{\kern -1pt} $\theta \! \in\! [0,\text{acos} ({\textstyle{1 \over 2}})]$,{\kern -1pt} letting us parameterize \Eq{30} by 
%===============================================================================
\begin{equation}%                  Equation 31
\theta  =\! \left\{\!\! {\begin{array}{*{20}l}
   {\text{asin}\!\left(\! {\sqrt {{\textstyle{{75} \over {106}}}\! +\! \sqrt {({\textstyle{{75} \over {106}}})^2  \!-\! {\textstyle{{423} \over {848}}}\Upsilon } } }\,\,\! \right)\!{\kern -0.5pt};} &\!\!\! {\Upsilon  \!\in\! [\Upsilon_{T},1]}  \\
   {\text{asin}\!\left(\! {\sqrt {{\textstyle{{75} \over {106}}}\! -\! \sqrt {({\textstyle{{75} \over {106}}})^2 \! -\! {\textstyle{{423} \over {848}}}\Upsilon } } }\,\,\! \right)\!{\kern -0.5pt};} &\!\!\! {\Upsilon \! \in\! [0,\Upsilon_{T}),}  \\
\end{array}} \right.
\label{eq:31}
\end{equation}
%===============================================================================
where the transition $\Upsilon_{T}  \equiv {\textstyle{{352} \over {423}}}$ is where the argument of $\text{asin}$ is $1$. The power of this method is that while calculating the entanglement of a nonTGX state is usually difficult, it is often simpler to get it from the TGX core.

Furthermore, we do \textit{not} need the $\bm{\theta}$ states of \Eq{23} to reach all entangled pure states.  The reason is that the single-parameter TGX states $|\Phi _j (\theta )\rangle$ always reach all entanglement, and since $|\Phi _j (\theta )\rangle$ is pure, then \textit{all other pure states of the same entanglement are unitarily related to $|\Phi _j (\theta )\rangle$, since all pure states are unitarily equivalent}.

The catch is that the unitary relating $|\Phi _j (\theta )\rangle$ to general states is nonlocal and \textit{of a different form than the} $U_{\text{EPU}}$ \textit{of \Eq{12}}.  To see why, note that applying a local unitary to \Eq{30} can always be absorbed into the local unitaries of $U_{\text{EPU}}$, \textit{but it cannot change the Schmidt numbers}. Therefore, the only way we can reach the general Schmidt numbers of \Eq{29} is by some nonlocal EPU $V_{\text{EPU}}$ that is more general than \Eq{12}.  Thus, to parameterize general states with a single-parameter TGX core, we must use
%===============================================================================
\begin{equation}%                  Equation 32
|\Phi _G (\bm{\theta })\rangle  = V_{\text{EPU}}(\bm{\theta }) |\Phi _{\text{TGX}} (\theta )\rangle ,
\label{eq:32}
\end{equation}
%===============================================================================
where the existence of $V_{\text{EPU}}\equiv V_{\text{EPU}}(\bm{\theta })$ is easily proven; given $|\Phi _G (\bm{\theta })\rangle $ and $|\Phi _{\text{TGX}} (\theta )\rangle$ such that \smash{$\Upsilon (\rho _{|\Phi _G (\bm{\theta })\rangle } ) =$} \smash{$ \Upsilon (\rho _{|\Phi _{\text{TGX}} (\theta )\rangle } )$} (where such $|\Phi _{\text{TGX}} (\theta )\rangle$ are guaranteed to exist by \Eq{22}), they are unitarily related by \smash{$V_{\text{EPU}}=$} \smash{$\epsilon _{|\Phi _G (\bm{\theta })\rangle } \epsilon _{|\Phi _{\text{TGX}} (\theta )\rangle }^{\dag} $}, where \smash{$\epsilon_{|A\rangle}$} is the unitary eigenvector matrix of $|A\rangle$ such that \smash{$\epsilon _{|A\rangle}^{\dag} |A\rangle\langle A|\epsilon _{|A\rangle}$} is diagonal.

Thus, we have shown how to achieve the multipartite Schmidt decomposition, and proven that we can always parameterize the entanglement of any pure state with a \textit{single}-parameter $\theta$ state of TGX form.
%                                 End of V.B
%-------------------------------------------------------------------------------
%                                 END of V
%*******************************************************************************
%*******************************************************************************
%              VI. Ent Vector, Ent Array, and Mixed-State Ent
\section{\label{sec:VI}Ent Vector, Ent Array, and Mixed-State Ent}
%_______________________________________________________________________________
\begin{figure}[H]%Not a figure; Puts hypertarget at top of column to fix problem
\centering
\vspace{-12pt}
\setlength{\unitlength}{0.01\linewidth}
\begin{picture}(100,0)
\put(1,30){\hypertarget{Sec:VI}{}}
\end{picture}
\end{figure}
\vspace{-39pt}
%_______________________________________________________________________________
To evaluate an $N$-partite state $\rho$ as a resource for $S$-partite entanglement where $S\!\in\! 2,\ldots,N$, one approach is to check the $S$-partite entanglement of all groups of $S$ modes.  If done for all values of $S$, we amass a list of all multipartite entanglement \textit{directly} accessible as a resource.  We call such a list an \textit{entanglement vector}, in loose analogy to the concurrence vector \cite{Woo2}.  In fact, more general definitions are also possible, as we will see.

However, we cannot simply apply the ent adapted to various mode groups without first tracing over the unused modes, which generally induces a \textit{mixed} state as input to the ent. Thus, first we focus on the case where the reduction is pure and then move to more general cases.
%-------------------------------------------------------------------------------
%                     VI.A. Modal Ent and the Ent Vector
\subsection{\label{sec:VI.A}Modal Ent and the Ent Vector}
First, let the \textit{modal ent} of \textit{pure} multipartite reductions \smash{$\redrho{\mathbf{m}}$} (see \App{App.B}) of \textit{pure} $N$-partite parent state $\rho$ be
%===============================================================================
\begin{equation}%                  Equation 33
\Upsilon^{(\mathbf{m})}\!\equiv\!\Upsilon^{(\mathbf{m})} (\redrho{\mathbf{m}}) \!\equiv\! \frac{1}{{M^{(\mathbf{m})} }}\!\left(\! {1\! -\! \frac{1}{S}\!\sum\limits_{m \in \mathbf{m}}\!\! {\frac{{n_m P(\redrho{m})\! -\! 1}}{{n_m \! -\! 1}}} }\! \right)\!,
\label{eq:33}
\end{equation}
%===============================================================================
where \smash{$\mathbf{n}\equiv{n_{1},\ldots,n_{N}}$}, and \smash{$\mathbf{m}\equiv (m_1 , \ldots ,m_S )$} where \smash{$S\equiv S^{(\mathbf{m})}\equiv \dim(\mathbf{m})\!\in\! 2,\ldots,N$} is a list of unique (but not necessarily consecutive) mode labels of interest, \smash{$\redrho{\mathbf{m}}$} must be pure (and \smash{$\redrho{m}$} are generally mixed), and
%===============================================================================
\begin{equation}%                  Equation 34
M^{(\mathbf{m})}\equiv M^{(\mathbf{m})}(L_{*}^{(\mathbf{m})}) \equiv 1 - \frac{1}{S}\sum\limits_{m \in \mathbf{m}}\! {\frac{{n_m P_{\text{MP}}^{(m)} (L_{*}^{(\mathbf{m})}) - 1}}{{n_m  - 1}}},
\label{eq:34}
\end{equation}
%===============================================================================
where \smash{$P_{\text{MP}}^{(m)} (L_{*}^{(\mathbf{m})})$} is from \Eq{4} with input \smash{$L_{*}^{(\mathbf{m})}$} given by
%===============================================================================
\begin{equation}%                  Equation 35
\begin{array}{*{20}l}
   {L_{*}^{(\mathbf{m})}  = \min \{ \mathbf{L}_{*}^{(\mathbf{m})} \} ;} &\!\! {}  \\
   {\mathbf{L}_{*}^{(\mathbf{m})}  \equiv \{ L_{*}^{(\mathbf{m})} \} ;\;\text{s.t.}} &\!\! {\mathop {\min }\limits_{L^{(\mathbf{m})} \in 2, \ldots, n_{\,\overline{{\kern -1.8pt}\max^{~^{~^{~}}}\!\!\!\!\!\!\!\!\!\!}}^{(\mathbf{m})} } (1 - M^{(\mathbf{m})}(L^{(\mathbf{m})})),}  \\
\end{array}
\label{eq:35}
\end{equation}
%===============================================================================
where \smash{$n_{\,\overline{{\kern -1.8pt}\max^{~^{~^{~}}}\!\!\!\!\!\!\!\!\!\!}}^{(\mathbf{m})}\equiv\frac{n^{(\mathbf{m})}}{n_{\max}^{(\mathbf{m})}}$}, where $n^{(\mathbf{m})}\equiv n_{\mathbf{m}}\equiv n_{m_{1}}\cdots n_{m_{S}}$, $n_{\max}^{(\mathbf{m})}\equiv \max\{n_{m_{1}},\ldots, n_{m_{S}}\}$, and we abbreviate $S\equiv S^{(\mathbf{m})}$ though keep in mind that $S$ depends on mode group $\mathbf{m}$.  Thus, in the case where \smash{$\redrho{\mathbf{m}}$} is pure, the modal ent uses the variables of the $N$-partite system to correctly compute the $S$-partite ent for the subsystem of interest.

Since \smash{$\redrho{\mathbf{m}}$} is generally mixed \textit{even if $\rho$ is pure}, we \textit{must} use the convex-roof extension of \smash{$\Upsilon^{(\mathbf{m})} (\redrho{\mathbf{m}})$} in general.  Although the smallest coincidence-relevant reductions in \smash{$\redrho{\mathbf{m}}$} may be mixed, that does \textit{not} mean we can claim that a mixed \smash{$\redrho{\mathbf{m}}$} is $S$-partite entangled, because if it has strong correlations with modes \textit{outside} of $\mathbf{m}$, then its own internal correlations are generally \textit{not fully available} as a resource for entanglement for any external system interacting with \smash{$\redrho{\mathbf{m}}$}.

Thus, in analogy to \Eq{C.24}, the \textit{modal ent of formation} is the convex-roof extension (see \App{App.J}) of \smash{$\Upsilon^{(\mathbf{m})}$} as
%===============================================================================
\begin{equation}%                  Equation 36
\hat{\Upsilon}^{(\mathbf{m})}\!\equiv\!\hat{\Upsilon}^{(\mathbf{m})}(\redrho{\mathbf{m}}) \equiv\! \mathop {\min }\limits_{\scriptstyle \;\,\{ p_j ,\redrhotiny{\mathbf{m}}_j \} \;\text{s.t.} \hfill \atop 
  \scriptstyle \redrhotiny{\mathbf{m}}  = \sum\nolimits_j p_j \redrhotiny{\mathbf{m}}_j  \hfill} \!\!\!\!\!\!\!\left(\, {\sum\nolimits_j p_j \Upsilon^{(\mathbf{m})}(\redrho{\mathbf{m}}_j )} \right)\!,
\label{eq:36}
\end{equation}
%===============================================================================
where $p_j \in [0,1]$ and $\sum_{j}p_{j}=1$.

Then define and represent the \textit{ent vector} as an inverted triangular matrix $\nabla$ (not a gradient) with elements
%===============================================================================
\begin{equation}%                  Equation 37
\nabla_{k,l} (\rho ) \equiv \hat{\Upsilon}^{((\text{nCk}(\mathbf{c},k))_{l, \cdots } )} (\redrho{(\text{nCk}(\mathbf{c},k))_{l, \cdots }} ),
\label{eq:37}
\end{equation}
%===============================================================================
where \smash{$\mathbf{c} \equiv (1, \ldots ,N)$}, \smash{$k\in 2,\ldots,N$}, and \smash{$l \in 1, \ldots ,\binom{N}{k}$} where \smash{$\binom{N}{k}\equiv\frac{N!}{k!(N-k)!}$}, and \smash{$\text{nCk}(\mathbf{v},k)$} is the vectorized $n$-choose-$k$ function yielding the matrix whose rows are each unique combinations of the elements of $\mathbf{v}$ chosen $k$ at a time, and \smash{$A_{l,\cdots}$} is the $l$th row of matrix $A$.

We can then define the \textit{net ent} as
%===============================================================================
\begin{equation}%                  Equation 38
\Upsilon_{\text{net}} (\rho ) = \frac{{||\nabla (\rho )||_1 }}{\max(||\nabla (\rho )||_{1})}\,,
\label{eq:38}
\end{equation}
%===============================================================================
where the maximum is over all pure parent states $\rho$, and is \textit{not} generally a sum of ones, and again $\rho$ is required to be \textit{pure} (see \Sec{VI.C} for mixed $\rho$).  Thus, the $N$-mode term \smash{$\nabla _{N,1} (\rho ) \equiv \hat{\Upsilon}^{(1, \ldots ,N)} (\redrho{1, \ldots ,N} )$} can be replaced with $\Upsilon (\rho )$ from \Eq{2} since \smash{$\redrho{1, \ldots ,N} =\rho$}.

Note that \Eq{38} is a number between $0$ and $1$ characterizing how much multipartite entanglement a state could have \textit{over all possible particular modal perspectives}, but \textit{it does not mean that all such entanglement is available simultaneously}.  Elements of $\nabla$ tell how much $S$-partite entanglement is available among groups of $S$ modes.

As an example, when $N=4$, the ent vector has form
%===============================================================================
\begin{equation}%                  Equation 39
\nabla (\rho )=\!\left(\!\! {\begin{array}{*{20}c}
   {\begin{array}{*{20}c}
   {\hat \Upsilon ^{(1,2)} } & {\hat \Upsilon ^{(1,3)} } & {\hat \Upsilon ^{(1,4)} } & {\hat \Upsilon ^{(2,3)} } & {\hat \Upsilon ^{(2,4)} } & {\hat \Upsilon ^{(3,4)} }  \\
\end{array}}  \\
   {\begin{array}{*{20}c}
   {\hat \Upsilon ^{(1,2,3)} } & {\hat \Upsilon ^{(1,2,4)} } & {\hat \Upsilon ^{(1,3,4)} } & {\hat \Upsilon ^{(2,3,4)} }  \\
\end{array}}  \\
   {\hat \Upsilon ^{(1,2,3,4)} }  \\
\end{array}}\!\! \right)\!,
\label{eq:39}
\end{equation}
%===============================================================================
where again \smash{$\hat{\Upsilon}^{(\mathbf{m})}  \equiv \hat{\Upsilon}(\redrho{\mathbf{m}} )$}.
%                                 End of VI.A
%-------------------------------------------------------------------------------
%-------------------------------------------------------------------------------
%                  VI.B. Partitional Ent and the Ent Array
\subsection{\label{sec:VI.B}Partitional Ent and the Ent Array}
The modes targeted by the ent vector, as in \Eq{39}, are still not the full story of multipartite entanglement. For example, we could also view any $S$-partite subsystem of modes $\mathbf{m}=(m_{1},\ldots,m_{S})$ with structure $(n_{m_{1}},\ldots,n_{m_{S}})$ and $n^{(\mathbf{m})}\equiv n_{m_{1}}\cdots n_{m_{S}}$ levels as a $T$-partite system of \textit{different structure} $(n_{1}',\ldots,n_{T}')$ of $n'\equiv n_{1}'\cdots n_{T}'=n^{(\mathbf{m})}$ levels, and $T\in 2,\ldots,S$.  In other words, we can look at all $T$-partitions of any subsystem of $S$ modes without removing any modes, \textit{and} without splitting the modes $m_k$ defined by the system's coincidence behavior.

Thus, for \textit{pure} parent states $\rho$, the \textit{partitional ent} of \textit{pure}{\kern -1pt} $S$-partite{\kern -1pt} reductions{\kern -1.5pt} \smash{$\redrho{\mathbf{m}^{(\mathbf{T})}}$}{\kern -1pt} with{\kern -1pt} $T${\kern -1pt} new{\kern -1pt} partitions{\kern -1pt} of modes $\mathbf{m}=(m_{1},\ldots,m_{S})$, where $\mathbf{T}\equiv (1,\ldots,T)$, is
%===============================================================================
\begin{equation}%                  Equation 40
\Upsilon ^{(\mathbf{m}^{(\mathbf{T})} )}  \equiv \frac{1}{\rule{0pt}{9.0pt}{M^{(\mathbf{m}^{(\mathbf{T})} )} }}\left( {1 - \frac{1}{T}\sum\limits_{q = 1}^T {\frac{{n_{q}' P(\redrho{\mathbf{m}^{(q)}} ) - 1}}{{n_{q}' - 1}}} } \right)\!,
\label{eq:40}
\end{equation}
%===============================================================================
where \smash{$\Upsilon ^{(\mathbf{m}^{(\mathbf{T})} )}  \equiv \Upsilon ^{(\mathbf{m}^{(\mathbf{T})} )} (\rho ^{(\mathbf{m}^{(\mathbf{T})} )} )$}, with new mode structure \smash{$\mathbf{m}^{(\mathbf{T})}  \equiv (\mathbf{m}^{(1)} | \ldots |\mathbf{m}^{(T)} )$}, where \smash{$T \equiv T^{(\mathbf{m}^{(\mathbf{T})} )}  =$} \smash{$ \dim (\mathbf{m}^{(\mathbf{T})} ) \in 2, \ldots ,S$} where $S\in 2, \ldots ,N$, and the $T$ new composite modes \smash{$\mathbf{m}^{(q)}$} have internal structures \smash{$\mathbf{m}^{(q)}  \equiv (m_1^{(q)} , \ldots ,m_G^{(q)} )$},\rule{0pt}{10pt} where \smash{$G \equiv G^{(q)}  = \dim (\mathbf{m}^{(q)} )$} and \smash{$n_{q}' \equiv n_{m_1^{(q)} } \cdots n_{m_G^{(q)} } $}. If $T=S$, then \smash{$\mathbf{m}^{(1, \ldots ,S)}  = \mathbf{m}$}, \smash{$\mathbf{m}^{(q)}  = m_q$}, and\rule{0pt}{10pt} \smash{$n_{q}'\, =\, n_{m_q }$}. Also,
%===============================================================================
\begin{equation}%                  Equation 41
M^{(\mathbf{m}^{(\mathbf{T})} )}  \equiv 1 - \frac{1}{T}\sum\limits_{q = 1}^T {\frac{{n_{q}' P_{\text{MP}}^{(\mathbf{m}^{(q)} )} (L_*^{(\mathbf{m}^{(\mathbf{T})} )} ) - 1}}{{n_{q}' - 1}}} ,
\label{eq:41}
\end{equation}
%===============================================================================
where \smash{$M^{(\mathbf{m}^{(\mathbf{T})})}  \equiv M^{(\mathbf{m}^{(\mathbf{T})})} (L_*^{(\mathbf{m}^{(\mathbf{T})})} )$} and \smash{$P_{\text{MP}}^{(\mathbf{m}^{(q)} )} (L_*^{(\mathbf{m}^{(\mathbf{T})} )} )$} is{\kern -2pt} given{\kern -2pt} by{\kern -2pt} \Eq{4}{\kern -2pt} with{\kern -2pt} \smash{$L_*\!\to\! L_*^{(\mathbf{m}^{(\mathbf{T})} )} $}{\kern -2pt} and{\kern -2pt} $n_m \!\!\!\to\!\!\! n_{q}'$,{\kern -2pt} where \smash{${L_*}{~^{{\kern -7.5pt}\LowerExponentOrSubScript{-1.0pt}{(\mathbf{m}^{(\mathbf{T})} )}}}$}{\kern -2pt} is{\kern -2pt} given{\kern -2pt} by{\kern -2pt} \Eq{35}{\kern -2pt} with{\kern -2pt} \smash{$\mathbf{m}\to\mathbf{m}^{(\mathbf{T})}$},{\kern -2pt}\rule{0pt}{8.5pt} where{\kern -2pt}~\smash{${\nmaxnot}{\rule{0pt}{7pt}^{{\kern -16pt}(\mathbf{m}\rule{0pt}{1pt}^{(\mathbf{T})})}}\!\equiv$}{\kern -2pt} \smash{$n{~^{{\kern -3.0pt}\LowerExponentOrSubScript{-2.0pt}{(\mathbf{m}^{(\mathbf{T})} )}}}/n{~_{{\kern -3pt}\LowerExponentOrSubScript{-1pt}{\max}}}{~^{{\kern -18.5pt}\LowerExponentOrSubScript{-0.5pt}{(\mathbf{m}^{(\mathbf{T})} )}}}$}, with \smash{$n{~_{{\kern -3pt}\LowerExponentOrSubScript{-1pt}{\max}}}{~^{{\kern -18.5pt}\LowerExponentOrSubScript{-0.5pt}{(\mathbf{m}^{(\mathbf{T})} )}}}\equiv\max\{n'_{1},\ldots,n'_{T}\}$}\rule{0pt}{9.5pt}, and \smash{$n{~^{{\kern -3.0pt}\LowerExponentOrSubScript{-2.0pt}{(\mathbf{m}^{(\mathbf{T})} )}}}\equiv$} \smash{$n'= n^{(\mathbf{m})}$}.

Generally, \smash{$\redrho{\mathbf{m}}$} is mixed, so any reorganization of it as \smash{$\redrho{\mathbf{m}^{(\mathbf{T})}}$} is too, requiring the \textit{partitional ent of formation},
%===============================================================================
\begin{equation}%                  Equation 42
\hat{\Upsilon}^{(\mathbf{m}^{(\mathbf{T})})}\!\equiv\! \mathop {\min }\limits_{\scriptstyle \;\;\;\;\,\{ p_j ,\redrhotiny{\mathbf{m}^{(\mathbf{T})}}_j \} \;\text{s.t.} \hfill \atop 
  \scriptstyle \redrhotiny{\mathbf{m}^{(\mathbf{T})}}  = \sum\nolimits_j p_j \redrhotiny{\mathbf{m}^{(\mathbf{T})}}_j  \hfill} \!\!\!\!\!\!\!\left(\, {\sum\nolimits_j p_j \Upsilon^{(\mathbf{m}^{(\mathbf{T})})}(\redrho{\mathbf{m}^{(\mathbf{T})}}_j )} \right)\!,
\label{eq:42}
\end{equation}
%===============================================================================
where \smash{$\hat{\Upsilon}^{(\mathbf{m}^{(\mathbf{T})})}\!\equiv\!\hat{\Upsilon}^{(\mathbf{m}^{(\mathbf{T})})}(\redrho{\mathbf{m}^{(\mathbf{T})}})$}. See \App{App.J} for details.

Note that for each group of $S$ modes, we can check all possible $T$-partitional ents for $T=2,\ldots,S$. Thus for mode-group $\mathbf{m}$, we use another inverted triangular matrix to represent and define the \textit{partitional ent vector},
%===============================================================================
\begin{equation}%                  Equation 43
\Xi ^{(\mathbf{m})}  \equiv \left( {\begin{array}{*{20}c}
   {\{ \hat{\Upsilon}^{(\mathbf{m}_h^{(\mathbf{2})} )} \} }  \\
    \vdots   \\
   {\{ \hat{\Upsilon}^{(\mathbf{m}_h^{(\mathbf{S})} )} \} }  \\
\end{array}} \right)\!,
\label{eq:43}
\end{equation}
%===============================================================================
where \smash{$\{ \hat{\Upsilon}^{(\mathbf{m}_h^{(\mathbf{T})} )} \}$} is the set of all possible $T$-partitional ents of a given reduction \smash{$\redrho{\mathbf{m}}$}, so that row $T-1$ of \smash{$\Xi ^{(\mathbf{m})}$} lists all possible $T$-partitional ents. For example,
%===============================================================================
\begin{equation}%                  Equation 44
\Xi ^{(1,2,4)} \! =\! \left(\!\! {\begin{array}{*{20}c}
   {\begin{array}{*{20}c}
   {\hat \Upsilon ^{(1|2,4)} } & {\hat \Upsilon ^{(2|1,4)} } & {\hat \Upsilon ^{(4|1,2)} }  \\
\end{array}}  \\
   {\hat \Upsilon ^{(1|2|4)} }  \\
\end{array}}\!\! \right)\!.
\label{eq:44}
\end{equation}
%===============================================================================

Since a partitional ent vector exists for each reduction, we can define the \textit{ent array} as the matrix whose elements are partitional ent vectors,
%===============================================================================
\begin{equation}%                  Equation 45
\widetilde{\nabla}_{k,l} (\rho ) \equiv \Xi ^{((\text{nCk}(\mathbf{c},k))_{l, \cdots } )},
\label{eq:45}
\end{equation}
%===============================================================================
with the definitions of \Eq{37}.  For the example in \Eq{39},
%===============================================================================
\begin{equation}%                  Equation 46
\widetilde{\nabla} (\rho ) =\! \left(\!\! {\begin{array}{*{20}c}
   {\begin{array}{*{20}c}
   {\Xi ^{(1,2)} } & {\Xi ^{(1,3)} } & {\Xi ^{(1,4)} } & {\Xi ^{(2,3)} } & {\Xi ^{(2,4)} } & {\Xi ^{(3,4)} }  \\
\end{array}}  \\
   {\begin{array}{*{20}c}
   {\Xi ^{(1,2,3)} } & {\Xi ^{(1,2,4)} } & {\Xi ^{(1,3,4)} } & {\Xi ^{(2,3,4)} }  \\
\end{array}}  \\
   {\Xi ^{(1,2,3,4)} }  \\
\end{array}}\!\! \right)\!,
\label{eq:46}
\end{equation}
%===============================================================================
where each bipartite reduction there has only one partition so \smash{$\Xi ^{(1,2)}  = \hat \Upsilon ^{(1|2)}$}, \smash{$\Xi ^{(1,3)}  = \hat \Upsilon ^{(1|3)}$}, \smash{$\Xi ^{(1,4)}  = \hat \Upsilon ^{(1|4)}$}, \smash{$\Xi ^{(2,3)}  = \hat \Upsilon ^{(2|3)}$}, \smash{$\Xi ^{(2,4)}  = \hat \Upsilon ^{(2|4)}$}, \smash{$\Xi ^{(3,4)}  = \hat \Upsilon ^{(3|4)}$}, and for each tripartite reduction there are four partitions (each similar to \Eq{44} so we omit them here), and for the $4$-partite reduction there are fourteen partitions so the rows of \smash{$\Xi ^{(1,2,3,4)}$} are \smash{$\Xi _{1, \cdots }^{(1,2,3,4)}\!\!=$} \smash{$ \{ \hat \Upsilon ^{(1|2,3,4)}$}, \smash{$\hat \Upsilon ^{(2|1,3,4)}$}, \smash{$\hat \Upsilon ^{(3|1,2,4)}$}, \smash{$\hat \Upsilon ^{(4|1,2,3)}$}, \smash{$\hat \Upsilon ^{(1,2|3,4)}$}, \smash{$\hat \Upsilon ^{(1,3|2,4)}$}, \smash{$\hat \Upsilon ^{(1,4|2,3)} \} $}, and \smash{$\Xi _{2, \cdots }^{(1,2,3,4)}  =$} \smash{$ \{ \hat \Upsilon ^{(1|2|3,4)}$}, \smash{$\hat \Upsilon ^{(1|3|2,4)}$}, \smash{$\hat \Upsilon ^{(1|4|2,3)}$}, \smash{$\hat \Upsilon ^{(2|3|1,4)}$}, \smash{$\hat \Upsilon ^{(2|4|1,3)}$}, \smash{$\hat \Upsilon ^{(3|4|1,2)} \}$}, and \smash{$\Xi _{3, \cdots }^{(1,2,3,4)} = \hat \Upsilon ^{(1|2|3|4)}$},{\kern 1.4pt} so{\kern 1.4pt} \smash{$\text{dim}(\widetilde{\nabla} \!(\rho )){\kern -0.8pt}={\kern -0.8pt}36$}. Elements of a \smash{$\Xi^{(\mathbf{m})}$} with $\text{dim}(\mathbf{m})\!=\!N$, such as \smash{$\hat \Upsilon ^{(1,2|3,4)}$} here, do not need convex-roof extension, but only if $\rho$ is pure, as stipulated in this section.

We can also define the \textit{absolute ent} as
%===============================================================================
\begin{equation}%                  Equation 47
\Upsilon _{\text{abs}} (\rho ) \equiv \frac{{||\widetilde{\nabla} ||_{1} }}{{\max (||\widetilde{\nabla} ||_{1})}}\,,
\label{eq:47}
\end{equation}
%===============================================================================
for a single number characterizing the entanglement available from all possible perspectives (though again, not all of this entanglement is available simultaneously).

Thus, the ent array gives us the most \textit{fine-grained} view of multipartite entanglement, telling us not only \textit{how much}, but \textit{where} (between which modes) such entanglement exists.  What we do with this information then depends on the application, as we will see next.
%...............................................................................
%              VI.B.1 Special Case: Genuine Multipartite (GM) Ent
\subsubsection*{\label{sec:VI.B.1}Special Case: Genuine Multipartite (GM) Ent}
A pure state is \textit{biseparable} if there exists \textit{any} bipartition of its modes such that it is a tensor product of pure states \cite{MCCS,HMGH,HSGS,Hor5}.  Thus, for pure $\rho$, the \textit{GM ent} is
%===============================================================================
\begin{equation}%                  Equation 48
\Upsilon _{\text{GM}} (\rho ) \equiv \min(\Xi_{1,\cdots}^{(\mathbf{N})})= \min \{ \Upsilon ^{(\mathbf{N}_h^{(\mathbf{2})} )} \},
\label{eq:48}
\end{equation}
%===============================================================================
where \smash{$\{ \Upsilon ^{(\mathbf{N}_h^{(\mathbf{2})} )} \}$} is the set of $N$-mode bipartitional ents.  So in $N\!=\!4$, \smash{$\Upsilon _{\text{GM}} (\rho )=\min\{\Upsilon ^{(1|2,3,4)},$} \smash{$\Upsilon ^{(2|1,3,4)},$} \smash{$\Upsilon ^{(3|1,2,4)},$} \smash{$\Upsilon ^{(4|1,2,3)},$} \smash{$\Upsilon ^{(1,2|3,4)},$} \smash{$\Upsilon ^{(1,3|2,4)},$} \smash{$\Upsilon ^{(1,4|2,3)} \}$}.

In fact, if we define \textit{$k$-separable} pure states as those for which \textit{any} $k$-partition of the $N$-mode state is a tensor product of $k$ pure states, then a measure of \textit{genuinely $k$-partite}{\kern -1pt} $(\text{GM}_{k})${\kern -1pt} \textit{entanglement}{\kern -1pt} would{\kern -0.5pt} be{\kern -0.5pt} the{\kern -0.5pt} $\text{GM}_k${\kern -0.5pt} \textit{ent},{\kern -0.5pt} as
%===============================================================================
\begin{equation}%                  Equation 49
\Upsilon _{\text{GM}_{k}} (\rho ) \equiv \min \{ \Upsilon ^{(\mathbf{N}_h^{(\mathbf{k})} )} \},
\label{eq:49}
\end{equation}
%===============================================================================
where \smash{$\{ \Upsilon ^{(\mathbf{N}_h^{(\mathbf{k})} )} \}$} is all $N$-mode $k$-partitional ents, where $k$ is partitions of $N$ just as $T$ partitioned $S$ in \Sec{VI.B}.

We could keep going, and define more generalized measures that check all $T$-partitions of each $S$-partite reduction; the ent array \smash{$\widetilde{\nabla}(\rho )$} of \Eq{45} gives us all we need.

However, $\text{GM}_{k}$ entanglement of any $k$ may not always be useful.  For example, in $N=4$ if we need two-qubit entanglement from a $2\!\times\! 2\!\times\! 2\!\times\! 2$ system, the GM entanglement may be from a bipartition such as $(2,2|2,2)$ having the lowest bipartite entanglement, but that does not tell us anything specific about the two-qubit entanglement.

Nevertheless, we have shown that the ent is a valuable tool for studying many multipartite entanglement definitions, regardless of their particular meaning.  
%                                End of VI.B.1
%...............................................................................
%                                End of VI.B
%-------------------------------------------------------------------------------
%-------------------------------------------------------------------------------
%                       VI.C. Ent for Mixed States
\subsection{\label{sec:VI.C}Ent for Mixed States}
\begin{figure}[H]%         FIGURE 6 (made with entPaperFig6.m)
\centering
\includegraphics[width=1.00\linewidth]{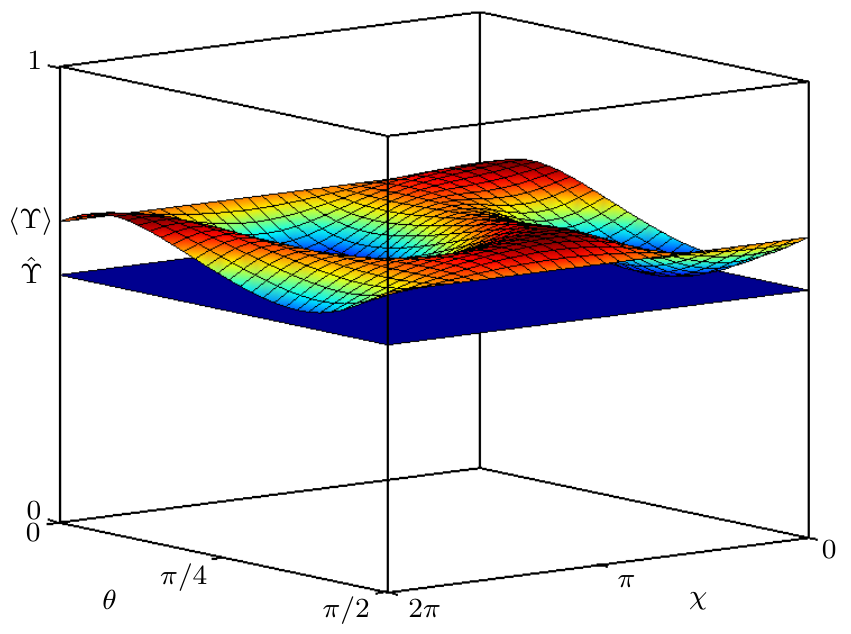}%
\vspace{-5pt}
\caption[]{(color online) The curved surface is the average ent $\langle\Upsilon\rangle$ ($N$-partite) for 30 divisions of each decomposition variable $\theta$ and $\chi$ (see \App{App.J}) for an arbitrary rank-2 mixed state $\rho$ in $2\times 2\times 3$, demonstrating a brute-force approximation of \smash{$\hat{\Upsilon}\equiv\hat{\Upsilon}^{(1,2,3)}(\rho)\approx 0.54$}, which is shown by the planar surface.}
\label{fig:6}
\end{figure}

The adaptation of the ent to mixed states $\rho$ requires the convex-roof extension (see \App{App.J}). For the modal ent or the partitional ent, this simply means using \Eq{36} or \Eq{42} \textit{always}, meaning \textit{no} exceptions occur when $\text{dim}(\mathbf{m})\!=\!N$ or \smash{$\text{dim}(\mathbf{m}^{(\mathbf{T})})\!=\!N$}.  For low-$n$ rank-2 states, brute-force approximation of the ent is feasible, as \Fig{6} shows.
%...............................................................................
%               VI.C.1 Example: Ent for Rank-2 Mixed States
\subsubsection{\label{sec:VI.C.1}Example: Ent for Rank-2 Mixed States}
\vspace{-11pt}
%_______________________________________________________________________________
%Unrounded ent value: \hat{\Upsilon} = 0.543158486929460
%_______________________________________________________________________________
However, for higher-rank states the brute-force method is not practical. Thus, finding a computable form of the ent for mixed states is a vital topic for future research.
%                               End of VI.C.1
%...............................................................................
%...............................................................................
%              VI.C.2 GM Ent and Comparison to GM Concurrence
\subsubsection{\label{sec:VI.C.2}GM Ent and Comparison to GM Concurrence}
The GM ent for mixed $\rho$ is, using \smash{$\Upsilon _{\text{GM}} (\rho )$} from \Eq{48},
%===============================================================================
\begin{equation}%                  Equation 50
\hat \Upsilon _{\text{GM}} (\rho ) \equiv \mathop {\min }\limits_{\scriptstyle \{ p_j ,\rho _j \} \text{s.t.} \hfill \atop 
  \scriptstyle \rho  = \sum\nolimits_j {p_j \rho _j }  \hfill}\! \left( {\sum\nolimits_j {p_j \Upsilon _{\text{GM}} (\rho _j )} } \right)\!.
\label{eq:50}
\end{equation}
%===============================================================================
The GM entanglement does not require the same bipartition to be used for each pure member $\rho_j$ of the decomposition of $\rho$, so this affects the physical meaning of GM entanglement for mixed states, by any measure.

A simple way to compare the GM ent to the GM concurrence \cite{MCCS} is to first find the \textit{optimal decomposition} that minimizes the average GM \textit{ent}, and then use that decomposition to compute the GM \textit{concurrence} to see if that is truly the minimum.  Since both measures are entanglement monotones, we expect agreement, so this is a nice check on the behavior of the ent.
%_______________________________________________________________________________
\begin{figure}[H]%         FIGURE 7 (made with entPaperFig7.m)
\centering
\includegraphics[width=1.00\linewidth]{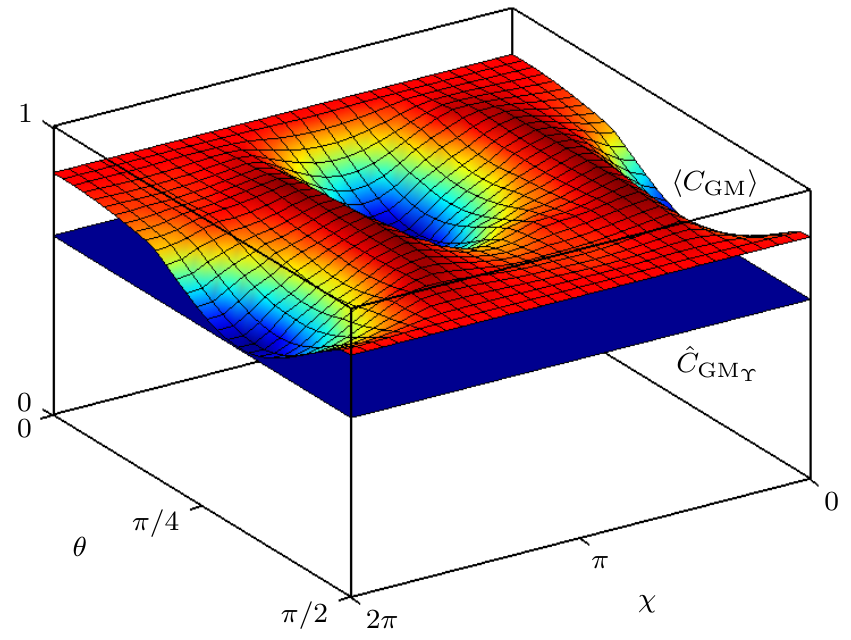}%
\vspace{-4pt}
\caption[]{(color online) The curved surface approximates the average GM concurrence $\langle C_{\text{GM}}\rangle\equiv\langle C_{\text{GM}}(\rho)\rangle$ \cite{MCCS} shown for 30 divisions of each decomposition variable $\theta$ and $\chi$ (see \App{App.J}) of an arbitrary rank-2 mixed state $\rho$ in $2\times 2\times 2$.  The planar surface approximates \smash{$\hat{C}_{\text{GM}_{\Upsilon}}\equiv \min(\langle C_{\text{GM}}\rangle_{\Upsilon})$}, the average $C_{\text{GM}}$ as computed using the optimal decomposition found from approximately minimizing the average GM ent $\Upsilon_{\text{GM}}$ (thus \smash{$\hat{C}_{\text{GM}_{\Upsilon}}$} is the ``ent-driven GM concurrence''). The agreement is excellent since \smash{$\hat{C}_{\text{GM}_{\Upsilon}}\approx 0.6209$} while the approximate target value is \smash{$\hat{C}_{\text{GM}}\equiv \min(\langle C_{\text{GM}}\rangle)\approx 0.6209$}.}
\label{fig:7}
\end{figure}
%Unrounded values:
%\hat{C}_{\text{GM}}            = 0.620912906097194
%\hat{C}_{\text{GM}_{\Upsilon}} = 0.620912906097194
%\hat{\Upsilon}=0.385532843142468 (correctly nonzero, unimportant in this plot)
%_______________________________________________________________________________
As shown in \Fig{7}, since the GM ent was able to find the decomposition that produced the minimum average GM concurrence over all decompositions of $\rho$, this shows that the GM ent qualitatively agrees with the GM concurrence about which decomposition is the least GM entangled, despite the two measures assigning different values to such entanglement.  This is because both are compatible with each other as entanglement monotones.
%                                End of VI.C.2
%...............................................................................
%                                 End of VI.C
%-------------------------------------------------------------------------------
%                                 END of VI
%*******************************************************************************
%*******************************************************************************
%                             VII. CONCLUSIONS
\section{\label{sec:VII}CONCLUSIONS}
%_______________________________________________________________________________
\begin{figure}[H]%Not a figure; Puts hypertarget at top of column to fix problem
\centering
\vspace{-12pt}
\setlength{\unitlength}{0.01\linewidth}
\begin{picture}(100,0)
\put(1,25){\hypertarget{Sec:VII}{}}
\end{picture}
\end{figure}
\vspace{-39pt}
%_______________________________________________________________________________
The ent, originally proposed in \cite{HedD}, has been shown here to be a highly useful entanglement measure, facilitating not only the quantification of multipartite entanglement but also \textit{the construction} of multipartite entangled states.  Furthermore, it can \textit{surpass} other measures, making up for their lack of scaling as shown for the logarithmic-negativity example.  The ent was even shown to agree with other entanglement monotones such as the GM concurrence, in its form as the GM ent.

The main feature that makes the ent so useful is its physical definition of what entanglement means in an $N$-partite setting.  Since the condition of $N$-partite separability for pure states is that they have \textit{product form} over all $N$ modes as $\rho  = \rho ^{(1)}  \otimes  \cdots  \otimes \rho ^{(N)}$ where $\rho ^{(m)}$ is \textit{pure} for all $m \in 1, \ldots ,N$, then \textit{maximal $N$-partite entanglement is when the reduced states of all of the $N$ modes all simultaneously have the lowest purity they can possibly have, given any pure parent state $\rho$}.  This definition captures the physics of the ent, meaning that the most nonlocal correlation between all $N$ modes happens when each reduction yields a state that is as mixed as possible.

The above clause ``given any pure parent $\rho$'' is due to the quirk of multipartite systems that if $\rho$ is pure, the minimal purity of each mode is not necessarily the ideal $\frac{1}{n_{m}}$ of an isolated $n_{m}$-level system.  This accounts for the ent's normalization factor.  

We \textit{could} have defined the ent differently, by defining the unitized purity in terms of the \textit{physically achievable minimum purity} as the \textit{contextually unitized reduction purity} \smash{$P_{\text{CU}}^{(m)}\equiv {{(P(\redrho{m} ) - P_{\text{MP}}^{(m)} )}}/{{(1 - P_{\text{MP}}^{(m)}) }}$}, where from \Eq{4}, $P_{\text{MP}}^{(m)}\equiv P_{\text{MP}}^{(m)}(L_{*})$, yielding the \textit{alternative ent} as \smash{$\widetilde{\Upsilon}  \equiv 1 - \frac{1}{N}\sum\nolimits_{m = 1}^N {{{(P(\redrho{m} ) - P_{\text{MP}}^{(m)} )}}/{{(1 - P_{\text{MP}}^{(m)}) }}} $}\rule{0pt}{9pt}, with the benefit that each \smash{$P_{\text{CU}}^{(m)}$}\rule{0pt}{10pt} could truly get down to $0$ since \smash{$P(\redrho{m} ) \in [P_{\text{MP}}^{(m)} ,1]$} for pure $\rho$, meaning no separate normalization factor is needed.  However, since \smash{$P_{\text{MP}}^{(m)}$} is calculated with the original unnormalized ent (see \Eq{D.32}) based on \textit{isolated} unitized reduction purities, it seemed more natural to let that motivation stand as part of the main definition in \Eq{2}.  Both definitions agree on ordering of states and on what constitutes maximal entanglement and separability, only their value scaling may differ.

The ent also lead to the parameterization of all pure entangled states, and a hypothetical multipartite Schmidt decomposition.  Furthermore, it enabled the \textit{proof} of the maximally entangled basis (MEB) theorem of \cite{HedX}.

The \textit{ent vector} of \Sec{VI.A} and \textit{ent array} of \Sec{VI.B} give us the tools to adapt the ent for \textit{any} other definition of multipartite entanglement, not just Werner's, as we showed by easily using \textit{partitional ent} (from the ent array) to define the GM ent.  As mentioned earlier, not all multipartite entanglement measures are necessarily useful for a given application.  Nevertheless, as an entanglement monotone, the ent can be adapted to agree with \textit{any} other monotone measure because all monotones agree with each other, and in fact are also equivalent to nearest-separable-state measures \cite{StKB}.

In closing, the ent is a powerful and versatile entanglement measure, and it has immediately yielded several new and interesting results about the structure of multipartite entangled states.  The main difficulty of the ent is that it has the same problem that all monotones share; while it can be applied to mixed states with convex-roof extension, that is only tractable for low-rank states.  Thus, a major goal for future work is obtaining a computable form of the ent for mixed states.  Nevertheless, it is hoped that the ent will help increase our understanding of many-body entanglement.
%                                END of VI
%*******************************************************************************
%*******************************************************************************
%                               ACKNOWLEDGEMENTS
\begin{acknowledgments}
Funding for much of this work came from the Innovation and Entrepreneurship Doctoral Fellowship at Stevens Institute of Technology.  Many thanks to Ting Yu for helpful feedback and discussions.
\end{acknowledgments}
%                            END of ACKNOWLEDGEMENTS
%*******************************************************************************
%*******************************************************************************
%                                   APPENDIX
\begin{appendix}
%-------------------------------------------------------------------------------
%                   App.A. Identifying the Relevant Reductions
\section{\label{sec:App.A}Identifying the Relevant Reductions}
%_______________________________________________________________________________
\begin{figure}[H]%Not a figure; Puts hypertarget at top of column to fix problem
\centering
\vspace{-12pt}
\setlength{\unitlength}{0.01\linewidth}
\begin{picture}(100,0)
\put(1,25){\hypertarget{Sec:VIII}{}\hypertarget{Sec:App.A}{}}
\end{picture}
\end{figure}
\vspace{-39pt}
%_______________________________________________________________________________
As proved in \App{App.C}, the necessary and sufficient condition for pure-state $N$-partite separability is that all reductions are pure.  But \textit{which} reductions are relevant?

\textit{The relevant reductions are determined by the coincidence behavior of the physical system}.

For example, consider two elementary particles X and Y, where X has two levels and Y has four levels.  This is a $2\times 4$ system, so $N=2$, since we cannot break Y down further; \textit{it has no internal coincidences}.  Every measurement of the pair X and Y yields a coincidence between some outcome of X and some outcome of Y; we always get \textit{two} numbers because this is a truly \textit{bipartite} system.

\textit{Mathematically}, we could expand Y in a bipartite basis as if it were a $2\times 2$ system, and even compute its ``entanglement properties.''   However that would have no relevance to Y as a physical object because Y \textit{only} exists in a single-particle manifold, meaning that if we look at its hypothetical reductions \textit{in any bipartite basis}, \textit{one of the reductions will always be a vacuum state}.  Thus, there can be no correlations between Alice-and-Bob results within Y, even if Y is ``maximally entangled,'' since Alice's measurements cause the only particle to be destroyed and Bob is left with vacuum in all cases.

Similarly if we mathematically treat a $2\times 2\times 2$ system as $2\times 4$, we would be ignoring the \textit{triple coincidence} behavior that is \textit{always} the reality for that physical system.

Thus, although we can mathematically use different multipartite perspectives, \textit{the proper multipartite definition is always clearly defined by the type of coincidences that describe the actual behavior of the physical system}.

Thus, the above is our motivation for focusing on $N$-partite separability in this paper. Measures based on \textit{biseparability} \cite{MCCS}, used in \Sec{VI.B.1} for the GM ent, do not guarantee separability between particular modes.  For example, a biseparability measure can report that a state like \smash{$\rho  = p_1 \rho _1^{(1)}  \otimes \rho _1^{(2,3)}\!  + p_2 \rho _2^{(1,2)}  \otimes \rho _2^{(3)}$} is ``separable,'' (really meaning \textit{biseparable}){\kern 1pt} even{\kern 1pt} if{\kern 1pt} \smash{$\rho _2^{(1,2)}$} were maximally entangled over modes $1$ and $2$.  In contrast, $N$-partite separability-based measures such as the ent would report \smash{$\rho  = p_1 \rho _1^{(1)}  \otimes \rho _1^{(2)} \otimes \rho _1^{(3)} + p_2 \rho _2^{(1)}  \otimes \rho _2^{(2)}\otimes \rho _2^{(3)}$} as separable.  (The partitional ent of \Sec{VI.B} allows us to be even more specific, but the principle is the same; separability between any set of modes means finding a decomposition where every pure state of the ensemble is in product states of those modes).  Therefore, we focus on the $N$-partite separability definition of \cite{Wern,ZHSL}.
%                                End of App.A
%-------------------------------------------------------------------------------
%-------------------------------------------------------------------------------
%                       App.B. Review of Reduced States
\section{\label{sec:App.B}Review of Reduced States}
%_______________________________________________________________________________
\begin{figure}[H]%Not a figure; Puts hypertarget at top of column to fix problem
\centering
\vspace{-12pt}
\setlength{\unitlength}{0.01\linewidth}
\begin{picture}(100,0)
\put(1,25){\hypertarget{Sec:App.B}{}}
\end{picture}
\end{figure}
\vspace{-39pt}
%_______________________________________________________________________________
%...............................................................................
%                     App.B.1 Reductions to a Single Mode
\subsection{\label{sec:App.B.1}Reductions to a Single Mode}
For an $N$-partite system in state $\rho$ where subsystem (mode) $m$ has $n_m$ levels so $\rho$ has $n=n_{1}\cdots n_{N}$ levels, the \textit{reduction for mode} $m$ (meaning the state we perceive if we ignore all modes except $m$) is \cite{HedD}
%===============================================================================
\begin{equation}%                  Equation B.1
\redrho{m}  \equiv \text{tr}_{\mbarsub} (\rho ) \equiv\!\! \sum\limits_{\mathbf{t}_{\mbarsub}  = \mathbf{1}_{\mbarsub} }^{\mathbf{n}_{\mbarsub} } \!\!\!{\langle \mathbf{t}_{\mbarsub} |\rho\, |\mathbf{t}_{\mbarsub} \rangle },
\label{eq:B.1}
\end{equation}
%===============================================================================
where the ``check'' in $\redrho{m}$ indicates that it is a \textit{reduction} of \textit{parent state} $\rho$ (and not merely an isolated system of same size as mode $m$), and the bar in $\mbar$ means ``not $m$'' to indicate that we are to trace over all subsystems whose labels are \textit{not} $m$, and $|\mathbf{t}_{\mbarsub} \rangle$ is the partial tracing basis,
%===============================================================================
\begin{equation}%                  Equation B.2
|\mathbf{t}_{\mbarsub} \rangle  \equiv \mathop  \otimes \limits_{k = 1}^N \left( {\begin{array}{*{20}l}
   {I^{(k)} } & {k = m}  \\
   {|t_k \rangle } & {k \ne m}  \\
\end{array}} \right)\!,
\label{eq:B.2}
\end{equation}
%===============================================================================
where $\{|t_k \rangle\}$ is any complete basis for mode $k$, and where $\mathbf{t}_{\mbarsub} \! \equiv\! t_1 , \ldots ,t_{m - 1} ,t_{m + 1} , \ldots ,t_N$, $\mathbf{1}_{\mbarsub} \! \equiv\! 1_1 , \ldots ,1_{m - 1} ,$ $1_{m + 1} , \ldots ,1_N$, and $\mathbf{n}_{\mbarsub}  \equiv n_1 , \ldots ,n_{m - 1}, n_{m + 1} , \ldots ,n_N$, and we use the convention that all ket labels start on $1$, so that $t_k  \in 1, \ldots ,n_k$, and are generic in the sense that $|1\rangle$ is not necessarily a Fock state, rather just the first basis state for a mode over which we are tracing.

Partial tracing reveals the reductions to be sums of elements of the parent state $\rho$, as
%===============================================================================
\begin{equation}%                  Equation B.3
(\redrho{m})_{a_m ,b_m }  =\! \sum\limits_{\mathbf{k}_{\mbarsub}  = \mathbf{1}_{\mbarsub} }^{\mathbf{n}_{\mbarsub} }\!\! {\left. {\rho _{R_\mathbf{a}^{\{ N,\mathbf{n}\} } ,R_\mathbf{b}^{\{ N,\mathbf{n}\} } } } \right|_{\scriptstyle \mathbf{a}_{\mbarsub}  = \mathbf{k}_{\mbarsub}  \hfill \atop 
  \scriptstyle \mathbf{b}_{\mbarsub}  = \mathbf{k}_{\mbarsub}  \hfill} },
\label{eq:B.3}
\end{equation}
%===============================================================================
where $\mathbf{n}\equiv n_{1},\ldots,n_{N}$, $\mathbf{a}\equiv a_{1},\ldots,a_{N}$, and indices of $\rho$ are computed with \textit{the indical register function},
%===============================================================================
\begin{equation}%                  Equation B.4
\begin{array}{*{20}l}
   {R_\mathbf{a}^{\{ N,\mathbf{n}\} } } &\!\! { \equiv 1 +\! \sum\limits_{m = 1}^N {\left( {(a_m  - 1)\!\!\!\!\prod\limits_{j = m + 1}^N \!\!\!\!{n_j } } \right)}}  \\
   {} &\!\! { = (a_1  - 1)(n_2  \cdots n_N )\! +\! (a_2  - 1)(n_3  \cdots n_N ) \!+  \cdots }  \\
   {} &\!\! {\;\;\;\, + (a_{N - 2}  - 1)(n_{N - 1} n_N )\! +\! (a_{N - 1}  - 1)n_N \! +\! a_N ,}  \\
\end{array}
\label{eq:B.4}
\end{equation}
%===============================================================================
where $a_m  \in 1, \ldots ,n_m \;\forall m$ so that \Eq{B.4} converts vector index $\mathbf{a}$, whose labels start on $1$, to a scalar index $a$ with the same convention in the parent system.  

Thus we can say that \Eq{B.3} is in \textit{row-column form} because the subscripts refer to row and column number of the matrix elements, so that $(\redrho{m} )_{a_m ,b_m }\equiv\langle a_m | \redrho{m} |b_m \rangle$ and $\rho_{a,b}\equiv \langle a|\rho |b\rangle$, where $a\equiv R_\mathbf{a}^{\{ N,\mathbf{n}\} }$ and $b\equiv R_\mathbf{b}^{\{ N,\mathbf{n}\} }$.

The relationship of the reduction elements to the parent elements is independent of the partial-tracing basis, and we include it here because the reduction purities play a crucial role in the entanglement of the parent state.
%                              End of App.B.1
%...............................................................................
%...............................................................................
%                      App.B.2 Multipartite Reductions
\subsection{\label{sec:App.B.2}Multipartite Reductions}
More generally, we can vectorize $m$ in \Eq{B.1} to handle \textit{multipartite reduction} to multiple potentially noncontiguous subsystems.  The $S$-partite reduction to a composite subsystem of modes $\mathbf{m} \equiv m_1 , \ldots ,m_S$ (when the label values of $m_{k}$ are in order) is the \textit{multipartite partial trace},
%===============================================================================
\begin{equation}%                  Equation B.5
\redrho{\mathbf{m}}  \equiv \text{tr}_{\mathbf{\mbarsub}} (\rho ) \equiv\!\! \sum\limits_{\mathbf{t}_{\mathbf{\mbarsub}}  = \mathbf{1}_{\mathbf{\mbarsub}} }^{\mathbf{n}_{\mathbf{\mbarsub}} } \!\!\!{\langle \mathbf{t}_{\mathbf{\mbarsub}} |\rho\, |\mathbf{t}_{\mathbf{\mbarsub}} \rangle },
\label{eq:B.5}
\end{equation}
%===============================================================================
where $\mathbf{\mbar}$ means ``not $\mathbf{m}$,'' and
%===============================================================================
\begin{equation}%                  Equation B.6
|\mathbf{t}_{\mathbf{\mbarsub}} \rangle  \equiv \mathop  \otimes \limits_{k = 1}^N \left( {\begin{array}{*{20}l}
   {I^{(k)} } & {k \in \mathbf{m}}  \\
   {|t_k \rangle } & {k \notin \mathbf{m}}  \\
\end{array}} \right)\!,
\label{eq:B.6}
\end{equation}
%===============================================================================
where all other definitions match \Eq{B.2}, except that here, quantities of the form{\kern 1.5pt} \smash{$\mathbf{f}_{\mathbf{\mbarsub}}$} are made by removing elements $\{ f_{m_1 } , \ldots ,f_{m_S } \}${\kern -1.8pt} from{\kern -1.8pt} $\{ f_1 , \ldots ,f_N \}$.

Here, the multipartite reduction elements in terms of the elements of parent-state $\rho$ are (in row-column form),
%===============================================================================
\begin{equation}%                  Equation B.7
(\redrho{\mathbf{m}})_{a,b}  =\! \sum\limits_{\mathbf{k}_{\mathbf{\mbarsub}}  = \mathbf{1}_{\mathbf{\mbarsub}} }^{\mathbf{n}_{\mathbf{\mbarsub}} }\!\!\! {\left. {\rho _{R_\mathbf{a}^{\{ N,\mathbf{n}\} } ,R_\mathbf{b}^{\{ N,\mathbf{n}\} } } } \right|_{\scalebox{0.80}{$\begin{array}{*{20}l}
 \mathbf{a}_{\mathbf{\mbarsub}}\! &\!\!\! = \mathbf{k}_{\mathbf{\mbarsub}} \\ 
 \mathbf{b}_{\mathbf{\mbarsub}}\! &\!\!\! = \mathbf{k}_{\mathbf{\mbarsub}} \\ 
 \mathbf{a}_{\mathbf{m}}\! &\!\!\! = \rule{0pt}{8pt}\mathbf{c}_a^{\{ S,\mathbf{n}_{\mathbf{m}} \} }\rule{0pt}{12.5pt} \\ 
 \mathbf{b}_{\mathbf{m}}\! &\!\!\! = \mathbf{d}_b^{\{ S,\mathbf{n}_{\mathbf{m}} \} } \\ 
 \end{array}$}} }\!\!\!,
\label{eq:B.7}
\end{equation}
%===============================================================================
where \smash{$R_\mathbf{a}^{\{ N,\mathbf{n}\} }$} is from \Eq{B.4}, and \smash{$\mathbf{c}_{\mathbf{m}}\equiv c_{m_{1}},\ldots,c_{m_{S}}$}, and \smash{$\mathbf{c}_a^{\{ N,\mathbf{n} \} }$} is the \textit{inverse indical register function} of \Eq{H.1} that maps scalar $a$ to vector $\mathbf{c}$.
%                              End of App.B.2
%...............................................................................
%...............................................................................
%           App.B.3 Multipartite Reductions with Mode Reordering
\subsection{\label{sec:App.B.3}Multipartite Reductions with Mode Reordering}
For quantities such as modal ent and partitional ent in \Sec{VI}, we generally need multipartite reductions for which elements of $\mathbf{m}$ have a \textit{different order} than the original mode labels $1,\ldots,N$.

A simple way to handle this is to first reduce to the intended modes in the original system order, and then convert to the new order, so that \Eq{B.5} generalizes to
%===============================================================================
\begin{equation}%                  Equation B.8
\redrho{\mathbf{m}}  \equiv \Pi^{(\mathbf{m})}\redrho{\mathbf{m}'}\Pi^{(\mathbf{m})\dag},
\label{eq:B.8}
\end{equation}
%===============================================================================
where the \textit{modal permutation unitary} is
%===============================================================================
\begin{equation}%                  Equation B.9
\Pi ^{(\mathbf{m})}  \equiv \sum\limits_{\mathbf{k}_{\mathbf{m}}  = \mathbf{1}_{\mathbf{m}} }^{\mathbf{n}_{\mathbf{m}} } \!{|\,\mathbf{k}_{\mathbf{m}} \rangle \langle \mathbf{k}_{\mathbf{m}'} |},
\label{eq:B.9}
\end{equation}
%===============================================================================
with new basis members $|\,\mathbf{k}_{\mathbf{m}} \rangle\!\equiv\! |k_{m_1 } \rangle  \otimes  \cdots  \otimes |k_{m_S } \rangle$ and old basis members $|\mathbf{k}_{\mathbf{m}'} \rangle\!\equiv \!|k_{m'_1 } \rangle  \otimes  \cdots  \otimes |k_{m'_S } \rangle$, all in the same local generic computational basis we have been using, where $\mathbf{n}_{\mathbf{m}}\equiv (n_{m_{1}},\ldots,n_{m_{S}})$, $\mathbf{n}_{\mathbf{m}'}\!\equiv \!(n_{m'_{1}},\ldots,n_{m'_{S}})$, $n_{\mathbf{m}}\equiv n_{m_{1}}\cdots n_{m_{S}}$, and $\mathbf{m}'\!\equiv\!\text{sort}_{\uparrow}(\mathbf{m}) \equiv (m'_1,\ldots,m'_S )$ is the set of label values of $\mathbf{m}$ in increasing order.  For example, if $\mathbf{m}\!=\!(2,1,3)$, then $\mathbf{m}'\!=\!(1,2,3)$.

In \Eq{B.9}, $\mathbf{n}$ has the \textit{original} mode order, because the mode reordering was only specified for the \textit{reduction modes} $\mathbf{m}$, and the structure of the reordered reduction is $\mathbf{n}_{\mathbf{m}}$, which allows $\redrho{\mathbf{m}}$ to mean the reduction for the modes of $\mathbf{m}$ \textit{for any order of the modes in} $\mathbf{m}$.
%                              End of App.B.3
%...............................................................................
%                                End of App.B
%-------------------------------------------------------------------------------
%-------------------------------------------------------------------------------
%               App.C. Proof: Ent is an Entanglement Monotone
\section{\label{sec:App.C}Proof: Ent is an Entanglement Monotone}
%_______________________________________________________________________________
\begin{figure}[H]%Not a figure; Puts hypertarget at top of column to fix problem
\centering
\vspace{-12pt}
\setlength{\unitlength}{0.01\linewidth}
\begin{picture}(100,0)
\put(1,30){\hypertarget{Sec:App.C}{}}
\end{picture}
\end{figure}
\vspace{-39pt}
%_______________________________________________________________________________
Based on \cite{HedD}, the first step to proving that the ent (or any function) is an entanglement monotone is to show that it is a necessary and sufficient measure of entanglement for \textit{pure} states.  We then extend it to mixed states.
%...............................................................................
%   App.C.1 Proof that the Ent is a Necessary and Sufficient Entanglement Measure for Pure States
\subsection{\label{sec:App.C.1}Proof that the Ent is a Necessary and Sufficient Entanglement Measure for Pure States}
First define the reduction product operator,
%===============================================================================
\begin{equation}%                  Equation C.1
\varsigma (\rho ) \equiv \redrho{1}  \otimes  \cdots  \otimes \redrho{N},
\label{eq:C.1}
\end{equation}
%===============================================================================
which will always be a physical state, where $\redrho{m}$ are the reductions as in \Eq{B.1}, and having purity
%===============================================================================
\begin{equation}%                  Equation C.2
P(\varsigma (\rho ))\! =\! \text{tr}((\redrho{1})^2  \otimes  \cdots  \otimes (\redrho{N})^2 )\! =\! P(\redrho{1}{\kern -0.5pt}) \cdots P(\redrho{N}{\kern -0.5pt}),
\label{eq:C.2}
\end{equation}
%===============================================================================
where each reduction purity generally obeys
%===============================================================================
\begin{equation}%                  Equation C.3
P(\redrho{m}) \in [{\textstyle{1 \over {n_m }}},1]
\label{eq:C.3}
\end{equation}
%===============================================================================
where $n_m  \ge 2$ (we will see that the lower limit in \Eq{C.3} can be higher than ${\textstyle{1 \over {n_m }}}$ depending on the parent state $\rho$, but here the positivity of the range and the upper limit are all that matter).

\textit{Separable} pure states $\rho_{S}$ always have product form,
%===============================================================================
\begin{equation}%                  Equation C.4
\rho _S  = \redrho{1}_S \otimes  \cdots  \otimes \redrho{N}_S = \varsigma (\rho _S ),
\label{eq:C.4}
\end{equation}
%===============================================================================
with purity $P(\rho_{S})\!=\!1$, so tracing the square of \Eq{C.4} gives
%===============================================================================
\begin{equation}%                  Equation C.5
1 = P(\redrho{1}_S ) \cdots P(\redrho{N}_S ) = P(\varsigma (\rho _S )).
\label{eq:C.5}
\end{equation}
%===============================================================================
Furthermore, given that each reduction purity obeys \Eq{C.3}, and is therefore a positive number less than or equal to $1$, then \textit{the only way} \Eq{C.5} can be true is if
%===============================================================================
\begin{equation}%                  Equation C.6
P(\redrho{m}_S ) = 1\;\,\forall m,
\label{eq:C.6}
\end{equation}
%===============================================================================
so for pure separable states, \textit{all} reductions must be pure.  Thus, we have proven that \Eq{C.6} is a \textit{necessary condition for pure-state separability}, since if a pure state is separable, then it \textit{must} have all pure reductions (where the relevant reductions are defined in \App{App.A}).

To show that \Eq{C.6} is \textit{also sufficient} for pure-state separability (meaning that if \Eq{C.6} is true, then the pure parent state must be separable), first observe that an \textit{entangled} pure state $\rho_E$ is defined by \textit{not} being separable, meaning that since it is pure, it cannot be a product state.  Thus, by definition,
%===============================================================================
\begin{equation}%                  Equation C.7
\rho _E  \ne \varsigma (\rho _E ),
\label{eq:C.7}
\end{equation}
%===============================================================================
which states that $\rho_E$ cannot be a product of its reductions.  Then, squaring \Eq{C.7} and tracing, we can use the fact that $P(\rho _E ) = 1$ to get
%===============================================================================
\begin{equation}%                  Equation C.8
1 \ne P(\varsigma (\rho _E )),
\label{eq:C.8}
\end{equation}
%===============================================================================
and expanding the right side of \Eq{C.8} gives
%===============================================================================
\begin{equation}%                  Equation C.9
1 \ne P(\redrho{1}_E ) \cdots P(\redrho{N}_E ).
\label{eq:C.9}
\end{equation}
%===============================================================================
Here, each reduction of $\rho_E$ still obeys $P(\redrho{m}_{E}) \in [{\textstyle{1 \over {n_m }}},1]$ as in \Eq{C.3}, yet the \textit{only} way \Eq{C.9} can be true is if
%===============================================================================
\begin{equation}%                  Equation C.10
P(\redrho{m}_E ) < 1\;\;\text{for at least one}\;\,m,
\label{eq:C.10}
\end{equation}
%===============================================================================
since that is necessary to avoid the case of all reductions being pure which would violate \Eq{C.9}. Thus, if a pure state is not separable, then \textit{at least one of its reductions is not pure}.  Since a state that is not separable is entangled by definition, then we have proven that \Eq{C.10} \textit{is a necessary condition for pure-state entanglement}.

Since all states are either separable or entangled, the \textit{failure of a necessary condition for entanglement is a sufficient condition for separability}.  The failure of \Eq{C.10} is when \textit{not even one reduction $m$ satisfies \Eq{C.10}, meaning that none of the reductions are mixed}. Since the case when no reductions are mixed is when all reductions are pure, then \Eq{C.6} \textit{is} the failure of \Eq{C.10}, and is therefore also a sufficient condition for pure-state separability.

Thus, for multipartite systems, we have proven that a \textit{necessary and sufficient condition for pure-state separability is that all reductions must be pure}, stated in \Eq{C.6}.

Then, since the ent for pure states maps all states satisfying \Eq{C.6} to $0$, while mapping all states that fail \Eq{C.6} to some nonzero number up to a maximum of $1$, then \textit{we have also proven that the ent is a necessary and sufficient measure of pure-state entanglement}.
%                               End of App.C.1
%...............................................................................
%...............................................................................
%   App.C.2 Proof that the Ent is an Entanglement Monotone Extendible to Mixed States
\subsection{\label{sec:App.C.2}Proof that the Ent is an Entanglement Monotone Extendible to Mixed States}
From \cite{Vida}, a necessary and sufficient condition for a function $E(\rho)$ to be an entanglement monotone on pure \textit{bipartite} states, where $E(\rho)$ is defined as either $E(\rho ) \equiv f(\text{tr}_1 (\rho )) = f(\redrho{2})$ or $E(\rho ) \equiv f(\text{tr}_2 (\rho )) = f(\redrho{1})$, is if
%===============================================================================
\begin{equation}%                  Equation C.11
\begin{array}{*{20}l}
   {1.} &\! {E(\rho ) \in \Re ^{[ + )}, }  \\
   {2.} &\! {E(\rho ) = 0\;\;\text{iff}\;\;\rho  \in \mathbb{S},}  \\
   {3.} &\! {f(U^{(m)}\redrho{m} U^{(m)\dag} ) = f(\redrho{m}),}  \\
   {4.} &\! {f(\sigma ^{(m)} ) \ge pf(\sigma _1 ^{(m)} ) + (1 - p)f(\sigma _2 ^{(m)} ),}  \\
\end{array}
\label{eq:C.11}
\end{equation}
%===============================================================================
where $\Re ^{[ + )}$ is the set of nonnegative real numbers, $\mathbb{S}$ is the set of separable states, \smash{$U^{(m)}$} is any $n_m$-level unitary that could act on mode $m$, $\redrho{m}$ is the reduction for mode $m$ as defined in \App{App.B}, $p\in[0,1]$, and \smash{$\sigma ^{(m)}  \equiv p\sigma _1 ^{(m)}  +$} \smash{$ (1 - p)\sigma _2 ^{(m)}$} is an $n_m$-level density operator in the space of mode $m$ where we can take \smash{$\sigma_{1}^{(m)}$} and \smash{$\sigma_{2}^{(m)}$} to be pure states without loss of generality, where $m\in\{1,2\}$.

Any function $E(\rho)$ defined as above and satisfying \Eq{C.11} is an entanglement monotone for pure states and can be extended to mixed states by the convex roof extension (which we will discuss later).  Condition 3 is \textit{local-unitary invariance} and Condition 4 is \textit{concavity}.

To extend these conditions to multipartite systems, note that \textit{bipartite} systems are a special case since the Schmidt decomposition guarantees that the eigenvalues of both reductions are the same. Paired with the local-unitary invariance property, that is why $E(\rho)$ can be defined as a function of \textit{either} reduction.

In multipartite systems, the eigenvalues of all the reductions are not necessarily the same.  However, it is easy to see that \textit{the bipartite case does not actually need to be defined as a function of exclusively one subsystem's reduction}.  Instead, a more general definition would be $E(\rho)\equiv f(\redrho{1},\redrho{2})$ such that local unitary invariance is maintained for all reductions and concavity as well.

Taking this idea further, we define
%===============================================================================
\begin{equation}%                  Equation C.12
E(\rho ) \equiv f(\redrho{1}, \ldots ,\redrho{N} ),
\label{eq:C.12}
\end{equation}
%===============================================================================
as a pure-state multipartite entanglement monotone if
%===============================================================================
\begin{equation}%                  Equation C.13
{\kern -4.0pt}\begin{array}{*{20}l}
   {1.} &\!\!\!\! {E(\rho ) \in \Re^{[ + )}, }  \\
   {2.} &\!\!\!\! {E(\rho ) = 0\;\;\text{iff}\;\;\rho  \in \mathbb{S},}  \\
   {3.} &\!\!\!\! {f(U^{(1)} \redrho{1} U^{(1)\dag}\!\! , \ldots ,U^{(N)} \redrho{N} U^{(N)\dag} )\! =\!\! f(\redrho{1}\!\!\!\!, \ldots ,\redrho{N}{ \kern -0.6pt}),}  \\
   {\!\!{\kern 0.5pt} \begin{array}{*{20}l}
   {4.}  \\
   {}  \\
\end{array}} &\!\!\!\!\!\! {\begin{array}{*{20}l}
   {f(\sigma ^{(1)} , \ldots ,\sigma ^{(N)} ) \ge } &\!\! {pf(\sigma _1 ^{(1)} , \ldots ,\sigma _1 ^{(N)} )}  \\
   {} &\!\! {\! + (1 - p)f(\sigma _2 ^{(1)} , \ldots ,\sigma _2 ^{(N)} ),}  \\
\end{array}}  \\
\end{array}\!\!
\label{eq:C.13}
\end{equation}
%===============================================================================
where $p\in[0,1]$, and \smash{$\sigma ^{(m)}  \equiv p\sigma _1 ^{(m)}  + (1 - p)\sigma _2 ^{(m)}$}, and \smash{$\sigma _1 ^{(m)}$} and \smash{$\sigma _2 ^{(m)}$} are pure.

To see that the ent satisfies the conditions in \Eq{C.13}, first define the candidate function,
%===============================================================================
\begin{equation}%                  Equation C.14
\begin{array}{*{20}l}
   {f(\redrho{1}, \ldots ,\redrho{N} )} &\!\! {\equiv \frac{1}{N}\sum\limits_{m = 1}^N {g(\redrho{m})},}  \\
\end{array}
\label{eq:C.14}
\end{equation}
%===============================================================================
with auxiliary function
%===============================================================================
\begin{equation}%                  Equation C.15
\begin{array}{*{20}l}
   {g(\redrho{m})} &\!\! {\equiv \frac{1}{M}(1 - \frac{n_m \text{tr}({\redrhotiny{m}}^{2} ) - 1}{n_m  - 1}),}  \\
\end{array}
\label{eq:C.15}
\end{equation}
%===============================================================================
where $M$ is a positive normalization constant.  Since \Eq{C.15} satisfies local-unitary invariance due to the basis-independence of the trace,
%===============================================================================
\begin{equation}%                  Equation C.16
\begin{array}{*{20}l}
   {g(U^{(m)} \redrho{m} U^{(m)\dag } )} &\!\! { = {\textstyle{1 \over M}}(1 - \frac{{n_m \text{tr}((U^{(m)} {\redrhotiny{m}} U^{(m)\dag } )^2 ) - 1}}{{n_m  - 1}})}  \\
   {} &\!\! { = {\textstyle{1 \over M}}(1 - \frac{{n_m \text{tr}({\redrhotiny{m}}^{2} U^{(m)\dag } U^{(m)} ) - 1}}{{n_m  - 1}})}  \\
   {} &\!\! { = {\textstyle{1 \over M}}(1 - \frac{{n_m \text{tr}({\redrhotiny{m}}^{2} ) - 1}}{{n_m  - 1}})}  \\
   {} &\!\! { = g(\redrho{m} ),}  \\
\end{array}
\label{eq:C.16}
\end{equation}
%===============================================================================
then \Eq{C.14} has local-unitary invariance in all arguments,
%===============================================================================
\begin{equation}%                  Equation C.17
\begin{array}{*{20}l}
   {f(U^{(1)} \redrho{1} U^{(1)\dag } , \ldots ,U^{(N)} \redrho{N} U^{(N)\dag } )}  \\
   {\begin{array}{*{20}l}
   {} &\;\; { = {\textstyle{1 \over N}}\sum\limits_{m = 1}^N {g(U^{(m)} \redrho{m} U^{(m) \dag } ) = {\textstyle{1 \over N}}\sum\limits_{m = 1}^N {g(\redrho{m} )} } }  \\
   {} &\;\; { = f(\redrho{1} , \ldots ,\redrho{N} ),}  \\
\end{array}}  \\
\end{array}
\label{eq:C.17}
\end{equation}
%===============================================================================
where we used \Eq{C.16}. Thus line 3 of \Eq{C.13} is satisfied.

Next, to prove concavity, first put \smash{$\sigma ^{(m)} \equiv p\sigma _1 ^{(m)}  +$} \smash{$ (1 - p)\sigma _2 ^{(m)}$} into \Eq{C.15} as
%===============================================================================
\begin{equation}%                  Equation C.18
\begin{array}{*{20}l}
   {g(\sigma ^{(m)} )} &\!\! { = g(p\sigma _1 ^{(m)}  + (1 - p)\sigma _2 ^{(m)} )}  \\
   {} &\!\! { = {\textstyle{1 \over M}}(1 - {\textstyle{{n_m \text{tr}((p\sigma _1 ^{(m)}  + (1 - p)\sigma _2 ^{(m)} )^2 ) - 1} \over {n_m  - 1}}})}  \\
   {} &\!\! {={\textstyle{1 \over M}}(1 - {\textstyle{{n_m (p^2  + 2p(1 - p)\text{tr}(\sigma _1 ^{(m)} \sigma _2 ^{(m)} ) + (1 - p)^2 ) - 1} \over {n_m  - 1}}})}  \\
   {} &\!\! {={\textstyle{1 \over M}}{\textstyle{{2n_m p(1 - p)(1 - \text{tr}(\sigma _1 ^{(m)} \sigma _2 ^{(m)} ))} \over {n_m  - 1}}}.}  \\
\end{array}
\label{eq:C.18}
\end{equation}
%===============================================================================
Meanwhile, the convex sum of this function acting on the constituent pure states is
%===============================================================================
\begin{equation}%                  Equation C.19
{\kern -1pt}\begin{array}{*{20}l}
   {pg(\sigma _1 ^{(m)} )\! +\! (1 \!-\! p)g(\sigma _2 ^{(m)} )}  \\
   {\begin{array}{*{20}l}
   {} & { = p{\textstyle{1 \over M}}(1\! -\! {\textstyle{{n_m \text{tr}(\sigma _1 ^{(m)2} ) - 1} \over {n_m  - 1}}})\! +\! (1\! -\! p){\textstyle{1 \over M}}(1\! -\! {\textstyle{{n_m \text{tr}(\sigma _2 ^{(m)2} ) - 1} \over {n_m  - 1}}})}  \\
   {} & { = 0.}  \\
\end{array}}  \\
\end{array}\!\!\!\!\!
\label{eq:C.19}
\end{equation}
%===============================================================================
Then, comparing \Eq{C.18} to \Eq{C.19} using the ``$\sim$'' symbol,
%===============================================================================
\begin{equation}%                  Equation C.20
\begin{array}{*{20}l}
   {g(\sigma ^{(m)} )} &\!\! \sim &\!\! {pg(\sigma _1 ^{(m)} )\! +\! (1\! -\! p)g(\sigma _2 ^{(m)} )}  \\
   {{\textstyle{1 \over M}}{\textstyle{{2n_m p(1 - p)(1 - \text{tr}(\sigma _1 ^{(m)} \sigma _2 ^{(m)} ))} \over {n_m  - 1}}}} &\!\! \sim &\!\! 0  \\
   {1 - \text{tr}(\sigma _1 ^{(m)} \sigma _2 ^{(m)} )} &\!\! \sim &\!\! 0.  \\
\end{array}
\label{eq:C.20}
\end{equation}
%===============================================================================
Since \smash{$\text{tr}(\sigma _1 ^{(m)} \sigma _2 ^{(m)} )$} is an overlap, its maximal value of $1$ only happens when \smash{$\sigma _1 ^{(m)} \! =\! \sigma _2 ^{(m)} $}, thus in general, \smash{$1 \!\ge\! \text{tr}(\sigma _1 ^{(m)} \sigma _2 ^{(m)} )$} so we replace $\sim$ with $\geq$ in \Eq{C.20} to get
%===============================================================================
\begin{equation}%                  Equation C.21
g(\sigma ^{(m)} ) \ge pg(\sigma _1 ^{(m)} ) + (1 - p)g(\sigma _2 ^{(m)} ).
\label{eq:C.21}
\end{equation}
%===============================================================================
Then, summing \Eq{C.21} over all $m$ and dividing by $N$ gives
%===============================================================================
\begin{equation}%                  Equation C.22
\begin{array}{*{20}l}
   {{\textstyle{1 \over N}}\sum\limits_{m = 1}^N {g(\sigma ^{(m)} )} } &\!\!  \ge  &\!\! {p{\textstyle{1 \over N}}\sum\limits_{m = 1}^N {g(\sigma _1 ^{(m)} )} }  \\
   {} &\!\! {} &\!\! { + (1 - p){\textstyle{1 \over N}}\sum\limits_{m = 1}^N {g(\sigma _2 ^{(m)} )} }  \\
   {f(\sigma ^{(1)} , \ldots ,\sigma ^{(N)} )} &\!\!  \ge  &\!\! {pf(\sigma _1 ^{(1)} , \ldots ,\sigma _1 ^{(N)} )}  \\
   {} &\!\! {} &\!\! { + (1 - p)f(\sigma _2 ^{(1)} , \ldots ,\sigma _2 ^{(N)} ),}  \\
\end{array}
\label{eq:C.22}
\end{equation}
%===============================================================================
where we used \Eq{C.14} on both sides, and which shows that the full candidate function obeys the multipartite concavity condition.  Furthermore, the fact that it also obeys it on a single-mode basis due to the relation in \Eq{C.21} is another illustration of the appropriateness of this generalization of the monotone definition.

Now, noting that the candidate $f$ \textit{is} the ent, since
%===============================================================================
\begin{equation}%                  Equation C.23
\begin{array}{*{20}l}
   {f(\redrho{1} , \ldots ,\redrho{N} )} &\!\! { = {\textstyle{1 \over N}}\sum\limits_{m = 1}^N {g(\redrho{m})} }  \\
   {} &\!\! { = {\textstyle{1 \over N}}\sum\limits_{m = 1}^N {{\textstyle{1 \over M}}(1 - {\textstyle{{n_m \text{tr}({\redrhotiny{m}}^{2} ) - 1} \over {n_m  - 1}}})} }  \\
   {} &\!\! { = {\textstyle{1 \over N}}{\textstyle{1 \over M}}(N - \sum\limits_{m = 1}^N {{\textstyle{{n_m \text{tr}({\redrhotiny{m}}^{2}) - 1} \over {n_m  - 1}}}} )}  \\
   {} &\!\! { = {\textstyle{1 \over M}}(1 - {\textstyle{1 \over N}}\sum\limits_{m = 1}^N {{\textstyle{{n_m P(\redrhotiny{m}) - 1} \over {n_m  - 1}}}} )}  \\
   {} &\!\! { = \Upsilon (\rho ),}  \\
\end{array}
\label{eq:C.23}
\end{equation}
%===============================================================================
if $M$ is defined as in \Eq{3}, and since we already proved in \App{App.C.1} that $\Upsilon(\rho)\in[0,1]$ and $\Upsilon(\rho)=0$ for all separable pure states, then these facts together with \Eq{C.17} and \Eq{C.22} prove that the ent is a multipartite entanglement monotone for pure states by the definition in \Eq{C.13}.

Then, since the ent is an entanglement monotone for \textit{pure} states, it can be adapted to all \textit{mixed states} by using the convex roof extension, as
%===============================================================================
\begin{equation}%                  Equation C.24
\hat{\Upsilon}(\rho ) \equiv \mathop {\min }\limits_{\scriptstyle \{ p_j ,\rho _j \} \;\;\text{s.t.} \hfill \atop 
  \scriptstyle \rho  = \sum\nolimits_j p_j \rho _j  \hfill} \!\left(\; {\sum\nolimits_j p_j \Upsilon (\rho _j )} \right).
\label{eq:C.24}
\end{equation}
%===============================================================================
Unfortunately, the number of parameters involved \Eq{C.24} makes it generally infeasible to find the decomposition that minimizes this measure over all possible decompositions.  Thus, while this proves that the ent is extendible to all mixed states, it does not provide a means of efficiently calculating it in this form.  See \App{App.J} for details.
%                               End of App.C.2
%...............................................................................
%                                End of App.C
%-------------------------------------------------------------------------------
%-------------------------------------------------------------------------------
%        App.D. Derivation of Conditions for Maximal Entanglement and Normalization of the Ent
\section{\label{sec:App.D}Derivation of Conditions for Maximal Entanglement and Normalization of the Ent}
%_______________________________________________________________________________
\begin{figure}[H]%Not a figure; Puts hypertarget at top of column to fix problem
\centering
\vspace{-12pt}
\setlength{\unitlength}{0.01\linewidth}
\begin{picture}(100,0)
\put(1,30){\hypertarget{Sec:App.D}{}}
\end{picture}
\end{figure}
\vspace{-39pt}
%_______________________________________________________________________________
Here we derive conditions that yield \Eqs{2}{5}. Since these results are not necessarily obvious, we will build up to them gradually, often with simple examples.   These examples are not the derivations, but rather they  \textit{illustrate} the derivations, so please be patient if an example or section does not seem sufficient; each section builds upon the last until all necessary and sufficient conditions are incorporated at the end.  See \App{App.G} for the \textit{application} of these ideas in an explicit generalized form to produce maximally entangled states.
%...............................................................................
%            App.D.1 True-Generalized X States (TGX States)
\subsection{\label{sec:App.D.1}True-Generalized X States (TGX States)}
First, we restrict ourselves to \textit{simple} states, defined in \cite{HedX} as states for which parent-state elements $\rho_{a,b}\equiv \langle a|\rho|b\rangle$ (in generic computational basis $\{|a\rangle\}\equiv\{|1\rangle,\ldots,|n\rangle\}$) appearing explicitly in the off-diagonal elements of the reductions (see \App{App.B}) are \textit{identically zero} (meaning that they do not merely add to zero).

For example, in $2\times 2$, the reductions of $\rho$ are
%===============================================================================
\begin{equation}%                  Equation D.1
\begin{array}{*{20}l}
   {\redrho{1} } &\!\! { = \left( {\begin{array}{*{20}c}
   {\rho _{1,1}  + \rho _{2,2} } & {\rho _{1,3}  + \rho _{2,4} }  \\
   {\rho _{3,1}  + \rho _{4,2} } & {\rho _{3,3}  + \rho _{4,4} }  \\
\end{array}} \right)}  \\
   {\redrho{2} } &\!\! { = \left( {\begin{array}{*{20}c}
   {\rho _{1,1}  + \rho _{3,3} } & {\rho _{1,2}  + \rho _{3,4} }  \\
   {\rho _{2,1}  + \rho _{4,3} } & {\rho _{2,2}  + \rho _{4,4} }  \\
\end{array}} \right)\!,}  \\
\end{array}
\label{eq:D.1}
\end{equation}
%===============================================================================
so a \textit{simple} parent state must have $\rho_{3,1}=0$, $\rho_{4,2}=0$, $\rho_{2,1}=0$, and $\rho_{4,3}=0$, so that simple parent states are
%===============================================================================
\begin{equation}%                  Equation D.2
\rho  = \!\left( {\begin{array}{*{20}c}
   {\rho _{1,1} } & 0 & 0 & {\rho _{1,4} }  \\
   0 & {\rho _{2,2} } & {\rho _{2,3} } & 0  \\
   0 & {\rho _{3,2} } & {\rho _{3,3} } & 0  \\
   {\rho _{4,1} } & 0 & 0 & {\rho _{4,4} }  \\
\end{array}} \right)\!,
\label{eq:D.2}
\end{equation}
%===============================================================================
which is an ``X state'' since its potentially nonzero elements are in the shape of an X \cite{YuEb}.  Extending the definition of \textit{simple} states to larger systems results in states that do not have an X shape yet are still generally sparse, so they were dubbed ``true generalized X (TGX) states'' in \cite{HedX}, in particular because for $n>4$ they seem to maintain entanglement properties that mere X states do not, yet have the same definition as \Eq{D.2}.  Specifically, \cite{HedX} proposed that TGX states are related to all general states by an entanglement-preserving unitary (EPU), giving strong numerical evidence of this in $2\times 2$ and $2\times 3$, showing that only TGX states could maintain this EPU equivalence in general.  Soon after, EPU equivalence of X states was proven for $2\times 2$ in \cite{MeMG}.

In this paper, we use \textit{TGX} to mean \textit{simple}, however, we clarify that simple states are only TGX \textit{candidates} because the ideal defining property of TGX states is that they contain the minimal set of states such that all general states can be unitarily transformed to TGX states while maintaining their entanglement.  Thus if simple states cannot accomplish this or are too general, we can reserve the name of TGX for the proper set if it exists.
%                               End of App.D.1
%...............................................................................
%...............................................................................
%                   App.D.2 Review of Purity Minimization
\subsection{\label{sec:App.D.2}Review of Purity Minimization}
One advantage of TGX states is that their reductions are all \textit{diagonal}, making it easier to minimize the reduction purities, which is a necessary and sufficient condition for maximal entanglement of pure states (see \App{App.C.1}).

To see how diagonality relates to purity $P\equiv P(\rho)\equiv\text{tr}(\rho^2)$, recall that the off-diagonals of all states obey
%===============================================================================
\begin{equation}%                  Equation D.3
0 \le |\rho _{a,b} | \le \sqrt {\rho _{a,a} \rho _{b,b} } \,,
\label{eq:D.3}
\end{equation}
%===============================================================================
for $a \!\ne\! b\! \in\! 1, \ldots ,n$. The purity in matrix-element form is
%===============================================================================
\begin{equation}%                  Equation D.4
P(\rho ) =\! \sum\limits_{a,b = 1,1}^{n,n} \!\!\!{|\rho _{a,b} |^2 }  = (\sum\limits_{c = 1}^n {\rho _{c,c} ^2 } ) + 2\!\!\!\sum\limits_{a,b = 2,1}^{n,a - 1}\!\!\! {|\rho _{a,b} |^2 }.
\label{eq:D.4}
\end{equation}
%===============================================================================
For pure states, \smash{$|\rho _{a,b} | = \sqrt {\rho _{a,a} \rho _{b,b} }$}, yielding $P=1$ in \Eq{D.4}.  Since mixed states by definition have purity less than $1$, a necessary condition for mixing is
%===============================================================================
\begin{equation}%                  Equation D.5
|\rho _{a,b} | < \sqrt {\rho _{a,a} \rho _{b,b} }\, ,\;\;\text{for at least one pair}\;\; (a \ne b).
\label{eq:D.5}
\end{equation}
%===============================================================================
However, \textit{diagonality is neither necessary nor sufficient for purity minimization of reduced states in general}, since reductions are not truly isolated, being constrained by their parent states.  For example, in $2\times 3$, when the parent is pure, the $3$-level system's reduction is at most rank $2$ by the Schmidt decomposition, which allows \textit{nondiagonal} reductions with the minimal physical purity of $\frac{1}{2}$.  Thus, diagonality is merely \textit{convenient} to achieve maximal mixing, as we will see.  Furthermore, there are \textit{diagonal pure states}, such as the computational basis states.

However, the basis-independence of the trace yields
%===============================================================================
\begin{equation}%                  Equation D.6
P(\rho ) = \sum\limits_{a = 1}^n {\lambda _a ^2 }\,,
\label{eq:D.6}
\end{equation}
%===============================================================================
where $\lambda _a$ are eigenvalues of $\rho$ such that $\lambda _a  \in [0,1]$ and \smash{$\sum\nolimits_{a = 1}^n \!{\lambda _a }\!  =\! 1$}.  {\kern -5pt}This{\kern -1pt} shows{\kern -1pt} that{\kern -1pt} \textit{a necessary condition for mixing is that a state must have more than one nonzero eigenvalue}.  Taking this idea further, it intuitively makes sense that diagonal states with greater numbers of nonzero eigenvalues are more mixed.

In fact, it is well-known that the minimal purity of an isolated $n$-level system is \smash{$P_{\min}=\frac{1}{n}$}, occurring when \smash{$\lambda_a=\frac{1}{n}\;\forall a\in 1,\ldots,n$}.  However, the proof of this is important and we sketch it here to support later results.

Given \Eq{D.6}, nonnegativity and normalization can be incorporated by writing the $\lambda _a$ as squared hyperspherical coordinates \cite{Schl} and then differentiating $P$ with respect to each hyperspherical angle, setting those partial derivatives to zero, and solving for the angle values in the first quadrant.  Induction then leads to the result that the eigenvalues that minimize the purity in all cases are
%===============================================================================
\begin{equation}%                  Equation D.7
\begin{array}{*{20}l}
   {\lambda _1 } &\!\! { = c_{\theta _1 }^2 }  \\
   {\lambda _2 } &\!\! { = s_{\theta _1 }^2 c_{\theta _2 }^2 }  \\
    \;\vdots  &\!\! {}  \\
   {\lambda _{n - 1} } &\!\! { = s_{\theta _1 }^2  \cdots s_{\theta _{n - 2} }^2 c_{\theta _{n - 1} }^2 }  \\
   {\lambda _n } &\!\! { = s_{\theta _1 }^2  \cdots s_{\theta _{n - 2} }^2 s_{\theta _{n - 1} }^2 }  \\
\end{array}\!\text{s.t.}\;\,\theta _k  \!=\! \text{atan} (\sqrt {n\! -\! k}\, ),
\label{eq:D.7}
\end{equation}
%===============================================================================
for $k\in 1,\ldots,n-1$, which yields the result that
%===============================================================================
\begin{equation}%                  Equation D.8
\lambda _a  = {\textstyle{1 \over n}}\;\;\forall a \in 1, \ldots ,n.
\label{eq:D.8}
\end{equation}
%===============================================================================
However, \Eq{D.8} is \textit{not true} for reduced states in general!  The problem is that reduced states inherit constraints from their parent states that isolated states do not have.

The above exercise shows \textit{why} \Eq{D.8} is the ideal to shoot for, and it is then easy to see that \textit{additional constraints will simply mean that we try to get as close to \Eq{D.8} as possible while still satisfying those additional constraints}.

Thus, to minimize the purity of \textit{diagonal} reduced states (such as for TGX states), \textit{the eigenvalues (the main-diagonal elements) must be as numerous as possible, and as evenly-distributed as possible}.
%                               End of App.D.2
%...............................................................................
%...............................................................................
%           App.D.3 Constraints On Reductions Due to Parent States
\subsection{\label{sec:App.D.3}Constraints On Reductions Due to Parent States}
As we saw in \App{App.B}, all reduced states can be expressed explicitly in terms of their parent states.  Since this is independent of the partial-tracing basis, some constraints imposed on reductions by their parent states are:
%+++++++++++++++++++++++++++++++++++++++++++++++++++++++++++++++++++++++++++++++
\begin{itemize}[leftmargin=*,labelindent=4pt]
%~~~~~~~~~~~~~~~~~~~~~~~~~~~~~~~~~~~~~~~~~~~~~~~~~~~~~~~~~~~~~~~~~~~~~~~~~~~~~~~
\item[\textbf{a.}]\hypertarget{constraint:a}{}\textit{Each} reduction contains \textit{each} parent-state main-diagonal element explicitly \textit{exactly once}. Thus the trace of each reduction is explicitly the same sum of matrix elements as the trace of the parent state.
%~~~~~~~~~~~~~~~~~~~~~~~~~~~~~~~~~~~~~~~~~~~~~~~~~~~~~~~~~~~~~~~~~~~~~~~~~~~~~~~
%~~~~~~~~~~~~~~~~~~~~~~~~~~~~~~~~~~~~~~~~~~~~~~~~~~~~~~~~~~~~~~~~~~~~~~~~~~~~~~~
\item[\textbf{b.}]\hypertarget{constraint:b}{}In each mode $m$, its reduction's matrix elements are sums of at most \smash{$n_{\mbarsub}\equiv\frac{n}{n_{m}}$} parent elements (proven by noting that the number of terms in the partial trace sum involves separate sums over every mode \textit{except} the one not being traced over yielding a total of \smash{$n_{1}\cdots n_{m-1} n_{m+1}\cdots n_{N}=n_{\mbarsub}$} terms).  Note that $n_{\mbarsub}$ can be more than, equal to, or less than $n_m$.
%~~~~~~~~~~~~~~~~~~~~~~~~~~~~~~~~~~~~~~~~~~~~~~~~~~~~~~~~~~~~~~~~~~~~~~~~~~~~~~~
%~~~~~~~~~~~~~~~~~~~~~~~~~~~~~~~~~~~~~~~~~~~~~~~~~~~~~~~~~~~~~~~~~~~~~~~~~~~~~~~
\item[\textbf{c.}]\hypertarget{constraint:c}{}If the parent state has $L$ outcomes in superposition (in the computational basis), then $L$ is the maximum number of nonzero terms available to distribute over all the main diagonals of each reduction towards making them as mixed as possible. For example, in \smash{$|\Phi^+\rangle=\frac{1}{\sqrt{2}}(|1\rangle+|4\rangle)$}, $L=2$.
%~~~~~~~~~~~~~~~~~~~~~~~~~~~~~~~~~~~~~~~~~~~~~~~~~~~~~~~~~~~~~~~~~~~~~~~~~~~~~~~
%~~~~~~~~~~~~~~~~~~~~~~~~~~~~~~~~~~~~~~~~~~~~~~~~~~~~~~~~~~~~~~~~~~~~~~~~~~~~~~~
\item[\textbf{d.}]\hypertarget{constraint:d}{}Since the parent's purity is $1$, then \smash{$|\rho _{a,b} | = \sqrt {\rho _{a,a} \rho _{b,b} }$} \textit{for all} parent elements.
%~~~~~~~~~~~~~~~~~~~~~~~~~~~~~~~~~~~~~~~~~~~~~~~~~~~~~~~~~~~~~~~~~~~~~~~~~~~~~~~
%~~~~~~~~~~~~~~~~~~~~~~~~~~~~~~~~~~~~~~~~~~~~~~~~~~~~~~~~~~~~~~~~~~~~~~~~~~~~~~~
\item[\textbf{e.}]\hypertarget{constraint:e}{}A necessary (but not sufficient) condition for entanglement is that a state must have some superposition, meaning that \textit{at least one off-diagonal element must be nonzero}.  In other words, superposition in the computational basis is a necessary condition for entanglement in that basis. Diagonal states have no entanglement since they are convex sums of product-basis states, making them separable by definition.
%~~~~~~~~~~~~~~~~~~~~~~~~~~~~~~~~~~~~~~~~~~~~~~~~~~~~~~~~~~~~~~~~~~~~~~~~~~~~~~~
\end{itemize}
%+++++++++++++++++++++++++++++++++++++++++++++++++++++++++++++++++++++++++++++++
%                               End of App.D.3
%...............................................................................
%...............................................................................
%            App.D.4 Consequences of the Parent-State Constraints
\subsection{\label{sec:App.D.4}Consequences of the Parent-State Constraints}
The effects of the above constraints have several consequences which we discuss here (labeled independently).
%~~~~~~~~~~~~~~~~~~~~~~~~~~~~~~~~~~~~~~~~~~~~~~~~~~~~~~~~~~~~~~~~~~~~~~~~~~~~~~~
%                        App.D.4.a. Compatibility Sets
\subsubsection{\label{sec:App.D.4.a}Compatibility Sets}
\textit{Only certain combinations of parent-state levels are pure-state compatible}. In TGX parent states, some off-diagonals are zero which limits the number of outcomes that can be involved for a state to be pure, since all off-diagonals of a pure state must obey
%===============================================================================
\begin{equation}%                  Equation D.9
|\rho _{a,b} | = \sqrt {\rho _{a,a} \rho _{b,b} } \,.
\label{eq:D.9}
\end{equation}
%===============================================================================
For candidate maximally entangled TGX states that definitely include a certain level designated as the \textit{starting level} $S_L$, such as $S_L =1$ for the first outcome $|1\rangle$, \textit{the set of all possible pure TGX states is the set of normalized matrix elements including \smash{$\rho_{S_{L},S_{L}}$} which all mutually obey \Eq{D.9} in their subspace of TGX space}. As a minimal requirement, these off-diagonals must not be identically zero in a general TGX state.

For example, in $2\times 3$, the reductions are
%===============================================================================
\begin{equation}%                  Equation D.10
\begin{array}{*{20}l}
   {\redrho{1}\! } &\!\! { =\! \left(\! {\begin{array}{*{20}c}
   {\rho _{1,1}  + \rho _{2,2}  + \rho _{3,3} } & {\rho _{1,4}  + \rho _{2,5}  + \rho _{3,6} }  \\
   {\rho _{4,1}  + \rho _{5,2}  + \rho _{6,3} } & {\rho _{4,4}  + \rho _{5,5}  + \rho _{6,6} }  \\
\end{array}}\! \right)}  \\
   {\redrho{2}\! } &\!\! { =\! \left(\! {\begin{array}{*{20}c}
   {\rho _{1,1}  + \rho _{4,4} } & {\rho _{1,2}  + \rho _{4,5} } & {\rho _{1,3}  + \rho _{4,6} }  \\
   {\rho _{2,1}  + \rho _{5,4} } & {\rho _{2,2}  + \rho _{5,5} } & {\rho _{2,3}  + \rho _{5,6} }  \\
   {\rho _{3,1}  + \rho _{6,4} } & {\rho _{3,2}  + \rho _{6,5} } & {\rho _{3,3}  + \rho _{6,6} }  \\
\end{array}}\! \right)\!,}  \\
\end{array}
\label{eq:D.10}
\end{equation}
%===============================================================================
so the TGX space is (representing zeros with dots),
%===============================================================================
\begin{equation}%                  Equation D.11
\rho  =\! \left( {\begin{array}{*{20}c}
   {\rho _{1,1} } &  \cdot  &  \cdot  &  \cdot  & {\rho _{1,5} } & {\rho _{1,6} }  \\
    \cdot  & {\rho _{2,2} } &  \cdot  & {\rho _{2,4} } &  \cdot  & {\rho _{2,6} }  \\
    \cdot  &  \cdot  & {\rho _{3,3} } & {\rho _{3,4} } & {\rho _{3,5} } &  \cdot   \\
    \cdot  & {\rho _{4,2} } & {\rho _{4,3} } & {\rho _{4,4} } &  \cdot  &  \cdot   \\
   {\rho _{5,1} } &  \cdot  & {\rho _{5,3} } &  \cdot  & {\rho _{5,5} } &  \cdot   \\
   {\rho _{6,1} } & {\rho _{6,2} } &  \cdot  &  \cdot  &  \cdot  & {\rho _{6,6} }  \\
\end{array}} \right)\!.
\label{eq:D.11}
\end{equation}
%===============================================================================
If $S_L =1$, then (from column $1$ of $\rho$) the maximum number of levels that can support pure states in this TGX space is $L_{*}=2$, because the only combinations of levels with compatible nonzero off-diagonals are $\{1,5\}$ and $\{1,6\}$. Although at first, column $1$ makes it seem like a $3$-level pure state would be possible involving levels $\{1,5,6\}$, we see that levels $5$ and $6$ are not mutually compatible for pure states since their mutual off-diagonal $\rho_{6,5}$ is identically zero in \Eq{D.11}, which would violate \Eq{D.9}.

Thus, the set of all possible combinations of levels that are mutually compatible with $S_L$ for supporting pure TGX states yields a set of candidates for maximally entangled TGX states involving $S_L$. For example, in $2\times 3$ with $S_L =1$, our candidates are superpositions of $\{|1\rangle,|5\rangle\}$ or $\{|1\rangle,|6\rangle\}$.

We can call the collection of all candidate sets of levels the \textit{compatibility sets}, which merely list all combinations of levels that could support pure TGX states, but do not necessarily represent maximally entangled TGX states.
%                              End of App.D.4.a
%~~~~~~~~~~~~~~~~~~~~~~~~~~~~~~~~~~~~~~~~~~~~~~~~~~~~~~~~~~~~~~~~~~~~~~~~~~~~~~~
%~~~~~~~~~~~~~~~~~~~~~~~~~~~~~~~~~~~~~~~~~~~~~~~~~~~~~~~~~~~~~~~~~~~~~~~~~~~~~~~
%            App.D.4.b. Even Distributions via the Occurrence Matrix
\subsubsection{\label{sec:App.D.4.b}Even Distributions via the Occurrence Matrix}
Since a maximally entangled parent state requires that we \textit{distribute the numerical values of the reduction eigenvalues as numerously and as evenly as possible in each reduction} (as concluded in \App{App.D.2}), then we need to look at how the main-diagonal parent elements are distributed amongst the diagonals of the reductions.  (The fact that reductions of TGX states are diagonal allows us to only look at the diagonals.)

To do this, we construct the \textit{occurrence matrix} by counting the explicit occurrences of each parent main-diagonal in each reduced main-diagonal.  

For example, in $2\times 2$ the occurrence matrix is
%===============================================================================
\begin{equation}%                  Equation D.12
\Omega  =\! \left( {\begin{array}{*{20}c}
   \Omega_{l,\cdots} &\vline & {\redrho{1}_{1,1} } & {\redrho{1}_{2,2} } &\vline & {\redrho{2}_{1,1} } & {\redrho{2}_{2,2} }  \\
   \hline
   \Omega_{1,\cdots} &\vline & 1 & 0 &\vline & 1 & 0  \\
   \Omega_{2,\cdots} &\vline & 1 & 0 &\vline & 0 & 1  \\
   \Omega_{3,\cdots} &\vline & 0 & 1 &\vline & 1 & 0  \\
   \Omega_{4,\cdots} &\vline & 0 & 1 &\vline & 0 & 1  \\
\end{array}} \right)\!,
\label{eq:D.12}
\end{equation}
%===============================================================================
where $\Omega$ is really the ``meat'' of \Eq{D.12}, and the leading row and column are only included here to indicate the structure, and $l$ gives the levels of $\rho$.  To see how to make the rows of $\Omega$, in \Eq{D.1} if we note that the level-$3$ parent{\kern -1pt} diagonal{\kern -1pt} $\rho_{3,3}${\kern -1pt} occurs{\kern -1pt} both{\kern -1pt} in{\kern -1pt} \smash{$\redrho{1}_{2,2}$}{\kern -1pt} and{\kern -1pt} \smash{$\redrho{2}_{1,1}$}, then the occurrence{\kern 1pt} vector{\kern 1pt} \smash{$\Omega_{3,\cdots}$}{\kern 1pt} for{\kern 1pt} $\rho_{3,3}${\kern 1pt} is{\kern 1pt} a $0$ in the \smash{$\redrho{1}_{1,1}$} slot, a $1${\kern 1.2pt} in{\kern 1.2pt} the{\kern 1.2pt} \smash{$\redrho{1}_{2,2}$}{\kern 1.2pt} slot,{\kern 1.2pt} a{\kern 1.2pt} $1${\kern 1.2pt} in{\kern 1.2pt} the{\kern 1.2pt} \smash{$\redrho{2}_{1,1}$}{\kern 1.2pt} slot, and a $0$ in the \smash{$\redrho{2}_{2,2}$} slot, yielding \smash{$\Omega_{3,\cdots}=(0,1|1,0)$}, as in \Eq{D.12}.

Note that each row of $\Omega$ is the concatenation of the transpose of the mode's computational basis vector, such as \smash{$\Omega_{1,\cdots}
\!=\!\left({\binom{1}{0}{\rule{0pt}{6pt}}^{T}\!,\binom{1}{0}{\rule{0pt}{6pt}}^{T}}\right)\!,$} and \smash{$\Omega_{2,\cdots}\!=\!\left({\binom{1}{0}{\rule{0pt}{6pt}}^{T}\!,\binom{0}{1}{\rule{0pt}{6pt}}^{T}}\right)\!,$} etc.  In general,{\kern -1pt} $\Omega${\kern -1pt} has{\kern -1pt} $n${\kern -1pt} rows{\kern -1pt} and{\kern -1pt} \smash{$\sum\nolimits_{m = 1}^N n_m $}{\kern -1pt} columns.

Note that $\Omega$ does not keep track of \textit{values} of the parent diagonals; it simply keeps track of their \textit{occurrences} in the reductions regarding \textit{location}.

Also, each row of $\Omega$ deposits an element of $\rho$ in each mode simultaneously, a consequence of \hyperlink{constraint:a}{Constraint a}.

The compatibility sets for a particular starting level $S_L$ limit which rows of $\Omega$ can be involved in the distribution process. For example, in $2\times 2$, \Eq{D.2} shows that the only compatibility set for $S_L =1$ is $\{1,4\}$, and therefore only rows $1$ and $4$ of $\Omega$ can be involved in constructing a candidate maximally entangled TGX state for this $S_L$.

To visualize all this, the occurrence vectors of compatibility set $\{1,4\}$ are $\Omega_{1,\cdots}=(1,0|1,0)$ and $\Omega_{4,\cdots}=(0,1|0,1)$, which, from \Eq{D.1}, cause a distribution of nonzero values in the reduction-element ``bins'' as
%===============================================================================
\begin{equation}%                  Equation D.13
\begin{array}{*{20}l}
   {\redrho{1} } &\!\! { = \left( {\begin{array}{*{20}c}
   \squareOne & \cdot  \\
   \cdot & \squareOne  \\
\end{array}} \right)\!\! \begin{array}{*{20}c}
   { \leftarrow \Omega _{1, \cdots } }  \\
   { \leftarrow \Omega _{4, \cdots } }  \\
\end{array}}  \\
   {\redrho{2} } &\!\! { = \left( {\begin{array}{*{20}c}
   \squareOne & \cdot  \\
   \cdot & \squareOne  \\
\end{array}} \right)\!\! \begin{array}{*{20}c}
   { \leftarrow \Omega _{1, \cdots } }  \\
   { \leftarrow \Omega _{4, \cdots } }  \\
\end{array}}  \\
\end{array}
\label{eq:D.13}
\end{equation}
%===============================================================================
where each black square represents the occurrence of a parent diagonal. Since the distribution in each reduction is already as wide as possible (as many elements are occupied as possible), then we can use differentiation to show that ideal maximal mixing can be achieved in each reduction if the parent state of superpositions of levels $\{1,4\}$ has balanced probability amplitudes, resulting in the maximally entangled TGX state \smash{$|\Phi\rangle=\frac{1}{\sqrt{2}}(|1\rangle+|4\rangle)$}.

For larger systems, there are more subtleties,\rule{0pt}{10.5pt} which we now develop.  First, for modes of unequal size, it is often only possible to maximize the distribution of parent elements within a \textit{subspace} of one of the modes.

For example, in $2\times 3$, where $S_L =1$ yielded compatibility sets $\{1,5\}$ and $\{1,6\}$, in the case of $\{1,5\}$, the occurrence vectors are $\Omega_{1,\cdots}=(1,0|1,0,0)$ and $\Omega_{5,\cdots}=(0,1|0,1,0)$, meaning that since $L_* =2$, the best distribution possible for the reductions has the form,
%===============================================================================
\begin{equation}%                  Equation D.14
\begin{array}{*{20}l}
   {\redrho{1} } &\!\! { = \left( {\begin{array}{*{20}c}
    \squareOne  &  \cdot   \\
    \cdot  &  \squareOne   \\
\end{array}} \right)\!\! \begin{array}{*{20}c}
   { \leftarrow \Omega _{1, \cdots } }  \\
   { \leftarrow \Omega _{5, \cdots } }  \\
\end{array}}  \\
   {\redrho{2} } &\!\! { = \left( {\begin{array}{*{20}c}
    \squareOne  &  \cdot  &  \cdot   \\
    \cdot  &  \squareOne  &  \cdot   \\
    \cdot  &  \cdot  & \squareZero  \\
\end{array}} \right)\!\! \begin{array}{*{20}c}
   { \leftarrow \Omega _{1, \cdots } }  \\
   { \leftarrow \Omega _{5, \cdots } }  \\
   {}  \\
\end{array}}  \\
\end{array}
\label{eq:D.14}
\end{equation}
%===============================================================================
so that mode 2 can never reach a state of ideal maximal mixing when its parent is a pure TGX state!  In fact, this can be proved for all bipartite systems using the Schmidt decomposition, but here we have duplicated that result from a different perspective.  The other set $\{1,6\}$ has the same feature but in a different order in \smash{$\redrho{2}$}.

Again, minimizing the purity of these reductions yields balanced superposition in the parent TGX state.

For larger systems, we will see examples of \textit{more than once occurrence} in each reduction diagonal.  For now, we explain how to \textit{use} the occurrence matrix.
%                              End of App.D.4.c
%~~~~~~~~~~~~~~~~~~~~~~~~~~~~~~~~~~~~~~~~~~~~~~~~~~~~~~~~~~~~~~~~~~~~~~~~~~~~~~~
%~~~~~~~~~~~~~~~~~~~~~~~~~~~~~~~~~~~~~~~~~~~~~~~~~~~~~~~~~~~~~~~~~~~~~~~~~~~~~~~
%        App.D.4.c. Goals Vector and Automatic Normalization of the Ent
\subsubsection{\label{sec:App.D.4.c}Goals Vector and Automatic Normalization of the Ent}
An important metric to see if a collection of levels is \textit{sufficient} for maximal entanglement in TGX states is the \textit{goals vector}, which is the sum of all occurrence vectors that give rise to the set of minimal reduction purities for a pure parent state.

To get an idea of what the goals vector would look like, for the example in \Eq{D.14}, we can place it beneath the occurrence matrix as
%===============================================================================
\begin{equation}%                  Equation D.15
\left( {\begin{array}{*{20}c}
   {\Omega _{l, \cdots } } &\vline &  {\redrho{1}_{1,1} } & {\redrho{1}_{2,2} } &\vline &  {\redrho{2}_{1,1} } & {\redrho{2}_{2,2} } & {\redrho{2}_{3,3} }  \\
\hline
   {\Omega _{1, \cdots } } &\vline &  1 & 0 &\vline &  1 & 0 & 0  \\
   {\Omega _{2, \cdots } } &\vline &  1 & 0 &\vline &  0 & 1 & 0  \\
   {\Omega _{3, \cdots } } &\vline &  1 & 0 &\vline &  0 & 0 & 1  \\
   {\Omega _{4, \cdots } } &\vline &  0 & 1 &\vline &  1 & 0 & 0  \\
   {\Omega _{5, \cdots } } &\vline &  0 & 1 &\vline &  0 & 1 & 0  \\
   {\Omega _{6, \cdots } } &\vline &  0 & 1 &\vline &  0 & 0 & 1  \\
\hline
   \mathbf{G} &\vline &  1 & 1 &\vline &  1 & 1 & 0  \\
\end{array}} \right)\!\!\begin{array}{*{20}c}
   {}  \\
   {\leftarrow}  \\
   {}  \\
   {}  \\
   {}  \\
   {\leftarrow}  \\
   {}  \\
   {}  \\
\end{array}
\label{eq:D.15}
\end{equation}
%===============================================================================
so the goals vector $\mathbf{G}$ for compatibility set $\{1,5\}$ is the sum of the occurrence vectors of that set as $\mathbf{G} = \Omega _{1, \cdots }  + \Omega _{5, \cdots }  = (1,1|1,1,0)$. Note that goals vectors for other compatibility sets like $\{1,6\}$ may look different, such as $\mathbf{G} = \Omega _{1, \cdots }  + \Omega _{6, \cdots }  = (1,1|1,0,1)$.

However the \textit{true} goals vector is not merely defined by summing over all occurrence vectors of any compatibility set (that only happens in simple examples). \textit{Instead, the goals vector is the sum of occurrence vectors for any parent TGX state of levels $L$ such that all reduction purities are minimized over all possible $L$}.

To get an explicit formula for the goals vector, we need to imagine the most general case of how parent diagonals can be distributed among the reduction diagonals.  Therefore, we now list all the possibilities:
%+++++++++++++++++++++++++++++++++++++++++++++++++++++++++++++++++++++++++++++++
\begin{itemize}[leftmargin=*,labelindent=4pt]\setlength\itemsep{-2pt}
%~~~~~~~~~~~~~~~~~~~~~~~~~~~~~~~~~~~~~~~~~~~~~~~~~~~~~~~~~~~~~~~~~~~~~~~~~~~~~~~
\item[\textbf{1.}]\hypertarget{possibility:1}{}All modes have equal size $n_{\min}\!=\!n_{\max}$.  Here, all mode sizes $n_m$ divide into {$\nmaxnot\!\equiv\!\frac{n}{n_{\max}}\!=\!\frac{n}{n_{m}}$} an integer number of times as $\nmaxnot\!=\! n_m ^N /n_{m}\! =\! n_m^{N - 1}\;\,\!\! \forall m \!\in\! 1, \ldots ,N$.  Thus, the parent diagonals can always be evenly distributed among the reduction diagonals, allowing ideal minimization of all reduction purities simultaneously.\vspace{2pt}
%~~~~~~~~~~~~~~~~~~~~~~~~~~~~~~~~~~~~~~~~~~~~~~~~~~~~~~~~~~~~~~~~~~~~~~~~~~~~~~~
%~~~~~~~~~~~~~~~~~~~~~~~~~~~~~~~~~~~~~~~~~~~~~~~~~~~~~~~~~~~~~~~~~~~~~~~~~~~~~~~
\item[\textbf{2.}]\hypertarget{possibility:2}{}Exactly one mode is larger than all the others; $n_{\min}<n_{\max}$ and $n_{m}\!=\!n_{\max}$ for only \textit{one} $m\!\in\! 1,\ldots,N$. Here, again the maximum number of parent terms in any reduction element is \smash{$\nmaxnot$}, but two quirks arise;
    %+++++++++++++++++++++++++++++++++++++++++++++++++++++++++++++++++++++++++++
    \begin{itemize}[leftmargin=12.5pt,labelindent=4pt]%
    %~~~~~~~~~~~~~~~~~~~~~~~~~~~~~~~~~~~~~~~~~~~~~~~~~~~~~~~~~~~~~~~~~~~~~~~~~~~
    \item[\textbf{a.}]\hypertarget{quirk:a}{}All modes smaller than the maximum are able to divide evenly into \smash{$\nmaxnot$} because it is a product of all of these modes; if mode $N$ is the largest, we can represent this as \smash{$\nmaxnot\!\equiv\!\frac{n}{n_{\max}}\!=\!{\nmaxnot}_{1}\!\cdots\!{\nmaxnot}_{N-1}$}, so that \textit{the size of all the modes together except the lone largest mode is always an integer multiple of all the smaller modes.} Therefore, in all of these modes, each of their reduction diagonals will contain at least one occurrence of a parent element, so each reduction can support ideal maximal mixing.
    %~~~~~~~~~~~~~~~~~~~~~~~~~~~~~~~~~~~~~~~~~~~~~~~~~~~~~~~~~~~~~~~~~~~~~~~~~~~
    %~~~~~~~~~~~~~~~~~~~~~~~~~~~~~~~~~~~~~~~~~~~~~~~~~~~~~~~~~~~~~~~~~~~~~~~~~~~
    \item[\textbf{b.}]\hypertarget{quirk:b}{}The second quirk is that the lone largest mode does not necessarily evenly divide into \smash{$\nmaxnot$}.  For example, in $2\times 3$, \smash{$n_{\max}=3$} and \smash{$\nmaxnot=2$}, so \smash{$n_{\max}>\nmaxnot$} and \smash{$\text{mod}(\nmaxnot,n_{\max})=\nmaxnot-$} \smash{$\text{floor}(\nmaxnot/n_{\max})n_{\max}=2\neq 0$}. But in $3\times 4\times 6$, \smash{$n_{\max}=6$} and \smash{$\nmaxnot=12$}, so \smash{$n_{\max}<\nmaxnot$} and \smash{$\text{mod}(\nmaxnot,n_{\max})=0$}.  And we can also have cases such as $2\times 3\times 6$ where \smash{$n_{\max}=\nmaxnot$} and \smash{$\text{mod}(\nmaxnot,n_{\max})=0$}, or like $2\times 3\times 5$ where \smash{$n_{\max}<\nmaxnot$} and \smash{$\text{mod}(\nmaxnot,n_{\max})=1\neq 0$}.  Thus, if \smash{$n_{\max}>\nmaxnot$}, then \smash{$\nmaxnot$} will never be an integer multiple of \smash{$n_{\max}$}, but if \smash{$n_{\max}\leq\nmaxnot$}, we only get integer division if \smash{$\text{mod}(\nmaxnot,n_{\max})= 0$}.
    %~~~~~~~~~~~~~~~~~~~~~~~~~~~~~~~~~~~~~~~~~~~~~~~~~~~~~~~~~~~~~~~~~~~~~~~~~~~
    \end{itemize}
    %+++++++++++++++++++++++++++++++++++++++++++++++++++++++++++++++++++++++++++
\vspace{-2pt}
Thus, for \hyperlink{possibility:2}{Possibility 2}, we can evenly distribute occurrences of the parent diagonals in all reductions smaller than the largest reduction, but our ability to distribute parent elements in the largest reduction depends on its size.  Soon we will unite all these cases, but there is one more possibility to consider.
%~~~~~~~~~~~~~~~~~~~~~~~~~~~~~~~~~~~~~~~~~~~~~~~~~~~~~~~~~~~~~~~~~~~~~~~~~~~~~~~
\vspace{4pt}
%~~~~~~~~~~~~~~~~~~~~~~~~~~~~~~~~~~~~~~~~~~~~~~~~~~~~~~~~~~~~~~~~~~~~~~~~~~~~~~~
\item[\textbf{3.}]\hypertarget{possibility:3}{}More than one mode has the maximum mode size; \smash{$n_{\min}<n_{\max}$} and \smash{$n_{m}\!=\!n_{\max}$} for \textit{more than one} $m\in 1,\ldots,N$.  Here, \smash{$\text{mod}(\nmaxnot,n_{\max})= 0$} \textit{always} because \smash{$\nmaxnot\!\equiv\!\frac{n}{n_{m}}$} always contains at least one factor of \smash{$n_{\max}$}, since there are at least two factors of \smash{$n_{\max}$} in $n$ and we are only dividing $n$ by one factor of \smash{$n_{\max}$}.  Therefore here, \smash{$\nmaxnot/n_{\max}$} is always an integer and we can always evenly distribute all the parent diagonals amongst the reduction diagonals.  So here, as in \hyperlink{possibility:1}{Possibility 1}, all reductions can reach ideal maximal mixing.
%~~~~~~~~~~~~~~~~~~~~~~~~~~~~~~~~~~~~~~~~~~~~~~~~~~~~~~~~~~~~~~~~~~~~~~~~~~~~~~~
\end{itemize}
%+++++++++++++++++++++++++++++++++++++++++++++++++++++++++++++++++++++++++++++++

To unite all these cases, we now consider the one with the least symmetry, \hyperlink{possibility:2}{Possibility 2}, with an example where $L_*$ is big enough to not just fill up the biggest mode once, but to wrap around and start filling it up again.

For example, in $2\times 2\times 3$, TGX space is
\vspace{-8pt}
%===============================================================================
\begin{equation}%                  Equation D.16
\scalebox{0.84}{$\begin{array}{l}
 \rho  =  \\ 
 \!\left( {\arraycolsep=1.0pt\def\arraystretch{1}
\begin{array}{*{20}c}
   {\rho _{1,1} } &  \cdot  &  \cdot  &  \cdot  & {\rho _{1,5} } & {\rho _{1,6} } &  \cdot  & {\rho _{1,8} } & {\rho _{1,9} } & {\rho _{1,10} } &\!\! {\rho _{1,11} } &\!\! {\rho _{1,12} }  \\
    \cdot  & {\rho _{2,2} } &  \cdot  & {\rho _{2,4} } &  \cdot  & {\rho _{2,6} } & {\rho _{2,7} } &  \cdot  & {\rho _{2,9} } & {\rho _{2,10} } &\!\! {\rho _{2,11} } &\!\! {\rho _{2,12} }  \\
    \cdot  &  \cdot  & {\rho _{3,3} } & {\rho _{3,4} } & {\rho _{3,5} } &  \cdot  & {\rho _{3,7} } & {\rho _{3,8} } &  \cdot  & {\rho _{3,10} } &\!\! {\rho _{3,11} } &\!\! {\rho _{3,12} }  \\
    \cdot  & {\rho _{4,2} } & {\rho _{4,3} } & {\rho _{4,4} } &  \cdot  &  \cdot  & {\rho _{4,7} } & {\rho _{4,8} } & {\rho _{4,9} } &  \cdot  &\!\! {\rho _{4,11} } &\!\! {\rho _{4,12} }  \\
   {\rho _{5,1} } &  \cdot  & {\rho _{5,3} } &  \cdot  & {\rho _{5,5} } &  \cdot  & {\rho _{5,7} } & {\rho _{5,8} } & {\rho _{5,9} } & {\rho _{5,10} } &\!\!  \cdot  &\!\! {\rho _{5,12} }  \\
   {\rho _{6,1} } & {\rho _{6,2} } &  \cdot  &  \cdot  &  \cdot  & {\rho _{6,6} } & {\rho _{6,7} } & {\rho _{6,8} } & {\rho _{6,9} } & {\rho _{6,10} } &\!\! {\rho _{6,11} } &\!\!  \cdot   \\
    \cdot  & {\rho _{7,2} } & {\rho _{7,3} } & {\rho _{7,4} } & {\rho _{7,5} } & {\rho _{7,6} } & {\rho _{7,7} } &  \cdot  &  \cdot  &  \cdot  &\!\! {\rho _{7,11} } &\!\! {\rho _{7,12} }  \\
   {\rho _{8,1} } &  \cdot  & {\rho _{8,3} } & {\rho _{8,4} } & {\rho _{8,5} } & {\rho _{8,6} } &  \cdot  & {\rho _{8,8} } &  \cdot  & {\rho _{8,10} } &\!\!  \cdot  &\!\! {\rho _{8,12} }  \\
   {\rho _{9,1} } & {\rho _{9,2} } &  \cdot  & {\rho _{9,4} } & {\rho _{9,5} } & {\rho _{9,6} } &  \cdot  &  \cdot  & {\rho _{9,9} } & {\rho _{9,10} } &\!\! {\rho _{9,11} } &\!\!  \cdot   \\
   {\rho _{10,1} } & {\rho _{10,2} } & {\rho _{10,3} } &  \cdot  & {\rho _{10,5} } & {\rho _{10,6} } &  \cdot  & {\rho _{10,8} } & {\rho _{10,9} } & {\rho _{10,10} } &\!\!  \cdot  &\!\!  \cdot   \\
   {\rho _{11,1} } & {\rho _{11,2} } & {\rho _{11,3} } & {\rho _{11,4} } &  \cdot  & {\rho _{11,6} } & {\rho _{11,7} } &  \cdot  & {\rho _{11,9} } &  \cdot  &\!\! {\rho _{11,11} } &\!\!  \cdot   \\
   {\rho _{12,1} } & {\rho _{12,2} } & {\rho _{12,3} } & {\rho _{12,4} } & {\rho _{12,5} } &  \cdot  & {\rho _{12,7} } & {\rho _{12,8} } &  \cdot  &  \cdot  &\!\!  \cdot  &\!\! {\rho _{12,12} }  \\
\end{array}}\! \right){\kern -3pt}. \\ 
 \end{array}$}
\label{eq:D.16}
\end{equation}
%===============================================================================
Here,{\kern -0.7pt} regardless{\kern -0.7pt} of{\kern -0.7pt} which{\kern -0.7pt} numbers{\kern -0.7pt} of{\kern -0.7pt} parent{\kern -0.7pt} levels{\kern -0.7pt} $L_*${\kern -0.7pt} cause{\kern -0.7pt} the{\kern -0.7pt} smallest{\kern -0.7pt} combination{\kern -0.7pt} of{\kern -0.7pt} reduction{\kern -0.7pt} purities,{\kern -0.7pt} since \smash{$n_{\max}=3$} and{\kern -1pt} \smash{$\nmaxnot=4$} so \smash{$\text{mod}(\nmaxnot,n_{\max})=1\neq 0$}, then the largest mode cannot achieve ideal minimum purity.  So we now look at three general cases for how different $L$ values affect the distribution of parent diagonals.
%+++++++++++++++++++++++++++++++++++++++++++++++++++++++++++++++++++++++++++++++
\begin{itemize}[leftmargin=*,labelindent=4pt]%
%~~~~~~~~~~~~~~~~~~~~~~~~~~~~~~~~~~~~~~~~~~~~~~~~~~~~~~~~~~~~~~~~~~~~~~~~~~~~~~~
\item[\textbf{1.}]\hypertarget{case:1}{}$2 \le L < n_{\max }$, so $L=2$ (minimum requirement for superposition in the parent state $\rho$). Here, we do not get a complete fill-up in the largest mode,
%===============================================================================
\begin{equation}%                  Equation D.17
\redrho{1} \! =\!\! \left(\! {\begin{array}{*{20}c}
    \squareOne  &\!\!\!  \cdot   \\
    \cdot  &\!\!\!  \squareOne   \\
\end{array}}\! \right)\!\!,\,\redrho{2} \! =\!\! \left(\! {\begin{array}{*{20}c}
    \squareOne  &\!\!\!  \cdot   \\
    \cdot  &\!\!\!  \squareOne   \\
\end{array}}\! \right)\!\!,\,\redrho{3}\! =\!\! \left(\! {\begin{array}{*{20}c}
    \squareOne  &\!\!\!  \cdot  &\!\!\!  \cdot   \\
    \cdot  &\!\!\!  \squareOne  &\!\!\!  \cdot   \\
    \cdot  &\!\!\!  \cdot  &\!\!\! {\squareZero}  \\
\end{array}}\! \right)\!\!,
\label{eq:D.17}
\end{equation}
%===============================================================================
such as for compatibility set (CS) $\{1,11\}$, and the minimal possible reduction purities are
%===============================================================================
\begin{equation}%                  Equation D.18
P_{\min}(\redrho{1} )\! =\! {\textstyle{1 \over 2}},\,P_{\min}(\redrho{2} ) \! =\! {\textstyle{1 \over 2}},\,P_{\min}(\redrho{3} ) \! =\! {\textstyle{1 \over 2}},
\label{eq:D.18}
\end{equation}
%===============================================================================
yielding minimal average \textit{unitized} reduction purity,
%===============================================================================
\begin{equation}%                  Equation D.19
\begin{array}{*{20}l}
   {\frac{1}{N}\sum\nolimits_{m = 1}^N \!{\frac{{n_m \min (P(\redrhotiny{m} )) - 1}}{{n_m  - 1}}} } &\!\! { = \frac{1}{{12}} \approx 0.0833},  \\
\end{array}
\label{eq:D.19}
\end{equation}
%===============================================================================
where \smash{${\textstyle{{n_m P(\redrhotiny{m} ) - 1} \over {n_m  - 1}}} \in [0,1]$} is the \textit{unitized purity} for the mode $m$ reduction, named for its range, $[0,1]$, scaled for isolated systems even though these are not.
%~~~~~~~~~~~~~~~~~~~~~~~~~~~~~~~~~~~~~~~~~~~~~~~~~~~~~~~~~~~~~~~~~~~~~~~~~~~~~~~
%~~~~~~~~~~~~~~~~~~~~~~~~~~~~~~~~~~~~~~~~~~~~~~~~~~~~~~~~~~~~~~~~~~~~~~~~~~~~~~~
\item[\textbf{2.}]\hypertarget{case:2}{}$L=n_{\max}$, so $L=3$, with parent occurrences,
%===============================================================================
\begin{equation}%                  Equation D.20
\redrho{1} \! =\!\! \left(\! {\begin{array}{*{20}c}
    \squareTwo  &\!\!\!  \cdot   \\
    \cdot  &\!\!\!  \squareOne   \\
\end{array}}\! \right)\!\!,\,\redrho{2} \! =\!\! \left(\! {\begin{array}{*{20}c}
    \squareOne  &\!\!\!  \cdot   \\
    \cdot  &\!\!\!  \squareTwo   \\
\end{array}}\! \right)\!\!,\,\redrho{3}\! =\!\! \left(\! {\begin{array}{*{20}c}
    \squareOne  &\!\!\!  \cdot  &\!\!\!  \cdot   \\
    \cdot  &\!\!\!  \squareOne  &\!\!\!  \cdot   \\
    \cdot  &\!\!\!  \cdot  &\!\!\! {\squareOne}  \\
\end{array}}\! \right)\!\!,
\label{eq:D.20}
\end{equation}
%===============================================================================
as for CS $\{1,5,12\}$, with minimal reduction purities
%===============================================================================
\begin{equation}%                  Equation D.21
P_{\min}(\redrho{1} )\! =\! {\textstyle{5 \over 9}},\,P_{\min}(\redrho{2} ) \! =\! {\textstyle{5 \over 9}},\,P_{\min}(\redrho{3} ) \! =\! {\textstyle{1 \over 3}},
\label{eq:D.21}
\end{equation}
%===============================================================================
and minimal average unitized reduction purity
%===============================================================================
\begin{equation}%                  Equation D.22
\begin{array}{*{20}l}
   {\frac{1}{N}\sum\nolimits_{m = 1}^N \!{\frac{{n_m \min (P(\redrhotiny{m} )) - 1}}{{n_m  - 1}}} } &\!\! { = \frac{2}{{27}} \approx 0.0741}.  \\
\end{array}
\label{eq:D.22}
\end{equation}
%===============================================================================
%~~~~~~~~~~~~~~~~~~~~~~~~~~~~~~~~~~~~~~~~~~~~~~~~~~~~~~~~~~~~~~~~~~~~~~~~~~~~~~~
%~~~~~~~~~~~~~~~~~~~~~~~~~~~~~~~~~~~~~~~~~~~~~~~~~~~~~~~~~~~~~~~~~~~~~~~~~~~~~~~
\item[\textbf{3.}]\hypertarget{case:3}{}$n_{\max } < L \leq \nmaxnot$, so $L=4$, with parent occurrences,
%===============================================================================
\begin{equation}%                  Equation D.23
\redrho{1} \! =\!\! \left(\! {\begin{array}{*{20}c}
    \squareTwo  &\!\!\!  \cdot   \\
    \cdot  &\!\!\!  \squareTwo   \\
\end{array}}\! \right)\!\!,\,\redrho{2} \! =\!\! \left(\! {\begin{array}{*{20}c}
    \squareTwo  &\!\!\!  \cdot   \\
    \cdot  &\!\!\!  \squareTwo   \\
\end{array}}\! \right)\!\!,\,\redrho{3}\! =\!\! \left(\! {\begin{array}{*{20}c}
    \squareTwo  &\!\!\!  \cdot  &\!\!\!  \cdot   \\
    \cdot  &\!\!\!  \squareOne  &\!\!\!  \cdot   \\
    \cdot  &\!\!\!  \cdot  &\!\!\! {\squareOne}  \\
\end{array}}\! \right)\!\!,
\label{eq:D.23}
\end{equation}
%===============================================================================
as for CS $\{1,5,9,10\}$, with minimal reduction purities
%===============================================================================
\begin{equation}%                  Equation D.24
P_{\min}(\redrho{1} )\! =\! {\textstyle{1 \over 2}},\,P_{\min}(\redrho{2} ) \! =\! {\textstyle{1 \over 2}},\, P_{\min}(\redrho{3} ) \! =\! {\textstyle{3 \over 8}},
\label{eq:D.24}
\end{equation}
%===============================================================================
and minimal average unitized reduction purity
%===============================================================================
\begin{equation}%                  Equation D.25
\begin{array}{*{20}l}
   {\frac{1}{N}\sum\nolimits_{m = 1}^N \!{\frac{{n_m \min (P(\redrhotiny{m} )) - 1}}{{n_m  - 1}}} } &\!\! { = \frac{1}{{48}} \approx 0.0208}.  \\
\end{array}
\label{eq:D.25}
\end{equation}
%===============================================================================
%~~~~~~~~~~~~~~~~~~~~~~~~~~~~~~~~~~~~~~~~~~~~~~~~~~~~~~~~~~~~~~~~~~~~~~~~~~~~~~~
\end{itemize}
%+++++++++++++++++++++++++++++++++++++++++++++++++++++++++++++++++++++++++++++++

Thus, \hyperlink{case:3}{Case 3} has the lowest average unitized reduction purity of all possible level numbers $L$, so $L_* =4$.  Note that there cannot be more levels than $\nmaxnot$ in pure TGX states because that is the maximum number of parent terms in each element of the largest mode, and is thus the maximum number of occurrences that can be distributed in that mode.

Now we can synthesize the above results to a general model of maximal entanglement.  First, looking at \Eq{D.17}, \Eq{D.20}, and \Eq{D.23}, we see that for modes $m$ where $L$ is larger than $n_m$, the number of extra single-element parent occurrences after the subsystem diagonals have been filled up completely as many times as possible is
%===============================================================================
\begin{equation}%                  Equation D.26
O_{\text{extra}}\equiv\text{mod}(L,n_{m}),
\label{eq:D.26}
\end{equation}
%===============================================================================
and therefore the number of remaining reduction diagonals that did \textit{not} get an extra occurrence is
%===============================================================================
\begin{equation}%                  Equation D.27
O_{\text{nonextra}}\equiv n_{m}-\text{mod}(L,n_{m}).
\label{eq:D.27}
\end{equation}
%===============================================================================

If the reduction purities are minimized when all occurrences are the same value due to balanced superposition in the parent TGX state (which we prove to be optimal in \App{App.D.5}), then that value must be \smash{$\frac{1}{L}$} for normalization.

The total value of each $O_{\text{nonextra}}$ reduction element is \smash{$\frac{1}{L}$} times the total number of times \textit{all} of that reduction's diagonals could be fully filled by one occurrence of \smash{$\frac{1}{L}$} each, which{\kern -1pt} is{\kern -1pt} \smash{$\text{floor}(\frac{L}{n_{m}})$} times for any given \smash{$L\in 2\ldots,\nmaxnot$}, so the{\kern 1.3pt} value{\kern 1.3pt} of{\kern 1.3pt} each \smash{$O_{\text{nonextra}}$} diagonal element is
%===============================================================================
\begin{equation}%                  Equation D.28
V_{\text{nonextra}}\equiv \text{floor}(\textstyle{{L}\over{n_{m}}})\textstyle{{1}\over{L}}.
\label{eq:D.28}
\end{equation}
%===============================================================================
The $O_{\text{extra}}$ elements represent the extra wrap-around, so they each have \textit{one} more occurrence of \smash{$\frac{1}{L}$} than the $O_{\text{nonextra}}$ diagonals, so the $O_{\text{extra}}$ diagonals have values,
%===============================================================================
\begin{equation}%                  Equation D.29
V_{\text{extra}}\equiv \textstyle{{1}\over{L}}+\text{floor}(\textstyle{{L}\over{n_{m}}})\textstyle{{1}\over{L}}.
\label{eq:D.29}
\end{equation}
%===============================================================================

Now that we know all the values and numbers of kinds of reduction diagonals for mode $m$, then since reductions are diagonal for TGX states, the mode-$m$ reduction purity is the sum of the number of each kind of element times the square of the value of that element,
%===============================================================================
\begin{equation}%                  Equation D.30
\begin{array}{*{20}l}
   {P_{\text{MP}}^{(m)} (L )} &\!\! {= O_{\text{extra}}V_{\text{extra}}^2 +O_{\text{nonextra}}V_{\text{nonextra}}^2 } \\
   {} &\!\! {=\bmod (L ,n_m )\left( {\frac{{1 + \text{floor}(L /n_m )}}{{L }}} \right)^2 \rule{0pt}{15pt}}  \\
   {} &\!\! {\;\;\;\,\, + (n_m  - \bmod (L ,n_m ))\left( {\frac{{\text{floor}(L /n_m )}}{{L }}} \right)^2\!\!\! ,}  \\
\end{array}
\label{eq:D.30}
\end{equation}
%===============================================================================
which is the minimum physical purity of $\redrho{m}$ of a pure TGX{\kern -1pt} parent{\kern -1pt} state{\kern -1pt} with{\kern -1pt} balanced{\kern -1pt} superposition{\kern -1pt} of{\kern -1pt} $L${\kern -1pt} levels.

Note that \Eq{D.30} is compatible with all cases of $L$ relative to $n_m$, even though we used \hyperlink{possibility:2}{Possibility 2} to derive it.  This is because when $L<n_{m}$, then $V_{\text{nonextra}}=$ \smash{$\text{floor}(\textstyle{{L}\over{n_{m}}})\textstyle{{1}\over{L}}=0$}, and we get cases such as \Eq{D.14} where \smash{$\redrho{2}$} does not fully fill up once with parent occurrences.

Now that we have \smash{$P_{\text{MP}}^{(m)} (L )$}, we need to find \textit{which} $L$ gives the lowest \textit{combination} of all mode purities.  The problem with purity for this purpose is that its minimum for an isolated system is dimension-dependent, since for an isolated $n$-level system, $P\in[\frac{1}{n},1]$.  As mentioned in \Eq{D.19}, we can get around this with the \textit{unitized purity},
%===============================================================================
\begin{equation}%                  Equation D.31
P_{U}\equiv\textstyle{{nP-1}\over{n-1}}\in[0,1],
\label{eq:D.31}
\end{equation}
%===============================================================================
which puts all reduction purities on more even footing for assessing entanglement. For example, in an $N$-qudit system, if the unitized purity of every reduction is $0$, then the pure parent state must be maximally entangled.

Since $N$-partite entanglement is how mixed all reductions are simultaneously, and since $P_{U}$ is nonnegative, then the sum of all $N$ reduction purities is a measure of simultaneous reduction purity.  Then, since there are $N$ unitized reduction purities each with at most a value of $1$, dividing by $N$ yields the average unitized reduction purity as a measure on $[0,1]$ for how simultaneously mixed all reductions are.  This suggests we define the ent as
%===============================================================================
\begin{equation}%                  Equation D.32
\begin{array}{*{20}l}
   {\Upsilon '} &\!\! {\equiv 1 - \frac{1}{N}\sum\limits_{m = 1}^N {\redP{\kern -4.2pt}{~}_{\!U}^{(m)} }  \equiv 1 - \langle \redP_{\!U} \rangle,}  \\
\end{array}
\label{eq:D.32}
\end{equation}
%===============================================================================
where \smash{$\langle \redP_{\!U} \rangle$} is the average unitized reduction purity over all modes relevant to coincidence outcomes of the parent state, and \smash{$\redP{\kern -4.2pt}{~}_{\!U}^{(m)}\equiv (n_{m}P(\redrho{m})-1)/(n_{m}-1)$} is the unitized reduction purity of mode $m$.  When all \smash{$\redP{\kern -4.2pt}{~}_{\!U}^{(m)}=1$}, their sum is $N$ and \smash{$\langle \redP_{\!U} \rangle=1$}, and $\Upsilon '=0$.  In systems where all reductions can reach their ideal minimum purities, all \smash{$\redP{\kern -4.2pt}{~}_{\!U}^{(m)}=0$}, and \smash{$\langle \redP_{\!U} \rangle=0$} and $\Upsilon '=1$. 

The need to normalize the ent further comes from the phenomenon that in some systems the ideal minimum purity is not attainable for some reductions when the parent state is maximally entangled (as we saw in $2\times 3$). 

Therefore, for a given $L$, the minimal value of \smash{$\langle \redP_{\!U} \rangle$} happens when each mode's reduction reaches its minimum of \smash{$P_{\text{MP}}^{(m)} (L )$}, producing the maximum of $\Upsilon '$ as
%===============================================================================
\begin{equation}%                  Equation D.33
\begin{array}{*{20}l}
   {M(L)} &\!\! { \equiv \max(\Upsilon ') = 1 - \frac{1}{N}\sum\limits_{m = 1}^N {\frac{{n_m P_{\text{MP}}^{(m)} (L) - 1}}{{n_m  - 1}}}. }  \\
\end{array}
\label{eq:D.33}
\end{equation}
%===============================================================================
However, \textit{not all} $L\in 2,\ldots,\nmaxnot$ will produce the physically achievable maximum of $\Upsilon '$. Therefore, we need to check each $L$ to see which ones cause \smash{$\langle \redP_{\!U} \rangle$} to be minimized.  Thus, we define the ``golden $L$'' values as
%===============================================================================
\begin{equation}%                  Equation D.34
\mathbf{L}_*  \equiv \{ L_* \} ;\;\;\text{s.t.}\;\!\mathop {\min }\limits_{L \in 2, \ldots, n_{\,\overline{{\kern -1.8pt}\max^{~^{~^{~}}}\!\!\!\!\!\!\!\!\!\!}} } (1 - M(L)),
\label{eq:D.34}
\end{equation}
%===============================================================================
since \smash{$\langle \redP_{\!U} \rangle=1 - M(L)$}, which is the result in \Eq{5} where we can arbitrarily pick $L_*  \equiv \min \{ \mathbf{L}_* \}$ as a representative value.  Then, we get an automatically normalized entanglement measure by defining
%===============================================================================
\begin{equation}%                  Equation D.35
\Upsilon\equiv{\textstyle{{1}\over{M(L_{*})}}}\Upsilon',
\label{eq:D.35}
\end{equation}
%===============================================================================
which is the result in \Eq{2}.

Now that we have derived \Eqs{2}{5}, we can understand the \textit{goals vector} $\mathbf{G}$, presented in \Eq{D.15}, as the sufficient condition needed to identify maximal entanglement, and which can be used to construct maximally entangled states, as detailed explicitly in \App{App.G}.

Basically, $\mathbf{G}$ is the sum of occurrence vectors for pure-state compatibility sets of $L_{*}\!\in\! \mathbf{L}_*$ levels that achieves the minimal average unitized reduction purity \smash{$\min(\langle \redP_{\!U} \rangle)$}.  We must define $\mathbf{G}$ this way because, given all compatibility sets of $L$ levels, \textit{not all} such sets minimize{\kern -1.5pt} \smash{$\langle \redP_{\!U} \rangle$}.

Thus, $\mathbf{G}$ can be constructed in a similar manner to which the minimal reduction purities were derived; see \hyperlink{step:3}{Step 3} in \App{App.G} for an explicit form of $\mathbf{G}$.  For an \textit{example} of $\mathbf{G}$, see \Eq{H.5}.  Then, \Eqs{H.6}{H.10} shows how to build the \textit{total goals matrix} $G$, which lists all possible distributions of parent elements in the reductions of pure TGX states that indicate maximal entanglement given balanced superposition in the parent.

We have now motivated all results except for the supposition of balanced superposition in the parent state, which we treat next.  Note that this is only a requirement for pure TGX states, and in general maximally entangled states can have different probability amplitudes, and reductions of minimal physical purity can be nondiagonal.
%                              End of App.D.4.c
%~~~~~~~~~~~~~~~~~~~~~~~~~~~~~~~~~~~~~~~~~~~~~~~~~~~~~~~~~~~~~~~~~~~~~~~~~~~~~~~
%                               End of App.D.4
%...............................................................................
%...............................................................................
%  App.D.5 Proof That Balanced Superposition in the Parent State is Necessary For Maximal Entanglement of TGX States
\subsection{\label{sec:App.D.5}Proof That Balanced Superposition is Necessary For Maximal Entanglement of TGX States}
As an illustration of how to start, from \Eq{D.10}, the reduction purities of TGX states in $2\times 3$ are
%===============================================================================
\begin{equation}%                  Equation D.36
\begin{array}{*{20}l}
   {P(\redrho{1} )} &\!\! { = (\rho _{1,1}  \! +\! \rho _{2,2} \! +\! \rho _{3,3} )^2  \! +\! (\rho _{4,4} \! +\! \rho _{5,5} \! +\! \rho _{6,6} )^2 }  \\
   {P(\redrho{2} )} &\!\! { = (\rho _{1,1} \! +\! \rho _{4,4} )^2 \! +\! (\rho _{2,2} \! +\! \rho _{5,5} )^2 \! +\! (\rho _{3,3} \! +\! \rho _{6,6} )^2 \!,}  \\
\end{array}
\label{eq:D.36}
\end{equation}
%===============================================================================
because reductions of TGX states are diagonal. Note that, for example, \smash{$P(\redrho{1} )$} can be rewritten as
%===============================================================================
\begin{equation}%                  Equation D.37
\begin{array}{*{20}l}
   {P(\redrho{1} ) = } &\!\! {\left( {\begin{array}{*{20}c}
   {\rho _{1,1} } &\!\! {\rho _{2,2} } &\!\! {\rho _{3,3} }  \\
\end{array}} \right)\!\!\left(\! {\begin{array}{*{20}c}
   1 & 1 & 1  \\
   1 & 1 & 1  \\
   1 & 1 & 1  \\
\end{array}}\! \right)\!\!\!\left(\! {\begin{array}{*{20}c}
   {\rho _{1,1} }  \\
   {\rho _{2,2} }  \\
   {\rho _{3,3} }  \\
\end{array}}\! \right)}  \\
   {} &\!\! { +\! \left( {\begin{array}{*{20}c}
   {\rho _{4,4} } &\!\! {\rho _{5,5} } &\!\! {\rho _{6,6} }  \\
\end{array}} \right)\!\!\left(\! {\begin{array}{*{20}c}
   1 & 1 & 1  \\
   1 & 1 & 1  \\
   1 & 1 & 1  \\
\end{array}}\! \right)\!\!\!\left(\! {\begin{array}{*{20}c}
   {\rho _{4,4} }  \\
   {\rho _{5,5} }  \\
   {\rho _{6,6} }  \\
\end{array}}\! \right)\!.}  \\
\end{array}
\label{eq:D.37}
\end{equation}
%===============================================================================

In general, there are $n_m$ such square terms in each reduction purity (one for each of the main diagonals of that reduction), and each one will have a matrix size of \smash{$n_{\mbarsub}$}, the number of parent terms appearing in each matrix element of that reduction.

However, due to the constraint that the parent be both pure and of TGX form, a number \textit{less} than \smash{$n_{\mbarsub}$} parent diagonals will be nonzero in each quadratic term of each reduction purity, which we call $L_k$ (which can never exceed \smash{$n_{\mbarsub}$}).  Therefore, each quadratic matrix (indexed by $k$) will always be the \smash{$L_k$}-level all-ones matrix,
%===============================================================================
\begin{equation}%                  Equation D.38
W^{[L_k]}  \equiv \mathbf{1}^{[L_k]} \mathbf{1}^{[L_k]T},
\label{eq:D.38}
\end{equation}
%===============================================================================
where \smash{$\mathbf{1}^{[L_k]T}\equiv (1_{1},\ldots,1_{L_k})$}. The spectral decomposition of \smash{$W^{[L_k]}$} can be written using an \smash{$L_k$}-level discrete Fourier matrix \smash{$F^{[L_k]}$} (see \Eq{20}).  Then, each of the $n_m$ quadratic groups, indexed by $k\in 1,\ldots,n_{m}$, becomes
%===============================================================================
\begin{equation}%                  Equation D.39
\bm{\rho}_k ^{\dag} W^{[L_k]} \bm{\rho}_k  = \bm{\rho}_k ^{\dag} F^{[L_k]} L_k |1\rangle \langle 1|F^{[L_k]\dag} \bm{\rho}_k \,,
\label{eq:D.39}
\end{equation}
%===============================================================================
where \smash{$\bm{\rho}_{k}^{\dag} \equiv((\rho_k)_{1},\ldots,(\rho_k)_{L_k})$}, where \smash{$(\rho_k)_{a_k}$} is the $a_k$th element of the $k$th quadratic group of parent elements in a reduction purity, grouped by reduction diagonal.  Note that the factor of $L_k$ in \Eq{D.39} is the eigenvalue of \smash{$W^{[L_k]}$}.

Doing several examples to find the effects of minimizing all reduction purities on the pure parent TGX state, it becomes clear that we only need to look at \textit{one} reduction to determine the possible parent amplitudes.  Therefore, we can pick the \textit{smallest} mode $m_{\text{min}}$ defined as the $m$ for which $n_{m} =\text{min}(\mathbf{n})\equiv n_{\text{min}}$, which ensures that its minimal physical reduction purity is the isolated minimum \smash{$\textstyle{1 \over n_{m}}$} (see \App{App.D.4.c} for details about why this is so).

{\kern 1.5pt}In mode $m=m_{\text{min}}$, every reduction diagonal will have an equal number of nonzero parent occurrences.  Therefore, every group $k$ has the same number of nonzero elements so that $L_k =L'\;\;\forall k\in 1,\ldots,n_{m}$.

We can then define a set of transformed vectors as
%===============================================================================
\begin{equation}%                  Equation D.40
\bm{\omega}_k  \equiv F^{[L']\dag } \bm{\rho}_k \,,
\label{eq:D.40}
\end{equation}
%===============================================================================
so that \Eq{D.39} becomes
%===============================================================================
\begin{equation}%                  Equation D.41
\bm{\rho}_k ^{\dag} W^{[L']} \bm{\rho}_k  = L' \bm{\omega}_k ^{\dag}  |1\rangle \langle 1|\bm{\omega}_k  = L' |(\bm{\omega}_k )_1 |^2.
\label{eq:D.41}
\end{equation}
%===============================================================================
The reduction purity for mode $m$ is then
%===============================================================================
\begin{equation}%                  Equation D.42
P(\redrho{m} ) = \sum\limits_{k = 1}^{n_m } {\bm{\rho}_k ^{\dag} W^{[L']} \bm{\rho}_k }  = L' \sum\limits_{k = 1}^{n_m } {|(\bm{\omega}_k )_1 |^2 \!.} 
\label{eq:D.42}
\end{equation}
%===============================================================================
Since this is just a sum of $n_m$ squares of nonnegative terms, we can use hyperspherical parameterization \cite{Schl} but allow nonunit radius $r_1$, so that
%===============================================================================
\begin{equation}%                  Equation D.43
(\bm{\omega} _k )_1  \equiv r_{1}^2 x_k ^2 ,
\label{eq:D.43}
\end{equation}
%===============================================================================
where $\{x_k\}$ are unit-hyperspherical coordinates for $n_m$ dimensions.  Then, if we set
%===============================================================================
\begin{equation}%                  Equation D.44
r_{1}^4  \equiv \textstyle{1 \over L'},
\label{eq:D.44}
\end{equation}
%===============================================================================
then \Eq{D.42} simplifies to
%===============================================================================
\begin{equation}%                  Equation D.45
P(\redrho{m}) = \sum\limits_{k = 1}^{n_m } {x_k ^4 } .
\label{eq:D.45}
\end{equation}
%===============================================================================
 
Therefore, for $m=m_{\min}$, the minimization of \Eq{D.45} follows the minimization of \App{App.D.2}, yielding
%===============================================================================
\begin{equation}%                  Equation D.46
x_k ^2  = {\textstyle{1 \over {n_m }}}\;\;\forall k \in 1, \ldots ,n_m ,
\label{eq:D.46}
\end{equation}
%===============================================================================
and since \smash{$r_{1}^2 =\textstyle{1 \over \sqrt{L' }}$} from \Eq{D.44}, that with \Eq{D.43} gives
%===============================================================================
\begin{equation}%                  Equation D.47
(\bm{\omega}_k )_1  = {\textstyle{1 \over {n_m \sqrt {L' } }}}.
\label{eq:D.47}
\end{equation}
%===============================================================================
Then, invert the coordinates to get
%===============================================================================
\begin{equation}%                  Equation D.48
\bm{\rho}_k  = F^{[L']} \bm{\omega}_k \,,
\label{eq:D.48}
\end{equation}
%===============================================================================
which yields, for each group $k\in 1,\ldots,n_{m}$,
%===============================================================================
\begin{equation}%                  Equation D.49
\left(\! {\begin{array}{*{20}c}
   {(\bm{\rho} _k )_1 }  \\
    \vdots   \\
   {(\bm{\rho}_k )_{L' } }  \\
\end{array}}\!\! \right) \!=\! \left(\! {\begin{array}{*{20}c}
   {{\textstyle{1 \over {\sqrt {L' } }}}} &  \cdots  & {{\textstyle{1 \over {\sqrt {L'} }}}}  \\
    \vdots  & * & *  \\
   {{\textstyle{1 \over {\sqrt {L'} }}}} & * & *  \\
\end{array}}\! \right)\!\!\!\left(\! {\begin{array}{*{20}c}
   {(\bm{\omega}_k )_1 }  \\
    \vdots   \\
   {(\bm{\omega}_k )_{L'} }  \\
\end{array}}\!\! \right)\!,
\label{eq:D.49}
\end{equation}
%===============================================================================
where the stars hide the general complex elements that maintain the unitarity of the discrete Fourier transform.

So far, we only know $(\bm{\omega}_k )_1 $, the first element of $\bm{\omega}_k$ from \Eq{D.47}. In fact, the remaining elements must always be zero, and to see why this is so, note that
%===============================================================================
\begin{equation}%                  Equation D.50
\bm{\omega}_k ^{\dag}  \bm{\omega}_k  = \bm{\rho}_k ^{\dag}  F^{[L' ]} F^{[L']\dag} \bm{\rho}_k  = \bm{\rho}_k ^{\dag}  \bm{\rho}_k \,,
\label{eq:D.50}
\end{equation}
%===============================================================================
which expands as
%===============================================================================
\begin{equation}%                  Equation D.51
|(\bm{\omega}_k )|_1 ^2  + \sum\limits_{a_k  = 2}^{L'} {|(\bm{\omega}_k )_{a_k }|^2 }  = \sum\limits_{a_k  = 1}^{L'} {|(\bm{\rho}_k )_{a_k }|^2 },
\label{eq:D.51}
\end{equation}
%===============================================================================
where we pulled out the known term on the left.  Then, summing over $k$ gives
%===============================================================================
\begin{equation}%                  Equation D.52
{\textstyle{1 \over {L' }}}P(\redrho{m} ) + \sum\limits_{k = 1}^{n_m } { \sum\limits_{a_k  = 2}^{L' } { (\bm{\omega}_k )_{a_k } ^2}}  = \sum_{k = 1}^{n_m } {\sum_{a_k  = 1}^{L' } {|(\bm{\rho}_k )_{a_k } |^2 } },
\label{eq:D.52}
\end{equation}
%===============================================================================
where we used \Eq{D.42} on the left.  Note that the right side is a sum of $n_{m}L'$ squared parent diagonals, which makes it a sum of $n_{m}L'$ hyperspherical coordinates to the fourth power, something we know how to minimize.

Thus, recalling that we have specified that $m=m_{\min}$, minimizing \Eq{D.52} gives
%===============================================================================
\begin{equation}%                  Equation D.53
{\textstyle{1 \over {L' }}}{\textstyle{1 \over {n_m }}} + \min \left( {\sum\limits_{k = 1}^{n_m } {\sum\limits_{a_k  = 2}^{L' } {|(\bm{\omega} _k )_{a_k } |^2 } } } \right) = {\textstyle{1 \over {n_{m}L'}}},
\label{eq:D.53}
\end{equation}
%===============================================================================
which reveals that,
%===============================================================================
\begin{equation}%                  Equation D.54
\min \left( {\sum\limits_{k = 1}^{n_m } {\sum\limits_{a_k  = 2}^{L' } {|(\bm{\omega} _k )_{a_k } |^2 } } } \right) =0,
\label{eq:D.54}
\end{equation}
%===============================================================================
and furthermore, as a sum of squares equaling zero, \Eq{D.54} means that, for the coordinates that minimize the reduction purity, each element of the transformed coordinates beyond the first is individually zero as well, so that
%===============================================================================
\begin{equation}%                  Equation D.55
(\bm{\omega}_k )_q  = 0,\;\;\forall k \in 1, \ldots ,n_m ,\;\;\forall q \in 2, \ldots ,L',
\label{eq:D.55}
\end{equation}
%===============================================================================
for the reduction-purity-minimizing coordinates.  Then, putting \Eq{D.47} and \Eq{D.55} into \Eq{D.49} we obtain
%===============================================================================
\begin{equation}%                  Equation D.56
\left(\! {\begin{array}{*{20}c}
   {(\bm{\rho}_k )_1 }  \\
    \vdots   \\
   {(\bm{\rho}_k )_{L' } }  \\
\end{array}}\! \right)\! =\! \left(\! {\begin{array}{*{20}c}
   {{\textstyle{1 \over {n_m L' }}}}  \\
    \vdots   \\
   {{\textstyle{1 \over {n_m L' }}}}  \\
\end{array}}\! \right)\;\forall k\in 1,\ldots,n_{m}.
\label{eq:D.56}
\end{equation}
%===============================================================================
Since \Eq{D.56} means that every single nonzero parent diagonal is equal, then together with the fact that the parent must be pure, this means that \textit{all reduction-purity-minimizing pure TGX states must have balanced superposition}.  Therefore, all maximally entangled TGX states must have balanced superposition.  

Thus we have proved the supposition of balanced superposition that we made in \Eqs{D.28}{D.30}.

Incidentally, since there are $n_{m}L'$ total nonzero parent elements in a maximally entangled TGX state, since we designate $L_*$ as this number, that means that the number of nonzero parent elements appearing in each diagonal of the reduction for the smallest mode is $L'=\frac{L_*}{n_{m}}$.
%                               End of App.D.5
%...............................................................................
%                                End of App.D
%-------------------------------------------------------------------------------
%-------------------------------------------------------------------------------
%                  App.E. Ent for Two-Mode Squeezed States
\section{\label{sec:App.E}Ent for Two-Mode Squeezed States}
%_______________________________________________________________________________
\begin{figure}[H]%Not a figure; Puts hypertarget at top of column to fix problem
\centering
\vspace{-12pt}
\setlength{\unitlength}{0.01\linewidth}
\begin{picture}(100,0)
\put(1,25){\hypertarget{Sec:App.E}{}}
\end{picture}
\end{figure}
\vspace{-39pt}
%_______________________________________________________________________________
To get the ent of two-mode squeezed state $|\xi \rangle _2 $ from \Eq{7} using \Eq{6}, first get its reductions,
%===============================================================================
\begin{equation}%                  Equation E.1
\begin{array}{*{20}l}
   {\redrho{m} } &\!\! { = {\textstyle{1 \over {\cosh ^2 (r)}}}\sum\limits_{n = 0}^\infty  {\tanh ^{2n} (r)|n^{(m)} \rangle \langle n^{(m)} |} ,}  \\
\end{array}
\label{eq:E.1}
\end{equation}
%===============================================================================
which are thermal states of mean particle number $\sinh ^2 (r)$, where $\{|n^{(m)} \rangle\}$ is the Fock basis in mode $m$, $n_1  = n_2  = \infty$, and $r$ is the magnitude of $\xi  \equiv re^{i\theta }$.

The purity of both reductions is
%===============================================================================
\begin{equation}%                  Equation E.2
\begin{array}{*{20}l}
   {P(\redrho{m})} &\!\! { = \text{tr}({\redrho{m}}^{2} ) = \sum\limits_{n = 0}^\infty  {{\textstyle{{\tanh ^{4n} (r)} \over {\cosh ^4 (r)}}}}}  \\
\end{array}\!,
\label{eq:E.2}
\end{equation}
%===============================================================================
which is \textit{almost} a geometric series, but the ratio $r'\equiv\tanh^{4}(r)$ fails the convergence requirement $|r'|<1$ at $r=\infty$, since then $r'=1$.  However, a nice trick is to use the coherent-state basis for the trace as
%===============================================================================
\begin{equation}%                  Equation E.3
\text{tr}(A) = {\textstyle{1 \over \pi }}\int {\langle \alpha |A|\alpha \rangle d^2 \alpha }\,,
\label{eq:E.3}
\end{equation}
%===============================================================================
where the coherent states are \cite{Schr,Gla1,Gla2}
%===============================================================================
\begin{equation}%                  Equation E.4
\begin{array}{*{20}l}
   {|\alpha \rangle } &\!\! { \equiv \sqrt {e^{ - |\alpha |^2 } } \sum\limits_{n = 0}^\infty  {{\textstyle{{\alpha ^n } \over {\sqrt{n!}}}}|n\rangle }, }  \\
\end{array}
\label{eq:E.4}
\end{equation}
%===============================================================================
and $|n\rangle$ are Fock states.  Thus, if $\alpha  = se^{i\phi }$, we get
%===============================================================================
\begin{equation}%                  Equation E.5
\begin{array}{*{20}l}
   {\text{tr}({\redrho{m}}^{2} )} &\!\! { = {\textstyle{1 \over \pi }}\int {\langle \alpha |{\redrho{m}}^{2} |\alpha \rangle d^2 \alpha } }  \\
   {P(\redrho{m} )} &\!\! { = {\textstyle{1 \over \pi }}\sum\limits_{n = 0}^\infty  {{\textstyle{{\tanh ^{4n} (r)} \over {\cosh ^4 (r)}}}\int {|\langle n^{(m)} |\alpha \rangle |^2 d^2 \alpha } } }  \\
   {} &\!\! { = {\textstyle{1 \over {\pi \cosh ^4 (r)}}}\int_0^{2\pi }\! {\int_0^\infty  \!{e^{ - s^2 }\! \sum\limits_{n = 0}^\infty \! {{\textstyle{{(s^2 \tanh ^4 (r))^n } \over {n!}}}sdsd\phi } } } }  \\
   {} &\!\! { = {\textstyle{2 \over {\cosh ^4 (r)}}}\int_0^\infty  {e^{ - s^2 (1 - \tanh ^4 (r))} sds}\,. }  \\
\end{array}
\label{eq:E.5}
\end{equation}
%===============================================================================
Then, recalling that \smash{$\int_0^\infty  {s^{2n + 1} e^{ - s^2 /a^2 } ds}  = {\textstyle{{n!} \over 2}}a^{2n + 2}$}, if we set \smash{$a^2  = {\textstyle{1 \over {1 - \tanh ^4 (r)}}}$}, and $n=0$, then \Eq{E.5} becomes
%===============================================================================
\begin{equation}%                  Equation E.6
P(\redrho{m}) = {\textstyle{1 \over {\cosh ^4 (r)}}}{\textstyle{1 \over {1 - \tanh ^4 (r)}}} = {\textstyle{1 \over {2\cosh ^2 (r) - 1}}}\,.
\label{eq:E.6}
\end{equation}
%===============================================================================
Thus, as an aside we get a closed form of the sum in \Eq{E.2},
%===============================================================================
\begin{equation}%                  Equation E.7
\begin{array}{*{20}l}
   {\sum\limits_{n = 0}^\infty \! {{\textstyle{{\tanh ^{4n} (r)} \over {\cosh ^4 (r)}}}} } &\!\! { =\! {\textstyle{1 \over {2\cosh ^2 (r) - 1}}}} \;\;\forall r\in[0,\infty], \\
\end{array}
\label{eq:E.7}
\end{equation}
%===============================================================================
so putting \Eq{E.6} into \Eq{6} yields the result in \Eq{8} as
%===============================================================================
\begin{equation}%                  Equation E.8
\Upsilon (\rho _{|\xi \rangle _2 } ) = 1 - {\textstyle{1 \over {2\cosh ^2 (r) - 1}}}\,.
\label{eq:E.8}
\end{equation}
%===============================================================================
%                                End of App.E
%-------------------------------------------------------------------------------
%-------------------------------------------------------------------------------
%     App.F. Application: Ent Provides a Gauge for Logarithmic Negativity
\section{\label{sec:App.F}Application: Ent Provides a Gauge for Logarithmic Negativity}
%_______________________________________________________________________________
\begin{figure}[H]%Not a figure; Puts hypertarget at top of column to fix problem
\centering
\vspace{-12pt}
\setlength{\unitlength}{0.01\linewidth}
\begin{picture}(100,0)
\put(1,30){\hypertarget{Sec:App.F}{}}
\end{picture}
\end{figure}
\vspace{-39pt}
%_______________________________________________________________________________
In entanglement measures such as logarithmic negativity $E_{\mathcal{N}} (\rho ) \equiv \log _2 ||\rho ^{T_1 } ||_1$ \cite{ViWe} which has a range of $E_{\mathcal{N}} (\rho ) \in [0,\infty )$ for infinite-dimensional systems, there is no way to tell from the measure how close the input state is to being maximally entangled.

For example, for a two-mode squeezed state \cite{Tian,BrvL},
%===============================================================================
\begin{equation}%                  Equation F.1
E_{\mathcal{N}} (\rho _{|\xi \rangle _2 } ) = {\textstyle{1 \over {\ln (\sqrt 2 )}}}r \in [0,\infty ).
\label{eq:F.1}
\end{equation}
%===============================================================================
Note that most values of $E_{\mathcal{N}} (\rho _{|\xi \rangle _2 } )$ are essentially \textit{zero} in comparison to its largest possible value of $\infty$, thus making it a difficult means of judging entanglement.

In contrast, since the ent maps as $\Upsilon (\rho _{|\xi \rangle _2 } ) \in [0,1]$, we always get a clear idea of how close a state is to maximal entanglement.  But more than that, the ent can help us gauge the value of measures such as $E_{\mathcal{N}} (\rho )$.

For example, for a desired ent value $\Upsilon _*$, inverting \Eq{8} gives the corresponding squeezing parameter value as
%===============================================================================
\begin{equation}%                  Equation F.2
r_*  \equiv r (\Upsilon _* ) = \cosh ^{ - 1} \left( {\sqrt {{\textstyle{1 \over 2}}({\textstyle{1 \over {1 - \Upsilon _* }}} + 1)}\, } \right).
\label{eq:F.2}
\end{equation}
%===============================================================================
Then, putting \Eq{F.2} into the value of the measure of interest, which is in this case \Eq{F.1}, gives
%===============================================================================
\begin{equation}%                  Equation F.3
E_{\mathcal{N}} (\rho _{|\xi \rangle _2 } ) = {\textstyle{1 \over {\ln (\sqrt 2 )}}}r_*  = {\textstyle{1 \over {\ln (\sqrt 2 )}}}\cosh ^{ - 1} \left( {\sqrt {{\textstyle{1 \over 2}}({\textstyle{1 \over {1 - \Upsilon _* }}} + 1)} \,} \right)\!.
\label{eq:F.3}
\end{equation}
%===============================================================================
Thus, if we want near-maximal entanglement such as $\Upsilon _* =0.999$, then \Eq{F.2} tells us we want $r_*\approx 3.80$, and \Eq{F.3} shows that this corresponds to $E_{\mathcal{N}} (\rho _{|\xi \rangle _2 } ) \approx 11.0 $, thus giving us a sense of what values of $E_{\mathcal{N}}$ are ``good.''
%                                End of App.F
%-------------------------------------------------------------------------------
%-------------------------------------------------------------------------------
%                       App.G. The 13-Step Algorithm
\section{\label{sec:App.G}The 13-Step Algorithm}
%_______________________________________________________________________________
\begin{figure}[H]%Not a figure; Puts hypertarget at top of column to fix problem
\centering
\vspace{-12pt}
\setlength{\unitlength}{0.01\linewidth}
\begin{picture}(100,0)
\put(1,25){\hypertarget{Sec:App.G}{}}
\end{picture}
\end{figure}
\vspace{-39pt}
%_______________________________________________________________________________
To construct a maximally entangled TGX state, follow the steps below.  For derivations, see \App{App.D}. For a full example of this algorithm, see \App{App.H.2}.
%+++++++++++++++++++++++++++++++++++++++++++++++++++++++++++++++++++++++++++++++
\begin{itemize}[leftmargin=*,labelindent=4pt]\setlength\itemsep{0pt}
%~~~~~~~~~~~~~~~~~~~~~~~~~~~~~~~~~~~~~~~~~~~~~~~~~~~~~~~~~~~~~~~~~~~~~~~~~~~~~~~
\item[\textbf{1.}]\hypertarget{step:1}{}Define the system by its mode-size composition as
%===============================================================================
\begin{equation}%                  Equation G.1
\mathbf{n} \equiv (n_1 , \ldots ,n_N ),\;\;N \equiv \dim (\mathbf{n}),\;\;n \equiv n_{1} \cdots n_{N}.
\label{eq:G.1}
\end{equation}
%===============================================================================
%~~~~~~~~~~~~~~~~~~~~~~~~~~~~~~~~~~~~~~~~~~~~~~~~~~~~~~~~~~~~~~~~~~~~~~~~~~~~~~~
%~~~~~~~~~~~~~~~~~~~~~~~~~~~~~~~~~~~~~~~~~~~~~~~~~~~~~~~~~~~~~~~~~~~~~~~~~~~~~~~
\item[\textbf{2.}]\hypertarget{step:2}{}Calculate the set $\mathbf{L}_* \equiv \{ L_* \}$ of numbers of levels of superposition that support maximal entanglement as
%===============================================================================
\begin{equation}%                  Equation G.2
\mathbf{L}_*  \equiv \{ L_* \} ;\;\;\text{s.t.}\;\mathop {\min }\limits_{L \in 2, \ldots, n_{\,\overline{{\kern -1.8pt}\max^{~^{~^{~}}}\!\!\!\!\!\!\!\!\!\!}} } (1 - M(L)),
\label{eq:G.2}
\end{equation}
%===============================================================================
where $M(L)$ is given in \Eq{3}, and $n_{\,\overline{{\kern -1.8pt}\max^{~^{~^{~}}}\!\!\!\!\!\!\!\!\!\!}} \equiv \frac{n}{{n_{\max } }}$, where $n_{\max}\equiv\max(\mathbf{n})$.  Pick one element of $\mathbf{L}_*$ to be $L_*$.
%~~~~~~~~~~~~~~~~~~~~~~~~~~~~~~~~~~~~~~~~~~~~~~~~~~~~~~~~~~~~~~~~~~~~~~~~~~~~~~~
%~~~~~~~~~~~~~~~~~~~~~~~~~~~~~~~~~~~~~~~~~~~~~~~~~~~~~~~~~~~~~~~~~~~~~~~~~~~~~~~
\item[\textbf{3.}]\hypertarget{step:3}{}Calculate the \textit{primary goals vector} as
%===============================================================================
\begin{equation}%                  Equation G.3
\mathbf{G} \equiv (\mathbf{G}^{(1)}| \cdots |\mathbf{G}^{(N)}),
\label{eq:G.3}
\end{equation}
%===============================================================================
where $\mathbf{G}^{(m)}\equiv L_* \text{diag}^T (\redrho{m} _{\text{MP}} )$, calculated as
%===============================================================================
\begin{equation}%                  Equation G.4
\begin{array}{*{20}l}
   {\mathbf{G}^{(m_{\,\overline{{\kern -1.8pt}\max^{~^{~^{~}}}\!\!\!\!\!\!\!\!\!\!}} )}}\! &\!\!\! {=\! \text{floor}({\textstyle{{L_* } \over {n_m }}})\mathbf{1}_{n_m }^T ,}  \\
   {\mathbf{G}^{(m_{\max } )} }\! &\!\!\! { =\! ((1\!\! +\! \text{floor}({\textstyle{{L_* } \over {n_m }}}))\mathbf{1}_{S_{m}}^T ,\text{floor}({\textstyle{{L_* } \over {n_m }}})\mathbf{1}_{R_m}^T ),}  \\
\end{array}\!\!\!\!
\label{eq:G.4}
\end{equation}
%===============================================================================
where $\mathbf{1}_{k}^T  \!\equiv \!(1_1 , \ldots ,1_{k} )$, $R_m \!\equiv\! n_m \! -\! S_m$, $S_{m}\!\equiv \!\text{mod}(L_* ,n_m )$, and $m_{\max } $ is the \textit{nominally largest subsystem}, meaning that if more than one subsystem is the largest, such as in $2 \times 3 \times 3$, we only pick \textit{one} of the subsystem labels as the \textit{nominally largest}, such as $m_{\max } \!=\!2$.  Thus $m_{\,\overline{{\kern -1.8pt}\max^{~^{~^{~}}}\!\!\!\!\!\!\!\!\!\!}}$\, is any mode label \textit{except} $m_{\max } $.
%~~~~~~~~~~~~~~~~~~~~~~~~~~~~~~~~~~~~~~~~~~~~~~~~~~~~~~~~~~~~~~~~~~~~~~~~~~~~~~~
%~~~~~~~~~~~~~~~~~~~~~~~~~~~~~~~~~~~~~~~~~~~~~~~~~~~~~~~~~~~~~~~~~~~~~~~~~~~~~~~
\item[\textbf{4.}]\hypertarget{step:4}{}List all possible arrangements of the goals in the nominally largest mode as an \smash{$\binom{n_{\max }}{\text{mod}(L_* ,n_{\max} )}\times n_{\max }$} matrix \smash{$G^{(m_{\max } )}$}, with elements
%===============================================================================
\begin{equation}%                  Equation G.5
G_{j,h} ^{(m_{\max } )}  = \left\{ {\begin{array}{*{20}l}
   {1 + \text{floor}(\frac{{L_* }}{{n_{\max } }});} & {h \in C_{j, \cdots }^{(m_{\max } )} }  \\
   {\text{floor}(\frac{{L_* }}{{n_{\max } }});} & {h \notin C_{j, \cdots }^{(m_{\max } )} }  \\
\end{array}} \right. ,
\label{eq:G.5}
\end{equation}
%===============================================================================
where{\kern -1.5pt} \smash{$\binom{a}{b}\equiv\frac{a!}{b!(a-b)!}$}, and \smash{$C_{j, \cdots }^{(m_{\max } )}$} is the $j$th row vector of{\kern 1.7pt} \smash{$\binom{n_{\max }}{\text{mod}(L_* ,n_{\max} )}\times \text{mod}(L_* ,n_{\max } )$} matrix
%===============================================================================
\begin{equation}%                  Equation G.6
C^{(m_{\max } )}  \equiv \text{nCk}(\mathbf{c}^{[n_{\max } ]T} ,\text{mod}(L_* ,n_{\max } )),
\label{eq:G.6}
\end{equation}
%===============================================================================
where $\mathbf{c}^{[n_{\max } ]T}  \equiv (1,2, \ldots ,n_{\max } )$, and $\text{nCk}(\mathbf{v},k)$ is the vectorized $n$-choose-$k$ function that gives the matrix whose rows are each unique combinations of the elements of $\mathbf{v}$ chosen $k$ at a time.
%~~~~~~~~~~~~~~~~~~~~~~~~~~~~~~~~~~~~~~~~~~~~~~~~~~~~~~~~~~~~~~~~~~~~~~~~~~~~~~~
%~~~~~~~~~~~~~~~~~~~~~~~~~~~~~~~~~~~~~~~~~~~~~~~~~~~~~~~~~~~~~~~~~~~~~~~~~~~~~~~
\item[\textbf{5.}]\hypertarget{step:5}{}Form the total goals matrix as
%===============================================================================
\begin{equation}%                  Equation G.7
G \equiv (G^{(1)} | \cdots |G^{(N)} ),
\label{eq:G.7}
\end{equation}
%===============================================================================
where $G^{(m \ne m_{\max } )} \! \equiv\! \mathbf{1}_{\binom{n_{\max }}{\text{mod}(L_* ,n_{\max } )}}\! \otimes\! \mathbf{G}^{(m)}$ where $\mathbf{1}_k \!\equiv\! (1_1 , \ldots ,1_k )^T$ and $\mathbf{G}^{(m)}$ is given in \Eq{G.4}, while $G^{(m = m_{\max } )}$ is from \hyperlink{step:4}{Step 4}.  Thus, $G$ is an $\binom{n_{\max }}{\text{mod}(L_* ,n_{\max } )}\times (\Sigma _{m = 1}^N n_m )$ matrix.
%~~~~~~~~~~~~~~~~~~~~~~~~~~~~~~~~~~~~~~~~~~~~~~~~~~~~~~~~~~~~~~~~~~~~~~~~~~~~~~~
%~~~~~~~~~~~~~~~~~~~~~~~~~~~~~~~~~~~~~~~~~~~~~~~~~~~~~~~~~~~~~~~~~~~~~~~~~~~~~~~
\item[\textbf{6.}]\hypertarget{step:6}{}Define the $n \times (\Sigma _{m = 1}^N n_m )$ occurrence matrix as
%===============================================================================
\begin{equation}%                  Equation G.8
\Omega  \equiv\! \left(\! {\begin{array}{*{20}c}
   {\Omega _{1, \cdots } }  \\
    \vdots   \\
   {\Omega _{n, \cdots } }  \\
\end{array}}\! \right)\!;\;\Omega _{v, \cdots }  \!\equiv \left({\langle a_1^{(1)} | ,  \ldots  , \langle a_N^{(N)} |}\right)\!,
\label{eq:G.8}
\end{equation}
%===============================================================================
so \smash{$\Omega _{k, \cdots }$} are concatenated vectors of computational mode-basis bras where $a_m  \in 1, \ldots ,n_m $ are given by the inverse indical register function of \Eq{H.1} as $\mathbf{a}\equiv (a_1 , \ldots ,a_N )\equiv\mathbf{a}_v^{\{N,\mathbf{n}\}}$. For example, in $2 \times 2$, \smash{$\Omega _{1, \cdots } =$} \smash{$(\langle 1^{(1)} | , \langle 1^{(2)} |) =\left({\binom{1}{0}{\rule{0pt}{5pt}}^{T},\binom{1}{0}{\rule{0pt}{5pt}}^{T}}\right)\!,$} \smash{$\Omega _{2, \cdots } = (\langle 1^{(1)} | , \langle 2^{(2)} |) =$} \smash{$\left({\binom{1}{0}{\rule{0pt}{5pt}}^{T},\binom{0}{1}{\rule{0pt}{5pt}}^{T}}\right)\!,$} etc.  By convention we label computational basis kets starting on $1$, as $|1\rangle,|2\rangle,\ldots$, to simplify formulas; these are \textit{not} necessarily Fock states.
%~~~~~~~~~~~~~~~~~~~~~~~~~~~~~~~~~~~~~~~~~~~~~~~~~~~~~~~~~~~~~~~~~~~~~~~~~~~~~~~
%~~~~~~~~~~~~~~~~~~~~~~~~~~~~~~~~~~~~~~~~~~~~~~~~~~~~~~~~~~~~~~~~~~~~~~~~~~~~~~~
\item[\textbf{7.}]\hypertarget{step:7}{}Make a ones matrix of the total TGX space as
%===============================================================================
\begin{equation}%                  Equation G.9
W_{\text{TGX}}^{[\mathbf{n}]}  \equiv W^{[\mathbf{n}]}  - W_{\,\overline{{\kern -0.5pt}\text{TGX}^{~^{~^{~}}}\!\!\!\!\!\!\!\!\!\!}}^{[\mathbf{n}]}\,\,,
\label{eq:G.9}
\end{equation}
%===============================================================================
where \smash{$W^{[\mathbf{n}]}$} is the $n\times n$ matrix of ones, and \smash{$W_{\,\overline{{\kern -0.5pt}\text{TGX}^{~^{~^{~}}}\!\!\!\!\!\!\!\!\!\!}}^{[\mathbf{n}]}$} is the anti-TGX ones matrix,
%===============================================================================
\begin{equation}%                  Equation G.10
W_{\,\overline{{\kern -0.5pt}\text{TGX}^{~^{~^{~}}}\!\!\!\!\!\!\!\!\!\!}}^{[\mathbf{n}]} \equiv\!\!\!\!\!{\kern -28pt}\!\! \sum\limits_{{\kern 34pt}m,a_m ,b_m ,\mathbf{k}_{\mbarsub}  \,\,=\, 1,1,1,\mathbf{1}_{\mbarsub} }^{{\kern 0pt}N,n_m ,n_m ,\mathbf{n}_{\mbarsub} }\!\!\!\!\!\!\!\!\!\!\!\!\!\!\!\!\!\!\!\!\!\!\!\!\!\!\!\!\!\!\! {(1\! -\! \delta _{a_m ,b_m } )\!\left. {E_{(R_\mathbf{a}^{\{ N,\mathbf{n}\} } ,R_\mathbf{b}^{\{ N,\mathbf{n}\} } )}^{[n]} } \right|_{\scriptstyle \mathbf{a}_{\mbarsub}  \,=\, \mathbf{k}_{\mbarsub}  \hfill \atop 
  \scriptstyle \mathbf{b}_{\mbarsub}  \,=\, \mathbf{k}_{\mbarsub}  \hfill} }\!,
\label{eq:G.10}
\end{equation}
%===============================================================================
where \smash{$E_{(a,b)}^{[n]}$} is the $n\times n$ elementary matrix with a $1$ in the row-$a$ column-$b$ element and $0$ elsewhere, and the vector-index definitions match those in \Eq{B.3}, and \smash{$R_{\mathbf{a}}^{\{ N,\mathbf{n}\} } $} is the indical register function of \Eq{B.4}.
%~~~~~~~~~~~~~~~~~~~~~~~~~~~~~~~~~~~~~~~~~~~~~~~~~~~~~~~~~~~~~~~~~~~~~~~~~~~~~~~
%~~~~~~~~~~~~~~~~~~~~~~~~~~~~~~~~~~~~~~~~~~~~~~~~~~~~~~~~~~~~~~~~~~~~~~~~~~~~~~~
\item[\textbf{8.}]\hypertarget{step:8}{}Pick a \textit{starting level}; the label (from $1$ to $n$) of the computational basis that is to definitely be included in the $L_*$-level maximally entangled TGX state, as
%===============================================================================
\begin{equation}%                  Equation G.11
S_L  \equiv l \in 1, \ldots ,n \,.
\label{eq:G.11}
\end{equation}
%===============================================================================
%~~~~~~~~~~~~~~~~~~~~~~~~~~~~~~~~~~~~~~~~~~~~~~~~~~~~~~~~~~~~~~~~~~~~~~~~~~~~~~~
%~~~~~~~~~~~~~~~~~~~~~~~~~~~~~~~~~~~~~~~~~~~~~~~~~~~~~~~~~~~~~~~~~~~~~~~~~~~~~~~
\item[\textbf{9.}]\hypertarget{step:9}{}Find the initial set of nonzero TGX levels that are pure-state compatible with $S_L$ as
%===============================================================================
\begin{equation}%                  Equation G.12
L_{\text{cand}} '' \equiv \{ R_{\text{cand}} ,C_{\text{cand}} \} ,
\label{eq:G.12}
\end{equation}
%===============================================================================
where $R_{\text{cand}}$ are the rows in column $S_L$ in which a $1$ exists in the lower-triangular part of the TGX-space matrix \smash{$W_{\text{TGX}}^{[\mathbf{n}]}$}, and $C_{\text{cand}}$ are the columns in row $S_L$ in which a $1$ exists in the lower-triangular part of \smash{$W_{\text{TGX}}^{[\mathbf{n}]}$},% both given by
%===============================================================================
\begin{equation}%                  Equation G.13
\begin{array}{*{20}l}
   {R_{\text{cand}} } &\!\! { \equiv \{ k\} |_{k = S_L  + 1}^n ;} &\!\! {\;\;\text{s.t.}\;\;(W_{\text{TGX}}^{[\mathbf{n}]} )_{k,S_L } } &\!\! { = 1}  \\
   {C_{\text{cand}} } &\!\! { \equiv \{ k\} |_{k = 1}^{S_L  - 1} ;} &\!\! {\;\;\text{s.t.}\;\;(W_{\text{TGX}}^{[\mathbf{n}]} )_{S_L ,k} } &\!\! { = 1}.  \\
\end{array}
\label{eq:G.13}
\end{equation}
%===============================================================================
The number of elements in $L_{\text{cand}} ''$ is
%===============================================================================
\begin{equation}%                  Equation G.14
\begin{array}{*{20}l}
   {\dim (L_{\text{cand}} '')} &\!\! {= {\textstyle{1 \over n}}N_{\text{TGX-off}}} &\!\! {= n\! -\! 1\! + \!N \!- \!\sum\limits_{m = 1}^N \!{n_m },}  \\
\end{array}
\label{eq:G.14}
\end{equation}
%===============================================================================
and the number of nonzero off-diagonals in \smash{$W_{\text{TGX}}^{[\mathbf{n}]}$} is
%===============================================================================
\begin{equation}%                  Equation G.15
\begin{array}{*{20}l}
   {N_{\text{TGX-off}}} &\!\! {= n^2  - n - \sum\limits_{m = 1}^N {n_{\mbarsub} \,(n_m^2  - n_m )}}  \\
   {} &\!\! {= n(n\! -\! 1\! +\! N\! -\! \sum\limits_{m = 1}^N \!{n_m } ).}  \\
\end{array}
\label{eq:G.15}
\end{equation}
%===============================================================================
%~~~~~~~~~~~~~~~~~~~~~~~~~~~~~~~~~~~~~~~~~~~~~~~~~~~~~~~~~~~~~~~~~~~~~~~~~~~~~~~
%~~~~~~~~~~~~~~~~~~~~~~~~~~~~~~~~~~~~~~~~~~~~~~~~~~~~~~~~~~~~~~~~~~~~~~~~~~~~~~~
\item[\textbf{10.}]\hypertarget{step:10}{}Construct the candidate compatibility sets for $S_L$ by listing all of the combinations of $L_{*}-1$ levels from the initial set $L_{\text{cand}} ''$, and prepending $S_L$ as
%===============================================================================
\begin{equation}%                  Equation G.16
L_{\text{cand}} ' \equiv \{ S_L \mathbf{1}_{\binom{\dim (L_{\text{cand}} '')}{L_*  - 1}} ,\text{nCk}(L_{\text{cand}} '',L_*  - 1)\} ,
\label{eq:G.16}
\end{equation}
%===============================================================================
which is an \smash{$\binom{\dim (L_{\text{cand}} '')}{L_*  - 1}\times L_{*}$} matrix where $\text{nCk}(\mathbf{v},k)$ is defined in \hyperlink{step:4}{Step 4}, and $\dim (L_{\text{cand}} '')$ is from \Eq{G.14}.
%~~~~~~~~~~~~~~~~~~~~~~~~~~~~~~~~~~~~~~~~~~~~~~~~~~~~~~~~~~~~~~~~~~~~~~~~~~~~~~~
%~~~~~~~~~~~~~~~~~~~~~~~~~~~~~~~~~~~~~~~~~~~~~~~~~~~~~~~~~~~~~~~~~~~~~~~~~~~~~~~
\item[\textbf{11.}]\hypertarget{step:11}{}Find rows of $L_{\text{cand}} '$ for which all the levels are \textit{mutually} pure-TGX-state compatible and arrange them as
%===============================================================================
\begin{equation}%                  Equation G.17
L_{\text{cand}}  \equiv\! \left(\! {\begin{array}{*{20}c}
   {(L_{\text{cand}} ')_{r_* , \cdots } }  \\
    \vdots   \\
\end{array}}\! \right)\!,
\label{eq:G.17}
\end{equation}
%===============================================================================
where rows $r_{*}$ of mutually compatible levels are
%===============================================================================
\begin{equation}%                  Equation G.18
\begin{array}{*{20}l}
   {\{ r_* \}\!\equiv\!} &\! {r \in 1, \ldots ,\binom{\dim (L_{\text{cand}} '')}{L_*  - 1}}  \\
   {\,\,\text{s.t.}} &\!\! {\!\! \begin{array}{*{20}l}
   {\sum\limits_{k = 1}^{L_*  - 1}\! {\sum\limits_{l = k + 1}^{L_* }\!\!\!\! {(W_{\text{TGX}}^{[\mathbf{n}]} )_{(L_{\text{cand}} ')_{r,l} ,(L_{\text{cand}} ')_{r,k} } } } } &\!\! { \!=\! \frac{L_* ^2  - L_* }{2},}  \\
\end{array}}  \\
\end{array}
\label{eq:G.18}
\end{equation}
%===============================================================================
where $\dim (L_{\text{cand}} '')$ is from \Eq{G.14}, \smash{$W_{\text{TGX}}^{[\mathbf{n}]}$} is from \Eq{G.9}, and \smash{$L_{\text{cand}} '$} is from \Eq{G.16}.
%~~~~~~~~~~~~~~~~~~~~~~~~~~~~~~~~~~~~~~~~~~~~~~~~~~~~~~~~~~~~~~~~~~~~~~~~~~~~~~~
%~~~~~~~~~~~~~~~~~~~~~~~~~~~~~~~~~~~~~~~~~~~~~~~~~~~~~~~~~~~~~~~~~~~~~~~~~~~~~~~
\item[\textbf{12.}]\hypertarget{step:12}{}For each row in \smash{$L_{\text{cand}}$}, find the sum of the goals vectors of every level in that row.  The rows whose goal sums exactly match any one of the goals vectors are then verified compatibility sets, which we arrange as
%===============================================================================
\begin{equation}%                  Equation G.19
L_{\text{ME}}  \equiv\! \left(\! {\begin{array}{*{20}c}
   {(L_{\text{cand}} )_{R_* , \cdots } }  \\
    \vdots   \\
\end{array}}\! \right)\!,
\label{eq:G.19}
\end{equation}
%===============================================================================
where the qualifying rows are
%===============================================================================
\begin{equation}%                  Equation G.20
\begin{array}{*{20}l}
   {\{ R_* \}\!  \equiv } &\!\! {R \in 1, \ldots ,\dim _{\text{rows}} (L_{\text{cand}} )}  \\
   {\,\,\text{s.t.}} &\!\! {\sum\limits_{c = 1}^{L_* } {\Omega _{((L_{\text{cand}} )_{R,c} ), \cdots } }  = G_{R_{\text{any}} , \cdots }\,, }  \\
\end{array}
\label{eq:G.20}
\end{equation}
%===============================================================================
where $\Omega$ is from \Eq{G.8}, $G$ is from \Eq{G.7}, and \smash{$R_{\text{any}}\in 1,\ldots,\binom{n_{\max}}{\text{mod}(L_{*},n_{\max})}$} means any row index corresponding to a row vector of $G$, meaning that as long as the sum of occurrence vectors equals at least one of the row vectors of the goals matrix, then row $R_{*}$ of \smash{$L_{\text{cand}}$} is a verified compatibility set.
%~~~~~~~~~~~~~~~~~~~~~~~~~~~~~~~~~~~~~~~~~~~~~~~~~~~~~~~~~~~~~~~~~~~~~~~~~~~~~~~
%~~~~~~~~~~~~~~~~~~~~~~~~~~~~~~~~~~~~~~~~~~~~~~~~~~~~~~~~~~~~~~~~~~~~~~~~~~~~~~~
\item[\textbf{13.}]\hypertarget{step:13}{}The rows of \smash{$L_{\text{ME}}$} form the set of all possible combinations of $L_*$ levels including $S_{L}$ for which balanced superposition yields maximally entangled TGX states.  Thus, the maximally entangled TGX states of zero relative phase produced by this algorithm are
%===============================================================================
\begin{equation}%                  Equation G.21
|\Phi _j \rangle  \equiv {\textstyle{1 \over {\sqrt {L_* } }}}\sum\limits_{k = 1}^{L_* } {|(L_{\text{ME}} )_{j,k} \rangle },
\label{eq:G.21}
\end{equation}
%===============================================================================
for $j\in 1,\ldots,\dim_{\text{rows}}(L_{\text{ME}})$. Infinite alternatives are
%===============================================================================
\begin{equation}%                  Equation G.22
|\Phi _j (\bm{\phi})\rangle  \equiv {\textstyle{1 \over {\sqrt {L_* } }}}\sum\limits_{k = 1}^{L_* } {e^{i\phi_k}|(L_{\text{ME}} )_{j,k} \rangle },
\label{eq:G.22}
\end{equation}
%===============================================================================
where $\phi_k \in[0,2\pi)$, with an unimportant global phase.  Thus, we can take the states of \Eq{G.21} $\forall j$ as the canonical output of the 13-step algorithm \smash{$\mathcal{A}_{13}$}, as in \Eq{9}.
%~~~~~~~~~~~~~~~~~~~~~~~~~~~~~~~~~~~~~~~~~~~~~~~~~~~~~~~~~~~~~~~~~~~~~~~~~~~~~~~
\end{itemize}
%+++++++++++++++++++++++++++++++++++++++++++++++++++++++++++++++++++++++++++++++

Thus, the 13-step algorithm finds all combinations of levels that support maximal entanglement in TGX states.  Again, for details about \textit{why} the above steps are correct, see \App{App.D}. To see a large number of example results as a quick reference, see \App{App.H.1}.  For a thorough example of this algorithm see \App{App.H.2}.
%                                End of App.G
%-------------------------------------------------------------------------------
%-------------------------------------------------------------------------------
%               App.H. Maximally Entangled TGX State Examples
\section{\label{sec:App.H}Maximally Entangled TGX State Examples}
%_______________________________________________________________________________
\begin{figure}[H]%Not a figure; Puts hypertarget at top of column to fix problem
\centering
\vspace{-12pt}
\setlength{\unitlength}{0.01\linewidth}
\begin{picture}(100,0)
\put(1,30){\hypertarget{Sec:App.H}{}}
\end{picture}
\end{figure}
\vspace{-39pt}
%_______________________________________________________________________________
\Appendix{App.H.1} gives many maximally entangled TGX-state examples, while \App{App.H.2} shows a step-by-step calculation in detail, both demonstrating the 13-step algorithm \smash{$\mathcal{A}_{13}$} from \Eq{9} and \App{App.G}.
%...............................................................................
%          App.H.1 Tables of Maximally Entangled TGX States
\subsection{\label{sec:App.H.1}Tables of Maximally Entangled TGX States}
Here we list many results of \smash{$\mathcal{A}_{13}$} from \Eq{9} and \App{App.G}, as a reference.  Due to the conceptual equivalence of different orders of modes, such as $2\times 3$ and $3\times 2$, we only show one kind for each system.
%_______________________________________________________________________________
%                                    TABLE 2
\begin{table}[H]
\caption{\label{tab:2}Sets of all possible $L_*$ values for TGX states of all multipartite systems up to $n=28$ dimensions, where $L_*$ is a number of nonzero probability amplitudes that can support maximal entanglement for a TGX state.  For example, in $2\times 2\times 2$, the entry $\{L_*\}=2,4$ means that this system can have maximally entangled TGX states of either \textit{two} or \textit{four} nonzero probability amplitudes.}
\begin{ruledtabular}
\begin{tabular}{|c|c|c|c|@{${\kern -4pt}$}c|c|c|}
$n$ & $n_{1}\times\cdots\times n_{N}$ & $\{L_{*}\}$ & &{\kern 2pt}$n$ & $n_{1}\times\cdots\times n_{N}$ & $\{L_{*}\}$ \\[0.5mm]
\cline{1-3} \cline{5-7}
%Table Elements:
$\vertsp{4}$ & $2\times 2$ & $2$ & &{\kern 2pt}$20$ & $2\times 10$ & $2$\\[0.5mm]
\cline{1-3} \cline{5-7}
$\vertsp{6}$ & $2\times 3$ & $2$ & &{\kern 2pt}$20$ & $4\times 5$ & $4$\\[0.5mm]
\cline{1-3} \cline{5-7}
$\vertsp{8}$ & $2\times 4$ & $2$ & &{\kern 2pt}$20$ & $2\times 2\times 5$ & $4$\\[0.5mm]
\cline{1-3} \cline{5-7}
$\vertsp{8}$ & $2\times 2\times 2$ &$2,4$ & &{\kern 2pt}$21$ & $3\times 7$ & $3$\\[0.5mm]
\cline{1-3} \cline{5-7}
$\vertsp{9}$ & $3\times 3$ & $3$ & &{\kern 2pt}$22$ & $2\times 11$ & $2$\\[0.5mm]
\cline{1-3} \cline{5-7}
$\vertsp{10}$ & $2\times 5$ & $2$ & &{\kern 2pt}$24$ & $2\times 12$ & $2$\\[0.5mm]
\cline{1-3} \cline{5-7}
$\vertsp{12}$ & $2\times 6$ & $2$ & &{\kern 2pt}$24$ & $3\times 8$ & $3$\\[0.5mm]
\cline{1-3} \cline{5-7}
$\vertsp{12}$ & $3\times 4$ & $3$ & &{\kern 2pt}$24$ & $4\times 6$ & $4$\\[0.5mm]
\cline{1-3} \cline{5-7}
$\vertsp{12}$ & $2\times 2\times 3$ & $4$ & &{\kern 2pt}$24$ & $2\times 2\times 6$ & $4$\\[0.5mm]
\cline{1-3} \cline{5-7}
$\vertsp{14}$ & $2\times 7$ & $2$ & &{\kern 2pt}$24$ & $2\times 3\times 4$ & $6$\\[0.5mm]
\cline{1-3} \cline{5-7}
$\vertsp{15}$ & $3\times 5$ & $3$ & &{\kern 2pt}$24$ & $2\times 2\times 2\times 3$ & $6$\\[0.5mm]
\cline{1-3} \cline{5-7}
$\vertsp{16}$ & $2\times 8$ & $2$ & &{\kern 2pt}$25$ & $5\times 5$ & $5$\\[0.5mm]
\cline{1-3} \cline{5-7}
$\vertsp{16}$ & $4\times 4$ & $4$ & &{\kern 2pt}$26$ & $2\times 13$ & $2$\\[0.5mm]
\cline{1-3} \cline{5-7}
$\vertsp{16}$ & $2\times 2\times 4$ & $4$ & &{\kern 2pt}$27$ & $3\times 9$ & $3$\\[0.5mm]
\cline{1-3} \cline{5-7}
$\vertsp{16}$ & $2\times 2\times 2\times 2$ & $2,4,6,8$ & &{\kern 2pt}$27$ & $3\times 3\times 3$ & $3,6,9$\\[0.5mm]
\cline{1-3} \cline{5-7}
$\vertsp{18}$ & $2\times 9$ & $2$ & &{\kern 2pt}$28$ & $2\times 14$ & $2$\\[0.5mm]
\cline{1-3} \cline{5-7}
$\vertsp{18}$ & $3\times 6$ & $3$ & &{\kern 2pt}$28$ & $4\times 7$ & $4$\\[0.5mm]
\cline{1-3} \cline{5-7}
$\vertsp{18}$ & $2\times 3\times 3$ & $6$ & &{\kern 2pt}$28$ & $2\times 2\times 7$ & $4$\\[0.5mm]
\end{tabular}
\end{ruledtabular}
\end{table}
%_______________________________________________________________________________
\Table{2} provides a list of all the possible \smash{$L_{*}$} values for several systems, which are numbers of nonzero probability amplitudes for maximally entangled TGX states, examples of which are seen in \Table{3} and \Table{4}.

Note that in \Table{3}, although real-valued states are the canonical representatives, there is actually full phase freedom in each state because they are TGX states. Thus, for example, in the $3\times 4$ row, we could also write \smash{$|\Phi \rangle  = {\textstyle{1 \over {\sqrt 3 }}}(|1\rangle  + e^{i\phi _{6|1} } |6\rangle  + e^{i\phi _{11|1} } |11\rangle )$}, where $\phi _{6|1}  \in [0,2\pi )$ and $\phi _{11|1}  \in [0,2\pi )$ are both relative phases that can take on any value without affecting the entanglement.

To convert these unipartite labels to multipartite labels, use the \textit{inverse indical register function} (see \Eq{B.4}),
%===============================================================================
\begin{equation}%                  Equation H.1
\mathbf{a}_v^{\{ N,\mathbf{n}\} }  = (a_1 , \ldots a_N );\;\;\left\{ {\begin{array}{*{20}l}
   {a_m } &\!\! { = \text{floor}(\frac{v_{m - 1}  - 1}{D_{m}})+1}  \\
   {v_m } &\!\! { = v_{m - 1}  - (a_m  - 1)D_m ,}  \\
\end{array}} \right.
\label{eq:H.1}
\end{equation}
%===============================================================================
for $m\in 1,\ldots,N$, where $v_0  \equiv v$, \smash{$D_m  \equiv \Pi_{q=m+1}^{N}n_{q}$}, and $v$ is the scalar value being converted into vector index $\mathbf{a}$.  Thus, applying \Eq{H.1} to the $3\times 4$ example above gives \smash{$|\Phi \rangle  = {\textstyle{1 \over {\sqrt 3 }}}(|1,1\rangle  +$} \smash{$ e^{i\phi _{6|1} } |2,2\rangle  + e^{i\phi _{11|1} } |3,3\rangle )$}, again where each subsystem's basis label starts on $1$ in our convention.

\Table{3} only gives one possible example for each system.  For more general results, \Table{4} gives \textit{all} unique canonical sets for the first nine systems.
%_______________________________________________________________________________
%                                    TABLE 3
\begin{table}[H]
\caption{\label{tab:3}Single ``canonical'' examples of maximally entangled TGX states for every system up to $n=28$ levels, where the canonical convention is to use starting level $S_{L}=1$, $L_{*}=\min\{\mathbf{L}_{*}\}$, and to pick the first state produced by \smash{$\mathcal{A}_{13}$}.  Entries are read as labels of kets in balanced superposition, such as for $2\times 3$, the entry $\{L_{\text{ME}}\}=1,5$ means \smash{$|\Phi \rangle  = {\textstyle{1 \over {\sqrt 2 }}}(|1\rangle  + |5\rangle )$} where the computational basis is $\{ |1\rangle , \ldots ,|n\rangle \}$, not to be confused with Fock states.  See \Eq{H.1} to convert to the coincidence basis.}
\begin{ruledtabular}
\begin{tabular}{|c|c|c|@{${\kern -3pt}$}c|c|}
%Column Titles:
$n_{1}\!\times\!\cdots\!\times\! n_{N}$ & $\{L_{\text{ME}}\}$ & &$\,n_{1}\!\times\!\cdots\!\times\! n_{N}$ & $\{L_{\text{ME}}\}$ \\[0.5mm]
\cline{1-2} \cline{4-5}
%Table Elements:
$\vertsp{2\!\times\! 2}$ & $1,\!4$ & &$2\!\times\! 10$ & $1,\!12$\\[0.5mm]
\cline{1-2} \cline{4-5}
$\vertsp{2\!\times\! 3}$ & $1,\!5$ & &$4\!\times\! 5$ & $1,\!7,\!13,\!19$\\[0.5mm]
\cline{1-2} \cline{4-5}
$\vertsp{2\!\times\! 4}$ & $1,\!6$ & &$2\!\times\! 2\!\times\! 5$ & $1,\!7,\!13,\!19$\\[0.5mm]
\cline{1-2} \cline{4-5}
$\vertsp{2\!\times\! 2\!\times\! 2}$ &$1,\!8$ & &$3\!\times\! 7$ & $1,\!9,\!17$\\[0.5mm]
\cline{1-2} \cline{4-5}
$\vertsp{3\!\times\! 3}$ & $1,\!5,\!9$ & &$2\!\times\! 11$ & $1,\!13$\\[0.5mm]
\cline{1-2} \cline{4-5}
$\vertsp{2\!\times\! 5}$ & $1,\!7$ & &$2\!\times\! 12$ & $1,\!14$\\[0.5mm]
\cline{1-2} \cline{4-5}
$\vertsp{2\!\times\! 6}$ & $1,\!8$ & &$3\!\times\! 8$ & $1,\!10,\!19$\\[0.5mm]
\cline{1-2} \cline{4-5}
$\vertsp{3\!\times\! 4}$ & $1,\!6,\!11$ & &$4\!\times\! 6$ & $1,\!8,\!15,\!22$\\[0.5mm]
\cline{1-2} \cline{4-5}
$\vertsp{2\!\times\! 2\!\times\! 3}$ & $1,\!5,\!8,\!12$ & &$2\!\times\! 2\!\times\! 6$ & $1,\!8,\!15,\!22$\\[0.5mm]
\cline{1-2} \cline{4-5}
$\vertsp{2\!\times\! 7}$ & $1,\!9$ & &$2\!\times\! 3\!\times\! 4$ & $1,\!6,\!11,\!14,\!17,\!24$\\[0.5mm]
\cline{1-2} \cline{4-5}
$\vertsp{3\!\times\! 5}$ & $1,\!7,\!13$ & &$2\!\times\! 2\!\times\! 2\!\times\! 3$ & $1,\!5,\!8,\!18,\!21,\!22$\\[0.5mm]
\cline{1-2} \cline{4-5}
$\vertsp{2\!\times\! 8}$ & $1,\!10$ & &$5\!\times\! 5$ & $1,\!7,\!13,\!19,\!25$\\[0.5mm]
\cline{1-2} \cline{4-5}
$\vertsp{4\!\times\! 4}$ & $1,\!6,\!11,\!16$ & &$2\!\times\! 13$ & $1,\!15$\\[0.5mm]
\cline{1-2} \cline{4-5}
$\vertsp{2\!\times\! 2\!\times\! 4}$ & $1,\!6,\!11,\!16$ & &$3\!\times\! 9$ & $1,\!11,\!21$\\[0.5mm]
\cline{1-2} \cline{4-5}
$\vertsp{2\!\times\! 2\!\times\! 2\!\times\! 2}$ & $1,\!16$ & &$3\!\times\! 3\!\times\! 3$ & $1,\!14,\!27$\\[0.5mm]
\cline{1-2} \cline{4-5}
$\vertsp{2\!\times\! 9$ & $1,\!11}$ & &$2\!\times\! 14$ & $1,\!16$\\[0.5mm]
\cline{1-2} \cline{4-5}
$\vertsp{3\!\times\! 6}$ & $1,\!8,\!15$ & &$4\!\times\! 7$ & $1,\!9,\!17,\!25$\\[0.5mm]
\cline{1-2} \cline{4-5}
$\vertsp{2\!\times\! 3\!\times\! 3}$ & $1,\!5,\!9,\!11,\!15,\!16$ & &$2\!\times\! 2\!\times\! 7$ & $1,\!9,\!17,\!25$\\[0.5mm]
\end{tabular}
\end{ruledtabular}
\end{table}
%_______________________________________________________________________________
%...............................................................................
% App.H.2 Example: Constructing Maximally Entangled TGX States in $2\times 2\times 3$
\subsection{\label{sec:App.H.2}Example: Constructing Maximally Entangled TGX States in $2\times 2\times 3$}
Here we give a step-by-step example in $2\times 2\times 3$ to show how the 13-step algorithm of \Eq{9} and \App{App.G} works.
%+++++++++++++++++++++++++++++++++++++++++++++++++++++++++++++++++++++++++++++++
\begin{itemize}[leftmargin=*,labelindent=4pt]\setlength\itemsep{0pt}
%~~~~~~~~~~~~~~~~~~~~~~~~~~~~~~~~~~~~~~~~~~~~~~~~~~~~~~~~~~~~~~~~~~~~~~~~~~~~~~~
\item[\textbf{1.}]\hypertarget{Exstep:1}{}Define the system:
%===============================================================================
\begin{equation}%                  Equation H.2
\mathbf{n} \equiv (2,2,3),\;\;N = 3,\;\;n = 12.
\label{eq:H.2}
\end{equation}
%===============================================================================
%~~~~~~~~~~~~~~~~~~~~~~~~~~~~~~~~~~~~~~~~~~~~~~~~~~~~~~~~~~~~~~~~~~~~~~~~~~~~~~~
%~~~~~~~~~~~~~~~~~~~~~~~~~~~~~~~~~~~~~~~~~~~~~~~~~~~~~~~~~~~~~~~~~~~~~~~~~~~~~~~
\item[\textbf{2.}]\hypertarget{Exstep:2}{}Calculate \smash{$\mathbf{L}_*  \equiv \{ L_* \}$}. First, \smash{$n_{\,\overline{{\kern -1.8pt}\max^{~^{~^{~}}}\!\!\!\!\!\!\!\!\!\!}} \equiv \frac{n}{{n_{\max } }}=4$}, so find $1-M(L)$ for \smash{$L \in 2, \ldots n_{\,\overline{{\kern -1.8pt}\max^{~^{~^{~}}}\!\!\!\!\!\!\!\!\!\!}}$}\,:
%===============================================================================
\begin{equation}%                  Equation H.3
\mathop {\min }\limits_{L \in 2, \ldots, 4}\! \left\{\!\! {\begin{array}{*{20}l}
   {1 - M(2)} &\!\! { = {\textstyle{1 \over {12}}}} &\!\! { = 0.08\overline{3}}  \\
   {1 - M(3)} &\!\! { = {\textstyle{2 \over {27}}}} &\!\! { = 0.\overline {074} }  \\
   {1 - M(4)} &\!\! { = {\textstyle{1 \over {48}}}} &\!\! { = 0.0208\overline{3}}  \\
\end{array}}\! \right\}\! \Rightarrow \mathbf{L}_* \! =\! \{ 4\}, 
\label{eq:H.3}
\end{equation}
%===============================================================================
so here there is only one minimizing $L$, so $L_* =4$.
%~~~~~~~~~~~~~~~~~~~~~~~~~~~~~~~~~~~~~~~~~~~~~~~~~~~~~~~~~~~~~~~~~~~~~~~~~~~~~~~
%~~~~~~~~~~~~~~~~~~~~~~~~~~~~~~~~~~~~~~~~~~~~~~~~~~~~~~~~~~~~~~~~~~~~~~~~~~~~~~~
\item[\textbf{3.}]\hypertarget{Exstep:3}{}Calculate the \textit{primary goals vector}: First $m_{\max}=3$, 
%_______________________________________________________________________________
%                                    TABLE 4
\begin{table}[H]
\caption{\label{tab:4}All sets of levels \smash{$\{L_{\text{ME}}\}_{u}$} for maximally entangled TGX states in the nine smallest systems, generated from the 13-step algorithm in \App{App.G}.  For example, the first row is read as \smash{$|\Phi _1 \rangle  = {\textstyle{1 \over {\sqrt 2 }}}(|1\rangle  + |4\rangle )$} and \smash{$|\Phi _2 \rangle  = {\textstyle{1 \over {\sqrt 2 }}}(|2\rangle  + |3\rangle )$}.  Use \Eq{H.1} to convert to the coincidence basis.}
\begin{ruledtabular}
\begin{tabular}{|c|c|c|}
%Column Titles:
$n$ & $n_{1}\!\times\!\cdots\!\times\! n_{N}$& $\{L_{\text{ME}}\}_{u}$\\[0.5mm]
\hline 
%Table Elements:
$\vertsp{4}$ & $2\times 2$ & $\{1,4\}, \{2,3\}\,\,$ \\
\hline
$\vertsp{6}$ & $2\times 3$ & $\begin{array}{*{20}l}
   {\{1,5\}, \{1,6\}, \{2,4\}, \{2,6\}, \{3,4\}, \{3,5\}\,\,}  \\
\end{array}$ \\
\hline
$\begin{array}{*{20}c}
   {\topsp{8}}  \\
   {}  \\
\end{array}$ & $\begin{array}{*{20}c}
   {\topsp{2\times 4}}  \\
   {}  \\
\end{array}$ & $\begin{array}{*{20}l}
   {\{1,6\}, \{1,7\}, \{1,8\}, \{2,5\}, \{2,7\}, \{2,8\},}  \\
   {\{3,5\}, \{3,6\}, \{3,8\}, \{4,5\}, \{4,6\}, \{4,7\} }  \\
\end{array}$ \\
\hline
$\begin{array}{*{20}c}
   {\topsp{8}}  \\
   {}  \\
\end{array}$ & $\begin{array}{*{20}c}
   {\topsp{2\times 2\times 2}}  \\
   {}  \\
\end{array}$ & $\begin{array}{*{20}l}
   {\{1,8\}, \{2,7\}, \{3,6\}, \{4,5\}, \{1,4,6,7\},}  \\
   {\{2,3,5,8\} }  \\
\end{array}$\,\,\,\,\,\, \\
\hline
$\begin{array}{*{20}c}
   {\topsp{9}}  \\
   {}  \\
\end{array}$ & $\begin{array}{*{20}c}
   {\topsp{3\times 3}}  \\
   {}  \\
\end{array}$ & $\begin{array}{*{20}l}
   {\{1,5,9\}, \{1,6,8\}, \{2,4,9\}, \{2,6,7\},}  \\
   {\{3,4,8\}, \{3,5,7\} }  \\
\end{array}${\kern 18pt} \\
\hline
$\!\!\begin{array}{*{20}c}
   {\topsp{10}}  \\
   {}  \\
   {}  \\
   {}  \\
\end{array}$ & $\begin{array}{*{20}c}
   {\topsp{2\times 5}}  \\
   {}  \\
   {}  \\
   {}  \\
\end{array}$ & ${\kern -4pt}\begin{array}{*{20}l}
   {\{1,7\}, \{1,8\}, \{1,9\}, \{1,10\}, \{2,6\}, \{2,8\},}  \\
   {\{2,9\}, \{2,10\}, \{3,6\}, \{3,7\}, \{3,9\}, \{3,10\},}  \\
   {\{4,6\}, \{4,7\}, \{4,8\}, \{4,10\}, \{5,6\}, \{5,7\},}  \\
   {\{5,8\}, \{5,9\}}  \\
\end{array}${\kern -4pt} \\
\hline
$\!\!\begin{array}{*{20}c}
   {\topsp{12}}  \\
   {}  \\
   {}  \\
   {}  \\
   {}  \\
\end{array}$ & $\begin{array}{*{20}c}
   {\topsp{2\times 6}}  \\
   {}  \\
   {}  \\
   {}  \\
   {}  \\
\end{array}$ & ${\kern -4pt}\begin{array}{*{20}l}
   {\{1,8\}, \{1,9\}, \{1,\!10\}, \{1,\!11\}, \{1,\!12\}, \{2,7\},}  \\
   {\{2,9\}, \{2,\!10\}, \{2,\!11\}, \{2,\!12\}, \{3,7\}, \{3,8\},}  \\
   {\{3,\!10\}, \{3,\!11\}, \{3,\!12\}, \{4,7\}, \{4,8\}, \{4,9\},}  \\
   {\{4,\!11\}, \{4,\!12\}, \{5,7\}, \{5,8\}, \{5,9\}, \{5,\!10\},}  \\
   {\{5,\!12\}, \{6,7\}, \{6,8\}, \{6,9\}, \{6,\!10\}, \{6,\!11\}}  \\
\end{array}${\kern -4pt} \\
\hline
$\!\!\begin{array}{*{20}c}
   {\topsp{12}}  \\
   {}  \\
   {}  \\
   {}  \\
   {}  \\
   {}  \\
\end{array}$ & $\begin{array}{*{20}c}
   {\topsp{3\times 4}}  \\
   {}  \\
   {}  \\
   {}  \\
   {}  \\
   {}  \\
\end{array}$ & $\begin{array}{*{20}l}
   {\{1,6,11\}, \{1,6,12\}, \{1,7,10\}, \{1,7,12\},}  \\
   {\{1,8,10\}, \{1,8,11\}, \{2,5,11\}, \{2,5,12\},}  \\
   {\{2,7,9\},  \{2,7,12\}, \{2,8,9\},  \{2,8,11\},}  \\
   {\{3,5,10\}, \{3,5,12\}, \{3,6,9\},  \{3,6,12\},}  \\
   {\{3,8,9\},  \{3,8,10\}, \{4,5,10\}, \{4,5,11\},}  \\
   {\{4,6,9\},  \{4,6,11\}, \{4,7,9\},  \{4,7,10\}}  \\
\end{array}${\kern -4pt} \\
\hline
$\!\!\begin{array}{*{20}c}
   {\topsp{12}}  \\
   {}  \\
   {}  \\
   {}  \\
\end{array}$ & $\begin{array}{*{20}c}
   {\topsp{2\times 2\times 3}}  \\
   {}  \\
   {}  \\
   {}  \\
\end{array}$ & $\begin{array}{*{20}l}
   {\{1,5,8,12\}, \{1,5,9,10\}, \{1,6,8,10\},}  \\
   {\{1,6,9,11\}, \{2,4,7,12\}, \{2,4,9,11\},}  \\
   {\{2,6,7,11\}, \{2,6,9,10\}, \{3,4,7,11\},}  \\
   {\{3,4,8,12\}, \{3,5,7,12\}, \{3,5,8,10\}}  \\
\end{array}${\kern -4pt} \\
\end{tabular}
\end{ruledtabular}
\end{table}
%_______________________________________________________________________________
%                               End of App.H.1
%...............................................................................
since mode $3$ is the largest.  Thus,
%===============================================================================
\begin{equation}%                  Equation H.4
\begin{array}{*{20}l}
   {\mathbf{G}^{(1)} } &\!\! { = \text{floor}({\textstyle{{L_* } \over {n_1 }}})\mathbf{1}_{n_1 }^T  = (\begin{array}{*{20}c}
   2 & 2  \\
\end{array}),}  \\
   {\mathbf{G}^{(2)} } &\!\! { = \text{floor}({\textstyle{{L_* } \over {n_2 }}})\mathbf{1}_{n_2 }^T  = (\begin{array}{*{20}c}
   2 & 2  \\
\end{array}),}  \\
   {\mathbf{G}^{(3)} } &\!\! { = ((1 + \text{floor}({\textstyle{{L_* } \over {n_3 }}}))\mathbf{1}_{S_3 }^T ,\text{floor}({\textstyle{{L_* } \over {n_3 }}})\mathbf{1}_{R_3 }^T )}  \\
   {} &\!\! { = ((2),(\begin{array}{*{20}c}
   1 & 1  \\
\end{array})) = (\begin{array}{*{20}c}
   2 & 1 & 1  \\
\end{array}),}  \\
\end{array}
\label{eq:H.4}
\end{equation}
%===============================================================================
where $R_3  = n_3  - S_3$ and $S_3  = \bmod (L_* ,n_3 )$, so then
%===============================================================================
\begin{equation}%                  Equation H.5
\mathbf{G} \equiv (\begin{array}{*{20}c}
   2 & 2  \\
\end{array}|\begin{array}{*{20}c}
   2 & 2  \\
\end{array}|\begin{array}{*{20}c}
   2 & 1 & 1  \\
\end{array}).
\label{eq:H.5}
\end{equation}
%===============================================================================
%~~~~~~~~~~~~~~~~~~~~~~~~~~~~~~~~~~~~~~~~~~~~~~~~~~~~~~~~~~~~~~~~~~~~~~~~~~~~~~~
%~~~~~~~~~~~~~~~~~~~~~~~~~~~~~~~~~~~~~~~~~~~~~~~~~~~~~~~~~~~~~~~~~~~~~~~~~~~~~~~
\item[\textbf{4.}]\hypertarget{Exstep:4}{}List all arrangements of the goals in the nominally largest subsystem: First, \smash{$\mathbf{c}^{[n_{\max } ]T}  \equiv (1,2,3)$}. Then,
%===============================================================================
\begin{equation}%                  Equation H.6
C^{(m_{\max } )}\!  \equiv \text{nCk}(\mathbf{c}^{[n_{\max } ]T}\!\! ,\bmod (L_* ,n_{\max } ))\! =\!\! \left(\! {\begin{array}{*{20}c}
   1  \\
   2  \\
   3  \\
\end{array}}\! \right)\!\!,
\label{eq:H.6}
\end{equation}
%===============================================================================
so the rows of \smash{$G^{(m_{\max})}$} are, from \Eq{G.5},
%===============================================================================
\begin{equation}%                  Equation H.7
\scalebox{0.90}{$\begin{array}{*{20}l}
   {G_{1,h} ^{(m_{\max } )} } &\!\! {\! =\! \left\{\! {\begin{array}{*{20}l}
   {1 + \text{floor}(L_* /n_{\max } ) = 2;} &\!\! {h \in C_{1, \cdots }^{(m_{\max } )} \! =\! 1}  \\
   {\text{floor}(L_* /n_{\max } ) = 1;} &\!\! {h \notin C_{1, \cdots }^{(m_{\max } )} \! =\! 1}  \\
\end{array}} \right.\!\!,}  \\
   {G_{2,h} ^{(m_{\max } )} } &\!\! {\! =\! \left\{\! {\begin{array}{*{20}l}
   {1 + \text{floor}(L_* /n_{\max } ) = 2;} &\!\! {h \in C_{2, \cdots }^{(m_{\max } )}  \!=\! 2}  \\
   {\text{floor}(L_* /n_{\max } ) = 1;} &\!\! {h \notin C_{2, \cdots }^{(m_{\max } )}  \!=\! 2}  \\
\end{array}} \right.\!\!,}  \\
   {G_{3,h} ^{(m_{\max } )} } &\!\! {\! =\! \left\{\! {\begin{array}{*{20}l}
   {1 + \text{floor}(L_* /n_{\max } ) = 2;} &\!\! {h \in C_{3, \cdots }^{(m_{\max } )}  \!=\! 3}  \\
   {\text{floor}(L_* /n_{\max } ) = 1;} &\!\! {h \notin C_{3, \cdots }^{(m_{\max } )}  \!=\! 3}  \\
\end{array}} \right.\!\!,}  \\
\end{array}$}
\label{eq:H.7}
\end{equation}
%===============================================================================
so the goals matrix in the nominally largest mode is
%===============================================================================
\begin{equation}%                  Equation H.8
G^{(3)}\!  =\! \left(\! {\begin{array}{*{20}c}
   {G_{1,1}^{(m_{\max } )} } &\! {G_{1,2}^{(m_{\max } )} } &\! {G_{1,3}^{(m_{\max } )} }  \\
   {G_{2,1}^{(m_{\max } )} } &\! {G_{2,2}^{(m_{\max } )} } &\! {G_{2,3}^{(m_{\max } )} }  \\
   {G_{3,1}^{(m_{\max } )} } &\! {G_{3,2}^{(m_{\max } )} } &\! {G_{3,3}^{(m_{\max } )} }  \\
\end{array}}\! \right) \!=\! \left(\! {\begin{array}{*{20}c}
   2 & 1 & 1  \\
   1 & 2 & 1  \\
   1 & 1 & 2  \\
\end{array}}\! \right)\!.
\label{eq:H.8}
\end{equation}
%===============================================================================
%~~~~~~~~~~~~~~~~~~~~~~~~~~~~~~~~~~~~~~~~~~~~~~~~~~~~~~~~~~~~~~~~~~~~~~~~~~~~~~~
%~~~~~~~~~~~~~~~~~~~~~~~~~~~~~~~~~~~~~~~~~~~~~~~~~~~~~~~~~~~~~~~~~~~~~~~~~~~~~~~
\item[\textbf{5.}]\hypertarget{Exstep:5}{}Form the total goals matrix:  First, using \Eq{H.4},
%===============================================================================
\begin{equation}%                  Equation H.9
\begin{array}{*{20}l}
   {G^{(1 \ne m_{\max } )} } &\!\! { \equiv\! \mathbf{1}_3  \otimes \mathbf{G}^{(1)} } &\!\! { =\! \left(\! {\begin{array}{*{20}c}
   1  \\
   1  \\
   1  \\
\end{array}}\! \right)\! \otimes (\!\begin{array}{*{20}c}
   2 & 2  \\
\end{array}\!)} &\!\! { =\! \left(\! {\begin{array}{*{20}c}
   2 & 2  \\
   2 & 2  \\
   2 & 2  \\
\end{array}}\! \right)\!\!,}  \\
   {G^{(2 \ne m_{\max } )} } &\!\! { \equiv\! \mathbf{1}_3  \otimes \mathbf{G}^{(2)} } &\!\! { =\! \left(\! {\begin{array}{*{20}c}
   1  \\
   1  \\
   1  \\
\end{array}}\! \right)\! \otimes (\!\begin{array}{*{20}c}
   2 & 2  \\
\end{array}\!)} &\!\! { =\! \left(\! {\begin{array}{*{20}c}
   2 & 2  \\
   2 & 2  \\
   2 & 2  \\
\end{array}}\! \right)\!\!,}  \\
\end{array}
\label{eq:H.9}
\end{equation}
%===============================================================================
so then, the total goals matrix is
%===============================================================================
\begin{equation}%                  Equation H.10
G \equiv (G^{(1)} |G^{(2)} |G^{(3)} ) =\! \left( {\begin{array}{*{20}c}
   2 & 2 &\vline &  2 & 2 &\vline &  2 & 1 & 1  \\
   2 & 2 &\vline &  2 & 2 &\vline &  1 & 2 & 1  \\
   2 & 2 &\vline &  2 & 2 &\vline &  1 & 1 & 2  \\
\end{array}} \right)\!,
\label{eq:H.10}
\end{equation}
%===============================================================================
where again the partitions are merely conceptual.
%~~~~~~~~~~~~~~~~~~~~~~~~~~~~~~~~~~~~~~~~~~~~~~~~~~~~~~~~~~~~~~~~~~~~~~~~~~~~~~~
%~~~~~~~~~~~~~~~~~~~~~~~~~~~~~~~~~~~~~~~~~~~~~~~~~~~~~~~~~~~~~~~~~~~~~~~~~~~~~~~
\item[\textbf{6.}]\hypertarget{Exstep:6}{}Define the occurrence matrix:
%===============================================================================
\begin{equation}%                  Equation H.11
\scalebox{0.9}{$\Omega  \equiv\! \left( {\begin{array}{*{20}c}
   1 & 0 & 1 & 0 & 1 & 0 & 0  \\
   1 & 0 & 1 & 0 & 0 & 1 & 0  \\
   1 & 0 & 1 & 0 & 0 & 0 & 1  \\
   1 & 0 & 0 & 1 & 1 & 0 & 0  \\
   1 & 0 & 0 & 1 & 0 & 1 & 0  \\
   1 & 0 & 0 & 1 & 0 & 0 & 1  \\
   0 & 1 & 1 & 0 & 1 & 0 & 0  \\
   0 & 1 & 1 & 0 & 0 & 1 & 0  \\
   0 & 1 & 1 & 0 & 0 & 0 & 1  \\
   0 & 1 & 0 & 1 & 1 & 0 & 0  \\
   0 & 1 & 0 & 1 & 0 & 1 & 0  \\
   0 & 1 & 0 & 1 & 0 & 0 & 1  \\
\end{array}} \right)\!.$}
\label{eq:H.11}
\end{equation}
%===============================================================================
%~~~~~~~~~~~~~~~~~~~~~~~~~~~~~~~~~~~~~~~~~~~~~~~~~~~~~~~~~~~~~~~~~~~~~~~~~~~~~~~
%~~~~~~~~~~~~~~~~~~~~~~~~~~~~~~~~~~~~~~~~~~~~~~~~~~~~~~~~~~~~~~~~~~~~~~~~~~~~~~~
\item[\textbf{7.}]\hypertarget{Exstep:7}{}Make a ones matrix of the total TGX space:
%===============================================================================
\begin{equation}%                  Equation H.12
\scalebox{0.90}{$W_{\text{TGX}}^{[\mathbf{n}]}  \equiv\! \left( {\begin{array}{*{20}c}
   1 &  \cdot  &  \cdot  &  \cdot  & 1 & 1 &  \cdot  & 1 & 1 & 1 & 1 & 1  \\
    \cdot  & 1 &  \cdot  & 1 &  \cdot  & 1 & 1 &  \cdot  & 1 & 1 & 1 & 1  \\
    \cdot  &  \cdot  & 1 & 1 & 1 &  \cdot  & 1 & 1 &  \cdot  & 1 & 1 & 1  \\
    \cdot  & 1 & 1 & 1 &  \cdot  &  \cdot  & 1 & 1 & 1 &  \cdot  & 1 & 1  \\
   1 &  \cdot  & 1 &  \cdot  & 1 &  \cdot  & 1 & 1 & 1 & 1 &  \cdot  & 1  \\
   1 & 1 &  \cdot  &  \cdot  &  \cdot  & 1 & 1 & 1 & 1 & 1 & 1 &  \cdot   \\
    \cdot  & 1 & 1 & 1 & 1 & 1 & 1 &  \cdot  &  \cdot  &  \cdot  & 1 & 1  \\
   1 &  \cdot  & 1 & 1 & 1 & 1 &  \cdot  & 1 &  \cdot  & 1 &  \cdot  & 1  \\
   1 & 1 &  \cdot  & 1 & 1 & 1 &  \cdot  &  \cdot  & 1 & 1 & 1 &  \cdot   \\
   1 & 1 & 1 &  \cdot  & 1 & 1 &  \cdot  & 1 & 1 & 1 &  \cdot  &  \cdot   \\
   1 & 1 & 1 & 1 &  \cdot  & 1 & 1 &  \cdot  & 1 &  \cdot  & 1 &  \cdot   \\
   1 & 1 & 1 & 1 & 1 &  \cdot  & 1 & 1 &  \cdot  &  \cdot  &  \cdot  & 1  \\
\end{array}} \right)\!.$}
\label{eq:H.12}
\end{equation}
%===============================================================================
%~~~~~~~~~~~~~~~~~~~~~~~~~~~~~~~~~~~~~~~~~~~~~~~~~~~~~~~~~~~~~~~~~~~~~~~~~~~~~~~
%~~~~~~~~~~~~~~~~~~~~~~~~~~~~~~~~~~~~~~~~~~~~~~~~~~~~~~~~~~~~~~~~~~~~~~~~~~~~~~~
\item[\textbf{8.}]\hypertarget{Exstep:8}{}Pick the starting level:
%===============================================================================
\begin{equation}%                  Equation H.13
S_L =1.
\label{eq:H.13}
\end{equation}
%===============================================================================
%~~~~~~~~~~~~~~~~~~~~~~~~~~~~~~~~~~~~~~~~~~~~~~~~~~~~~~~~~~~~~~~~~~~~~~~~~~~~~~~
%~~~~~~~~~~~~~~~~~~~~~~~~~~~~~~~~~~~~~~~~~~~~~~~~~~~~~~~~~~~~~~~~~~~~~~~~~~~~~~~
\item[\textbf{9.}]\hypertarget{Exstep:9}{}Find the initial set of nonzero TGX levels compatible with $S_L$:  First, from \Eq{G.13},
%===============================================================================
\begin{equation}%                  Equation H.14
\begin{array}{*{20}l}
   {R_{\text{cand}}  = } &\!\! {\{ k\} |_{k = S_L  + 1}^n } &\!\! { = \{ 5,6,8,9,10,11,12\} ;}  \\
   {} &\!\! {\text{s.t.}\;\;(W_{\text{TGX}}^{[\mathbf{n}]} )_{k,S_L }  = 1} &\!\! {}  \\
   {C_{\text{cand}}  = } &\!\! {\{ k\} |_{k = 1}^{S_L  - 1} } &\!\! { = \{ \} ;}  \\
   {} &\!\! {\text{s.t.}\;\;(W_{\text{TGX}}^{[\mathbf{n}]} )_{S_L ,k}  = 1,} &\!\! {}  \\
\end{array}
\label{eq:H.14}
\end{equation}
%===============================================================================
so then concatenating these gives the initial set,
%===============================================================================
\begin{equation}%                  Equation H.15
L_{\text{cand}} '' \equiv \{ R_{\text{cand}} ,C_{\text{cand}} \}  = \{ 5,6,8,9,10,11,12\}.
\label{eq:H.15}
\end{equation}
%===============================================================================
%~~~~~~~~~~~~~~~~~~~~~~~~~~~~~~~~~~~~~~~~~~~~~~~~~~~~~~~~~~~~~~~~~~~~~~~~~~~~~~~
%~~~~~~~~~~~~~~~~~~~~~~~~~~~~~~~~~~~~~~~~~~~~~~~~~~~~~~~~~~~~~~~~~~~~~~~~~~~~~~~
\item[\textbf{10.}]\hypertarget{Exstep:10}{}Build the candidate compatibility sets for $S_L$: First,
%===============================================================================
\begin{equation}%                  Equation H.16
\text{nCk}(L_{\text{cand}} '',L_*  - 1) = \text{nCk}((5,6,8,9,10,11,12),3),
\label{eq:H.16}
\end{equation}
%===============================================================================
which is a $35\times 3$ matrix whose rows are unique sets of $L_{\text{cand}}''$ taken in groups of $L_* -1=3$. Then prepend a column of $S_L$ to the left to get the $35\times 4$ matrix,
%===============================================================================
\begin{equation}%                  Equation H.17
L_{\text{cand}} ' \equiv \{ S_L \mathbf{1}_{\binom{\dim (L_{\text{cand}} '')}{L_* -1}} ,\text{nCk}(L_{\text{cand}} '',L_*  - 1)\},
\label{eq:H.17}
\end{equation}
%===============================================================================
which has rows such as $(1,5,6,8)$ and $(1,5,6,9)$, etc.
%~~~~~~~~~~~~~~~~~~~~~~~~~~~~~~~~~~~~~~~~~~~~~~~~~~~~~~~~~~~~~~~~~~~~~~~~~~~~~~~
%~~~~~~~~~~~~~~~~~~~~~~~~~~~~~~~~~~~~~~~~~~~~~~~~~~~~~~~~~~~~~~~~~~~~~~~~~~~~~~~
\item[\textbf{11.}]\hypertarget{Exstep:11}{}Find only the rows of \smash{$L_{\text{cand}} '$} for which all the levels are mutually pure-TGX-state compatible by using the test in \Eq{G.18}. For example, the first row that passes the test is row $7$, which is \smash{$(L_{\text{cand}} ')_{7, \cdots }  = (1,5,8,10)$}, because (abbreviating with \smash{$L_{\text{c}}'\equiv L_{\text{cand}} '$}),
%===============================================================================
\begin{equation}%                  Equation H.18
\scalebox{0.95}{$\begin{array}{*{20}l}
   {\sum\limits_{k = 1}^{L_* -1}\sum\limits_{l = k + 1}^{L_* }\!\! {(W_{\text{TGX}}^{[\mathbf{n}]})_{(L_{\text{cand}} ')_{7,l} ,(L_{\text{cand}} ')_{7,k} } } }  \\
   {\;\;\;\; = \!\left(\! \begin{array}{l}
 (W_{\text{TGX}}^{[\mathbf{n}]})_{(L_{\text{c}} ')_{7,2} ,(L_{\text{c}} ')_{7,1} }  + (W_{\text{TGX}}^{[\mathbf{n}]})_{(L_{\text{c}} ')_{7,3} ,(L_{\text{c}} ')_{7,1} }  \\ 
  + (W_{\text{TGX}}^{[\mathbf{n}]})_{(L_{\text{c}} ')_{7,4} ,(L_{\text{c}} ')_{7,1} }  + (W_{\text{TGX}}^{[\mathbf{n}]})_{(L_{\text{c}} ')_{7,3} ,(L_{\text{c}} ')_{7,2} }  \\ 
  + (W_{\text{TGX}}^{[\mathbf{n}]})_{(L_{\text{c}} ')_{7,4} ,(L_{\text{c}} ')_{7,2} }  + (W_{\text{TGX}}^{[\mathbf{n}]})_{(L_{\text{c}} ')_{7,4} ,(L_{\text{c}} ')_{7,3} }  \\
 \end{array}\! \right)}  \\
   {\;\;\;\; =\! \left(\! \begin{array}{l}
 (W_{\text{TGX}}^{[\mathbf{n}]})_{5,1}  + (W_{\text{TGX}}^{[\mathbf{n}]})_{8,1}  \\ 
  + (W_{\text{TGX}}^{[\mathbf{n}]})_{10,1}  + (W_{\text{TGX}}^{[\mathbf{n}]})_{8,5}  \\ 
  + (W_{\text{TGX}}^{[\mathbf{n}]})_{10,5}  + (W_{\text{TGX}}^{[\mathbf{n}]})_{10,8}  \\ 
 \end{array}\! \right) = 6 = {\textstyle{{L_* ^2  - L_* } \over 2}}.}  \\
\end{array}$}
\label{eq:H.18}
\end{equation}
%===============================================================================
The collection of candidate sets that pass this test is
%===============================================================================
\begin{equation}%                  Equation H.19
L_{\text{cand}}  =\! \left( {\begin{array}{*{20}c}
   1 & 5 & 8 & {10}  \\
   1 & 5 & 8 & {12}  \\
   1 & 5 & 9 & {10}  \\
   1 & 6 & 8 & {10}  \\
   1 & 6 & 9 & {10}  \\
   1 & 6 & 9 & {11}  \\
\end{array}} \right)\!,
\label{eq:H.19}
\end{equation}
%===============================================================================
the rows of which represent states that can both support a pure parent TGX state as well as the number of levels $L_* =4$ that enable the average unitized reduction purity to be minimized.
%~~~~~~~~~~~~~~~~~~~~~~~~~~~~~~~~~~~~~~~~~~~~~~~~~~~~~~~~~~~~~~~~~~~~~~~~~~~~~~~
%~~~~~~~~~~~~~~~~~~~~~~~~~~~~~~~~~~~~~~~~~~~~~~~~~~~~~~~~~~~~~~~~~~~~~~~~~~~~~~~
\item[\textbf{12.}]\hypertarget{Exstep:12}{}For each candidate pure-TGX compatibility set for $S_L$ (the rows of \smash{$L_{\text{cand}}$}),  test whether the sum of the goals vectors of every level in each set exactly equals any of the total goals vectors: For example, the first working set in \smash{$L_{\text{cand}}$} is row $2$, since that is a row for which the sum of rows of \Eq{H.11} picked by row $2$ of \smash{$L_{\text{cand}}$} is
%===============================================================================
\begin{equation}%                  Equation H.20
\begin{array}{*{20}l}
   {\sum\limits_{c = 1}^{L_* } {\Omega _{(L_{\text{cand}} )_{2,c} , \cdots } } } &\!\! { = \Omega _{(L_{\text{cand}} )_{2,1} , \cdots }  + \Omega _{(L_{\text{cand}} )_{2,2} , \cdots } }  \\
   {} &\!\! {{\kern 8pt} + \Omega _{(L_{\text{cand}} )_{2,3} , \cdots }  + \Omega _{(L_{\text{cand}} )_{2,4} , \cdots } }  \\
   {} &\!\! { = \Omega _{1, \cdots }  + \Omega _{5, \cdots }  + \Omega _{8, \cdots }  + \Omega _{12, \cdots } }  \\
   {} &\!\! { =\! \left(\! \begin{array}{r}
 (\begin{array}{*{20}c}
   1 & 0 & 1 & 0 & 1 & 0 & 0  \\
\end{array}) \\ 
  + (\begin{array}{*{20}c}
   1 & 0 & 0 & 1 & 0 & 1 & 0  \\
\end{array}) \\ 
  + (\begin{array}{*{20}c}
   0 & 1 & 1 & 0 & 0 & 1 & 0  \\
\end{array}) \\ 
  + (\begin{array}{*{20}c}
   0 & 1 & 0 & 1 & 0 & 0 & 1  \\
\end{array}) \\ 
 \end{array}\! \right)}  \\
   {} &\!\! { ={\kern 16.3pt} (\begin{array}{*{20}c}
   2 & 2 & 2 & 2 & 1 & 2 & 1  \\
\end{array}),}  \\
\end{array}
\label{eq:H.20}
\end{equation}
%===============================================================================
which exactly matches row $2$ of the total goals matrix in \Eq{H.10}, so row $2$ of \Eq{H.19} is a verified compatibility set for maximally entangled TGX states.  The full set of verified compatibility sets for this example is
%===============================================================================
\begin{equation}%                  Equation H.21
L_{\text{ME}}  =\! \left( {\begin{array}{*{20}c}
   1 & 5 & 8 & {12}  \\
   1 & 5 & 9 & {10}  \\
   1 & 6 & 8 & {10}  \\
   1 & 6 & 9 & {11}  \\
\end{array}} \right)\!.
\label{eq:H.21}
\end{equation}
%===============================================================================
%~~~~~~~~~~~~~~~~~~~~~~~~~~~~~~~~~~~~~~~~~~~~~~~~~~~~~~~~~~~~~~~~~~~~~~~~~~~~~~~
%~~~~~~~~~~~~~~~~~~~~~~~~~~~~~~~~~~~~~~~~~~~~~~~~~~~~~~~~~~~~~~~~~~~~~~~~~~~~~~~
\item[\textbf{13.}]\hypertarget{Exstep:13}{}The list of verified compatibility sets $L_{\text{ME}}$ for $S_L$ is the set of all possible combinations of $L_*$ levels for which balanced superposition yields maximally entangled states: Thus, \Eq{H.21} represents the set,
%===============================================================================
\begin{equation}%                  Equation H.22
\begin{array}{*{20}l}
   {|\Phi _1 \rangle } &\!\! { = {\textstyle{1 \over {\sqrt 4 }}}(|1\rangle  + |5\rangle  + |8\rangle  + |12\rangle )}  \\
   {|\Phi _2 \rangle } &\!\! { = {\textstyle{1 \over {\sqrt 4 }}}(|1\rangle  + |5\rangle  + |9\rangle  + |10\rangle )}  \\
   {|\Phi _3 \rangle } &\!\! { = {\textstyle{1 \over {\sqrt 4 }}}(|1\rangle  + |6\rangle  + |8\rangle  + |10\rangle )}  \\
   {|\Phi _4 \rangle } &\!\! { = {\textstyle{1 \over {\sqrt 4 }}}(|1\rangle  + |6\rangle  + |9\rangle  + |11\rangle ),}  \\
\end{array}
\label{eq:H.22}
\end{equation}
%===============================================================================
where since these are TGX states, we can insert any relative unit-magnitude phase factors without affecting the maximal entanglement.  Each state in \Eq{H.22} has ent $\Upsilon=1$, and their reduction purities are as minimal as it is possible for them to be given a pure parent state in this system.  Use of \Eq{H.1} reveals that
%===============================================================================
\begin{equation}%                  Equation H.23
\begin{array}{*{20}l}
   {|\Phi _1 \rangle } &\!\! { = {\textstyle{1 \over {\sqrt 4 }}}(|1,1,1\rangle  + |1,2,2\rangle  + |2,1,2\rangle  + |2,2,3\rangle )}  \\
   {|\Phi _2 \rangle } &\!\! { = {\textstyle{1 \over {\sqrt 4 }}}(|1,1,1\rangle  + |1,2,2\rangle  + |2,1,3\rangle  + |2,2,1\rangle )}  \\
   {|\Phi _3 \rangle } &\!\! { = {\textstyle{1 \over {\sqrt 4 }}}(|1,1,1\rangle  + |1,2,3\rangle  + |2,1,2\rangle  + |2,2,1\rangle )}  \\
   {|\Phi _4 \rangle } &\!\! { = {\textstyle{1 \over {\sqrt 4 }}}(|1,1,1\rangle  + |1,2,3\rangle  + |2,1,3\rangle  + |2,2,2\rangle ).}  \\
\end{array}
\label{eq:H.23}
\end{equation}
%===============================================================================
This is not an exhaustive list; other starting levels yield other states.  \Table{4} reveals that there are exactly $12$ unique combinations of $4$ levels that support maximal entanglement for this system, and that \textit{is} an exhaustive list, since it was created doing this example for each of the $12$ different starting levels.
%~~~~~~~~~~~~~~~~~~~~~~~~~~~~~~~~~~~~~~~~~~~~~~~~~~~~~~~~~~~~~~~~~~~~~~~~~~~~~~~
\end{itemize}
%+++++++++++++++++++++++++++++++++++++++++++++++++++++++++++++++++++++++++++++++
%                               End of App.H.2
%...............................................................................
%                                End of App.H
%-------------------------------------------------------------------------------
%-------------------------------------------------------------------------------
%                App.I. Schmidt Decomposition and Reversal
\section{\label{sec:App.I}Schmidt Decomposition and Reversal}
%_______________________________________________________________________________
\begin{figure}[H]%Not a figure; Puts hypertarget at top of column to fix problem
\centering
\vspace{-12pt}
\setlength{\unitlength}{0.01\linewidth}
\begin{picture}(100,0)
\put(1,25){\hypertarget{Sec:App.I}{}}
\end{picture}
\end{figure}
\vspace{-39pt}
%_______________________________________________________________________________
For any bipartite $n_1 \times n_2$ pure state,
%===============================================================================
\begin{equation}%                  Equation I.1
\begin{array}{*{20}l}
   {|\psi \rangle} &\!\! {= \sum\limits_{j,k = 1,1}^{n_1 ,n_2 } {A_{j,k} |j^{(1)} \rangle  \otimes |k^{(2)} \rangle },}  \\
\end{array}
\label{eq:I.1}
\end{equation}
%===============================================================================
where \smash{$\{ |j^{(1)} \rangle \}$} and \smash{$\{ |k^{(2)} \rangle \}$} are complete basis sets for subsystems 1 and 2 (not necessarily Fock states, and we use the convention of starting labels on 1), with complex state coefficients $A_{j,k}$, the Schmidt decomposition allows us to rewrite the state in ``Schmidt-diagonal form'' as
%===============================================================================
\begin{equation}%                  Equation I.2
\begin{array}{*{20}l}
   {|\psi \rangle} &\!\! {= \sum\limits_{l = 1}^{n_S } {\lambda _l |u_l ^{(1)} \rangle  \otimes |v_l ^{(2)} \rangle },}  \\
\end{array}
\label{eq:I.2}
\end{equation}
%===============================================================================
where $n_S  \equiv \min \{ n_1 ,n_2 \}$, and $\lambda_l$ are the nonzero singular values of the matrix $A$ formed by \textit{deliberately misinterpreting} the indices of $A_{j,k}$ as indices of matrix elements, so that the singular value decomposition (SVD) is $A = U\Sigma V^{\dag}$, where $U$ is an $n_1$-level unitary, $V$ is an $n_2$-level unitary, and $\Sigma$ is the diagonal matrix of singular values $\lambda_l$ with $n_1$ rows and $n_2$ columns such that the singular values obey $\lambda _1  \ge  \cdots  \ge \lambda _r  \ge 0$, which are the \textit{Schmidt numbers}.  The basis states are then given by \smash{$|u_l ^{(1)} \rangle  \equiv \sum\nolimits_{j = 1}^{n_1 } {U_{j,l} |j^{(1)} \rangle }$} and \smash{$|v_l ^{(2)} \rangle  \equiv \sum\nolimits_{k = 1}^{n_2 } {V^* _{k,l} |k^{(2)} \rangle }$}.

The Schmidt numbers tell us that
%===============================================================================
\begin{equation}%                  Equation I.3
\begin{array}{*{20}l}
   {|\psi \rangle  \in \mathbb{S}} & {\text{iff}} & {\lambda _1  = 1\;\;\text{and}\;\;\lambda _{l \ge 2}  = 0}  \\
   {|\psi \rangle  \in \mathbb{E}} & {\text{iff}} & {\text{more than one}\;\;\lambda _l  \ne 0}  \\
   {|\psi \rangle  \in \mathbb{E}_{\max } } & {\text{iff}} & {\lambda _l  = {\textstyle{1 \over {\sqrt {n_S } }}}\;\;\forall l \in 1, \ldots ,n_S },  \\
\end{array}
\label{eq:I.3}
\end{equation}
%===============================================================================
where $\mathbb{S}$ is the set of separable states, $\mathbb{E}$ is the set of entangled states, and $\mathbb{E}_{\max }$ is the set of maximally entangled states (all of which include pure states only here).  Since entanglement is basis-dependent, this refers to entanglement in the basis of \Eq{I.1}, but also holds in \Eq{I.2}.

\textit{Reverse-Schmidt Decomposition} is simply the process of choosing a set of nonzero (positive) singular values according to the rules in \Eq{I.3} to form $\Sigma$ and picking unitaries $U$ and $V$ to then calculate matrix elements $A_{j,k}  = (U\Sigma V^{\dag}  )_{j,k} $ which we then \textit{deliberately misinterpret} as the state coefficients of the expansion in \Eq{I.1}.  Thus, we can parameterize bipartite pure states by their entanglement in terms of Schmidt numbers.

For example, maximally entangled $2\times 2$ states can be made by defining
%===============================================================================
\begin{equation}%                  Equation I.4
U \equiv \!\left(\! {\begin{array}{*{20}r}
   {a_1 } & {b_1 }  \\
   { - b_1 ^* } & {a_1 ^* }  \\
\end{array}}\! \right)\!,\;\Sigma  \equiv\! \left(\! {\begin{array}{*{20}c}
   {{\textstyle{1 \over {\sqrt 2 }}}} & 0  \\
   0 & {{\textstyle{1 \over {\sqrt 2 }}}}  \\
\end{array}}\! \right)\!,\;V^{\dag}   \equiv\! \left(\! {\begin{array}{*{20}r}
   {a_2 } & {b_2 }  \\
   { - b_2 ^* } & {a_2 ^* }  \\
\end{array}}\! \right)\!,
\label{eq:I.4}
\end{equation}
%===============================================================================
where $|a_1 |^2  + |b_1 |^2  = 1$ and $|a_2 |^2  + |b_2 |^2  = 1$. So then computing $A_{j,k}  = (U\Sigma V^{\dag}  )_{j,k} $ and using \Eq{I.1} yields
%===============================================================================
\begin{equation}%                  Equation I.5
\scalebox{0.95}{$|\psi \rangle  = {\textstyle{1 \over {\sqrt 2 }}}\!\left(\! {\begin{array}{*{20}r}
   {a_1 a_2  - b_1 b_2 ^* }  \\
   {a_1 b_2  + b_1 a_2 ^* }  \\
   { - b_1 ^* a_2  - a_1 ^* b_2 ^* }  \\
   { - b_1 ^* b_2  + a_1 ^* a_2 ^* }  \\
\end{array}} \right)\!,$}
\label{eq:I.5}
\end{equation}
%===============================================================================
which is maximally entangled for all parameter values.

To see how the proposed multipartite reverse-Schmidt decomposition from \Eq{11} and \Eq{12} compares to the above result, pick the core maximally entangled TGX state as \smash{$|\Phi _{\text{TGX}} \rangle  \equiv {\textstyle{1 \over {\sqrt 2 }}}(|1,1\rangle  + |2,2\rangle )$}, and let\\
\vspace{-15pt}
%===============================================================================
\begin{equation}%                  Equation I.6
\scalebox{0.95}{$U^{(1)}\!  \equiv \!\left(\! {\begin{array}{*{20}c}
   a & b  \\
   { - b^* } & {a^* }  \\
\end{array}}\! \right),U^{(2)}\!  \equiv\! \left(\! {\begin{array}{*{20}c}
   c & d  \\
   { - d^* } & {c^* }  \\
\end{array}}\! \right)\!,D \equiv\! \text{diag}\!\left\{\!\! {\begin{array}{*{20}c}
   {e^{i\eta _1 } }  \\
   {e^{i\eta _2 } }  \\
   {e^{i\eta _3 } }  \\
   {e^{i\eta _4 } }  \\
\end{array}}\!\! \right\}\!,$}
\label{eq:I.6}
\end{equation}
%===============================================================================
where $|a|^2  + |b|^2  = 1$ and $|c|^2  + |d|^2  = 1$.  Then, setting $a = c_1$, $b = d_1$, $c = c_2$, $d =  - d_2 ^*$ and $\eta\equiv\eta_4 -\eta_1$, we find that $|\Phi _G \rangle  \equiv (U^{(1)}  \otimes U^{(2)} )D|\Phi _{\text{TGX}} \rangle$ gives\\
\vspace{-5pt}
%===============================================================================
\begin{equation}%                  Equation I.7
\scalebox{0.95}{$|\Phi _G \rangle  = {\textstyle{e^{i\eta_{1} } \over {\sqrt 2 }}}\!\left(\! {\begin{array}{*{20}r}
   {c_1 c_2  - d_1 d_2 ^* e^{i\eta } }  \\
   {c_1 d_2  + d_1 c_2 ^* e^{i\eta } }  \\
   { - d_1 ^* c_2  - c_1 ^* d_2 ^* e^{i\eta } }  \\
   { - d_1 ^* d_2  + c_1 ^* c_2 ^* e^{i\eta } }  \\
\end{array}} \right)\!,$}
\label{eq:I.7}
\end{equation}
%===============================================================================
which has the same form as \Eq{I.5} except for the irrelevant global phase factor $e^{i\eta_{1}}$, and $\eta$.

So have we discovered something more general than the Schmidt decomposition because of $\eta$?  To answer this, note that we can achieve \Eq{I.7} from the usual reverse-Schmidt decomposition if we modify it as
%===============================================================================
\begin{equation}%                  Equation I.8
\scalebox{0.95}{$A'\! \equiv \! U'\Sigma' V^{\prime\dag} \! \equiv \!\left(\! {\begin{array}{*{20}r}
   {c_1 } & {d_1 }  \\
   { - d_1 ^* } & {c_1 ^* }  \\
\end{array}}\! \right)\!\!\left(\!\! {\begin{array}{*{20}c}
   {{\textstyle{1 \over {\sqrt 2 }}}e^{i\eta_1 }} &\!\!\! 0  \\
   0 &\!\!\! {{\textstyle{1 \over {\sqrt 2 }}}e^{i\eta_4 } }  \\
\end{array}}\!\! \right)\!\!\left(\! {\begin{array}{*{20}r}
   {c_2 } & {d_2 }  \\
   { - d_2 ^* } & {c_2 ^* }  \\
\end{array}}\! \right)\!.$}
\label{eq:I.8}
\end{equation}
%===============================================================================
for a state given by \smash{$|\Phi_{G}\rangle  = \sum\nolimits_{j = 1}^{n_1}\!\sum\nolimits_{k = 1}^{n_2 }\! {A'_{j,k} |j^{(1)} \rangle  \otimes |k^{(2)} \rangle }$}, where{\kern -1.2pt} $|c_1 |^2 \!{\kern -0.5pt} +\! |d_1 |^2 \!{\kern -0.5pt} =\! 1${\kern -1.5pt} and{\kern -1.5pt} $|c_2 |^2 \!{\kern -0.5pt} +\! |d_2 |^2 \! =\! 1$.  {\kern 1.0pt}Then, since singular values are always real and positive, we must factor $e^{i\eta_1}$ and $e^{i\eta_4}$ out of \smash{$\Sigma'$}.  Therefore, let
%===============================================================================
\begin{equation}%                  Equation I.9
A' \equiv U'\Sigma 'V^{\prime\dag}   = e^{i\eta _1 } \widetilde{U}\Sigma \widetilde{V}^{\dag} , 
\label{eq:I.9}
\end{equation}
%===============================================================================
with new unitaries (keeping it as general as possible),
%===============================================================================
\begin{equation}%                  Equation I.10
\widetilde{U} \equiv U'\widetilde{D}^x ,\;\;\widetilde{V}^{\dag} \equiv \widetilde{D}^{1 - x} V^{\prime\dag} ,\;\;\widetilde{D}  \equiv \text{diag}\{ 1,e^{i\eta } \}, 
\label{eq:I.10}
\end{equation}
%===============================================================================
for any real $x$.  Then, using parameterization $(c_1 ,d_1 ) \equiv (c_\theta  e^{i\alpha } ,s_\theta  e^{i\beta } )$ and $(c_2 ,d_2 ) \equiv (c_\varepsilon  e^{i\gamma } ,s_\varepsilon  e^{i\delta } )$, we get
%===============================================================================
\begin{equation}%                  Equation I.11
\begin{array}{*{20}l}
   {\widetilde{U}} &\!\! { = e^{i{\textstyle{{x\eta } \over 2}}}\! \left(\!\! {\begin{array}{*{20}r}
   {c_\theta  e^{i(\alpha  - {\textstyle{{x\eta } \over 2}})} } & {s_\theta  e^{i(\beta  + {\textstyle{{x\eta } \over 2}})} }  \\
   { - s_\theta  e^{ - i(\beta  + {\textstyle{{x\eta } \over 2}})} } & {c_\theta  e^{ - i(\alpha  - {\textstyle{{x\eta } \over 2}})} }  \\
\end{array}}\!\! \right)}  \\
   {\widetilde{V}^{\dag}  } &\!\! { = e^{i{\textstyle{{(1 - x)\eta } \over 2}}} \!\left(\!\! {\begin{array}{*{20}r}
   {c_\varepsilon  e^{i(\gamma  - {\textstyle{{(1 - x)\eta } \over 2}})} } & {s_\varepsilon  e^{i(\delta  - {\textstyle{{(1 - x)\eta } \over 2}})} }  \\
   { - s_\varepsilon  e^{ - i(\delta  - {\textstyle{{(1 - x)\eta } \over 2}})} } & {c_\varepsilon  e^{ - i(\gamma  - {\textstyle{{(1 - x)\eta } \over 2}})} }  \\
\end{array}}\!\! \right)\!,}  \\
\end{array}
\label{eq:I.11}
\end{equation}
%===============================================================================
where $c_{\theta}\equiv\cos(\theta)$, $s_{\theta}\equiv\sin(\theta)$, and $\theta,\varepsilon,\alpha,\beta,\gamma,\delta\in\Re$.  Then implicitly define \textit{new} phases as
%===============================================================================
\begin{equation}%                  Equation I.12
\begin{array}{*{20}l}
   \alpha  &\!\! { \equiv {\textstyle{{x\eta } \over 2}} + \alpha '} &\;\; \gamma  &\!\! { \equiv {\textstyle{{(1 - x)\eta } \over 2}} + \gamma '}  \\
   \beta  &\!\! { \equiv  - {\textstyle{{x\eta } \over 2}} + \beta '} &\;\; \delta  &\!\! { \equiv {\textstyle{{(1 - x)\eta } \over 2}} + \delta ',}  \\
\end{array}
\label{eq:I.12}
\end{equation}
%===============================================================================
where $\{\alpha ',\beta',\gamma',\delta'\}$ can be \textit{any} real numbers, which lets \Eq{I.11} simplify and be reparameterized as
%===============================================================================
\begin{equation}%                  Equation I.13
\begin{array}{*{20}l}
   {\widetilde{U}} &\!\! { = e^{i{\textstyle{{x\eta } \over 2}}}\! \left(\!\! {\begin{array}{*{20}r}
   {c_\theta  e^{i\alpha'} } &\!\! {s_\theta  e^{i\beta'} }  \\
   { - s_\theta  e^{ - i\beta'} } &\!\! {c_\theta  e^{ - i\alpha'} }  \\
\end{array}}\!\! \right)} &\!\! { = e^{i{\textstyle{{x\eta } \over 2}}}\! \left(\!\! {\begin{array}{*{20}r}
   {a_1 } &\!\! {b_1}  \\
   { - b_1^* } &\!\! a_1^*  \\
\end{array}}\!\! \right)} \\
   {\widetilde{V}^{\dag}  } &\!\! { = e^{i{\textstyle{{(1 - x)\eta } \over 2}}} \!\left(\!\! {\begin{array}{*{20}r}
   {c_\varepsilon  e^{i\gamma'} } &\!\! {s_\varepsilon  e^{i\delta'} }  \\
   { - s_\varepsilon  e^{ - i\delta'} } &\!\! {c_\varepsilon  e^{ - i\gamma'} }  \\
\end{array}}\!\! \right)\!} &\!\! { = e^{i{\textstyle{{(1 - x)\eta } \over 2}}} \!\left(\!\! {\begin{array}{*{20}r}
   {a_2 } &\!\! {b_2}  \\
   { - b_2^* } &\!\! a_2^*  \\
\end{array}}\!\! \right)\!,} \\
\end{array}
\label{eq:I.13}
\end{equation}
%===============================================================================
with \smash{$(a_1 ,b_1 ) \!\equiv\! (c_\theta  e^{i\alpha' }\!\! ,s_\theta  e^{i\beta' } )$} and \smash{$(a_2 ,b_2 ) \!\equiv\! (c_\varepsilon  e^{i\gamma' }\!\! ,s_\varepsilon  e^{i\delta' } )$}.  Then, putting \Eq{I.13} into \Eq{I.9} gives
%===============================================================================
\begin{equation}%                  Equation I.14
\scalebox{0.95}{$A' = e^{i\eta _1 } e^{i{\textstyle{\eta  \over 2}}} \left(\! {\begin{array}{*{20}r}
   {a_1 } & {b_1 }  \\
   { - b_1 ^* } & {a_1 ^* }  \\
\end{array}}\! \right)\!\!\left(\! {\begin{array}{*{20}c}
   {{\textstyle{1 \over {\sqrt 2 }}}} & 0  \\
   0 & {{\textstyle{1 \over {\sqrt 2 }}}}  \\
\end{array}}\! \right)\!\!\left(\! {\begin{array}{*{20}r}
   {a_2 } & {b_2 }  \\
   { - b_2 ^* } & {a_2 ^* }  \\
\end{array}}\! \right)\!,$}
\label{eq:I.14}
\end{equation}
%===============================================================================
resulting in the Schmidt-form state,\vspace{-6pt}
%===============================================================================
\begin{equation}%                  Equation I.15
\scalebox{0.95}{$|\Phi_G \rangle  = e^{i(\eta _1  + {\textstyle{\eta  \over 2}})}{\textstyle{1 \over {\sqrt 2 }}}\!\left(\! {\begin{array}{*{20}r}
   {a_1 a_2  - b_1 b_2 ^* }  \\
   {a_1 b_2  + b_1 a_2 ^* }  \\
   { - b_1 ^* a_2  - a_1 ^* b_2 ^* }  \\
   { - b_1 ^* b_2  + a_1 ^* a_2 ^* }  \\
\end{array}} \right)\!,$}
\label{eq:I.15}
\end{equation}
%===============================================================================
where we can discard the irrelevant global phase factor, thus showing that \Eq{I.15} has the same form as \Eq{I.5}.  Since the primed phases were defined in \Eq{I.12} as being free, the \textit{only} part of $|\Phi_{G} \rangle$ that depends on the $\{\eta_k\}$ is the global phase; all other parameters are independent of $\{\eta_k\}$.

An interesting quirk of the proposed multipartite Schmidt method of \Eq{11} is that in some cases, as seen in \Table{2}, different numbers of levels $L_*$ can give rise to maximal entanglement in the same system. While larger-$L_*$ $|\Phi_{\text{TGX}}\rangle$ may seem like they could involve more degrees of freedom (DOF), the unitary equivalence of all pure states means that the versions of \Eq{11} using the \textit{smallest}-$L_*$ $|\Phi_{\text{TGX}}\rangle$ capture \textit{all} the DOF necessary to reach all maximally entangled states (for the most general $U_{\text{EPU}}$, whether it is our hypothesized form in \Eq{12} or something more general).
%                                End of App.I
%-------------------------------------------------------------------------------
%-------------------------------------------------------------------------------
%            App.J. Decomposition Freedom of {\normalsize $\rho$}
\section{\label{sec:App.J}Decomposition Freedom of {\normalsize $\rho$}}
%_______________________________________________________________________________
\begin{figure}[H]%Not a figure; Puts hypertarget at top of column to fix problem
\centering
\vspace{-12pt}
\setlength{\unitlength}{0.01\linewidth}
\begin{picture}(100,0)
\put(1,25){\hypertarget{Sec:App.J}{}}
\end{picture}
\end{figure}
\vspace{-39pt}
%_______________________________________________________________________________
From \cite{Woot}, any density matrix $\rho$ can be expanded as
%===============================================================================
\begin{equation}%                  Equation J.1
\begin{array}{*{20}l}
   {\rho} &\!\! {= \sum_{j = 1}^D {|\widetilde{w}_j \rangle \langle \widetilde{w}_j |},}  \\
\end{array}
\label{eq:J.1}
\end{equation}
%===============================================================================
where $D \in [R, \ldots ,\infty )$, $R = \text{rank}(\rho )$, and \smash{$|\widetilde{w}_j \rangle$} are ``subnormalized'' decomposition states,
%===============================================================================
\begin{equation}%                  Equation J.2
\begin{array}{*{20}l}
   {|\widetilde{w}_j \rangle} &\!\! {\equiv \sum_{k = 1}^R {U_{j,k} \sqrt {\lambda _k } |e_k \rangle } ,}  \\
\end{array}
\label{eq:J.2}
\end{equation}
%===============================================================================
where $\lambda_k$ are eigenvalues of $\rho$ with eigenstates $|e_k \rangle$, and $U$ is \textit{any} $D$-level unitary matrix.

To adapt \Eq{J.1} for convex-roof extension, rewrite it as
%===============================================================================
\begin{equation}%                  Equation J.3
\begin{array}{*{20}l}
   {\rho} &\!\! {= \sum_{j = 1}^D {p_j |w_j \rangle \langle w_j |}\equiv \sum_{j = 1}^D {p_j \rho_{j}},}  \\
\end{array}
\label{eq:J.3}
\end{equation}
%===============================================================================
where $\rho_{j}\!\equiv\!|w_j \rangle \langle w_j |$ are normalized decomposition states,
%===============================================================================
\begin{equation}%                  Equation J.4
\begin{array}{*{20}l}
   {|w_j \rangle} &\!\! {\equiv {\textstyle{1 \over {\sqrt {p_j } }}}|\widetilde{w}_j \rangle ,}  \\
\end{array}
\label{eq:J.4}
\end{equation}
%===============================================================================
with probabilities
%===============================================================================
\begin{equation}%                  Equation J.5
\begin{array}{*{20}l}
   {p_j} &\!\! {\equiv \langle \widetilde{w}_j |\widetilde{w}_j \rangle  = \sum_{k = 1}^R {\lambda _k |U_{j,k} |^2 }.}  \\
\end{array}
\label{eq:J.5}
\end{equation}
%===============================================================================
All of this is possible since elements of $U$ obey
%===============================================================================
\begin{equation}%                  Equation J.6
\begin{array}{*{20}l}
   {\sum_{j = 1}^D {U_{j,k}^* U_{j,l} }} &\!\! {= \delta _{k,l}} & {\text{and}} & {\sum_{k = 1}^D {U_{j,k}^* U_{l,k} }} &\!\! {= \delta _{j,l} .} \\
\end{array}
\label{eq:J.6}
\end{equation}
%===============================================================================

Thus, the convex-roof extension of any pure-state entanglement monotone $E(\rho)$ (see \App{App.C.2}) is
%===============================================================================
\begin{equation}%                  Equation J.7
\begin{array}{*{20}l}
   {\hat E(\rho )} &\!\! {\equiv \mathop {\min }\limits_{\{ U\} } \left( {\sum\nolimits_{j = 1}^D {p_j E(\rho _j )} }\! \right)\!,}  \\
\end{array}
\label{eq:J.7}
\end{equation}
%===============================================================================
where the $U$-dependence can be seen from
%===============================================================================
\begin{equation}%                  Equation J.8
\begin{array}{*{20}l}
   {\rho _j} &\!\! {= \frac{1}{p_{j}}\sum\limits_{k,l, = 1,1}^{R,R}\! {U_{j,k} U_{j,l}^* \sqrt {\lambda _k \lambda _l } } |e_k \rangle \langle e_l |,}  \\
\end{array}
\label{eq:J.8}
\end{equation}
%===============================================================================
and the $p_j$ depend on $U$ as seen in \Eq{J.5}.  Thus, the convex-roof extension of $E$ is the minimum average of $E$ over all unitaries $U$, and the decomposition constraint that $\rho=\sum\nolimits_{j}p_{j}\rho_{j}$ is automatically built-in, through $U$.

The minimum size of $U$ is $D\!=\!R$. Since \smash{$U_{j,k} U_{j,l}^*$} appear in $\rho_{j}$, this limits the degrees of freedom (DOF) added by $U$. Since the DOF of $R$ outer products of normalized orthogonal $R$-level pure states is the DOF of a pure $R$-level density matrix minus the $R\!-\!1$ DOF of its eigenvalues, the DOF of the smallest $U$ is $R^2 \!-\!1 \!-\!(R\!-\!1)\!=\!R^2 \! -\!R$.  Thus, \Eq{J.7} has a minimum of $R^2 -R$ variables to search.

For example, for $R\!=\!2$, let \smash{$U\equiv\binom{\;a\;\;b}{-b^* \; a^*}$} with $(a,b)\!\equiv\!(\cos(\theta),\sin(\theta)e^{i\chi})$,{\kern -1.5pt} and{\kern -1.5pt} $\theta  \!\in\! [0,{\textstyle{\pi  \over 2}}]${\kern -1.5pt} and{\kern -2pt} $\chi \! \in\! [0,2\pi )$.  Thus, we only need to search $\theta$ and $\chi$ at an acceptable resolution to find which decomposition minimizes the average $E$.

Rank-$3$ states have $6$ DOF, which is computationally intractable to search.  Thus, while convex-roof extension is well-defined, it is generally not practical to compute.
%                                End of App.J
%-------------------------------------------------------------------------------
\end{appendix}
%                               END of APPENDIX
%*******************************************************************************
%*******************************************************************************
%                                 BIBLIOGRAPHY
%
%                             END of BIBLIOGRAPHY
%*******************************************************************************
\end{document}